\documentclass[a4paper,11pt]{article}
\usepackage[utf8]{inputenc}
\usepackage{geometry}
\usepackage{amsmath}
\usepackage{amssymb}
\usepackage{graphicx}
\usepackage{booktabs}
\usepackage{graphicx} 
\usepackage{amsmath}  
\usepackage{cite} 
\usepackage{algorithm}
\usepackage{algpseudocode}
\usepackage{amsmath}
\usepackage{multicol}
\usepackage{subcaption}
\usepackage{amsthm}
\newtheorem{proposition}{Proposition}
\graphicspath{{Figures/}}
\usepackage{graphicx}
\usepackage{hyperref} 
\hypersetup{
    colorlinks=true,
    linkcolor=blue,
    filecolor=magenta,
    urlcolor=cyan,
    citecolor=blue,
}

\usepackage{authblk} 

\title{\textbf{Equation-Free Digital Twins for Nonlinear Structural Dynamics}}

\author[1]{Mohammad Mahdi Abaei\thanks{Corresponding author: mohammad.abaei@aalto.fi}}
\author[1]{Ahmad BahooToroody}
\author[1]{Arttu Polojärvi}
\author[1]{Heikki Remes}
\author[2]{Ulf Tyge Tygesen}
\author[1]{Mikko Suominen}
\author[3]{Michael Beer}

\affil[1]{Department of Mechanical Engineering, Marine Technology and Arctic Group, Aalto University, Finland}

\affil[2]{Independent Expert, Digital Twin Technology, Ribe, Denmark}
\affil[3]{Institute for Risk and Reliability, Leibniz University Hannover, Germany}

\date{}

\begin{document}

\maketitle

\begin{abstract}
Monitoring high-dimensional engineering structures in extreme environments is limited not only by nonlinearity but also by the combination of non-stationary excitation, nonlinear structural kinematics and complex stochastic forcing. Traditional model-based and black-box data-driven frameworks often fail to resolve such dynamic behaviour in real time or lead to the curse of dimensionality during sensor failure. This paper introduces a rank-optimized digital twin framework based on Koopman operator theory. It uses Hankel-matrix embeddings and dynamic mode decomposition to link high-fidelity physics with real-time  processing. By lifting operational data into a linear invariant subspace, the model gives an output-only reconstruction of the structural state without \textit{a priori} mass or stiffness matrices. A sufficient identifiability condition is derived for the modal amplitudes recovered from sparse, delay-embedded observations. It shows how the time-delay structure compensates for missing spatial information. The amplitudes are estimated by a Tikhonov-regularized projection. As a stochastic extension, the same identified operator is used in a Kalman form in which the unmeasured forcing enters as process noise.

The framework was validated on an NREL 5MW spar-buoy FOWT model, which is a demanding case of coupled aero-hydro-servo-elastic dynamics. It was also verified on basin-test data of a 1:70-scale IEA 15MW semi-submersible floating turbine. In the simulations, the identified tower modes agree with stochastic subspace identification and frequency-domain decomposition (MAC $\ge 0.99$). The deterministic 3P rotor harmonic is identified as a distinct DMD pole whose frequency changes proportionally with rotor speed. Identification from the output correlation functions gives more reliable damping estimates than response-domain identification. The simulated sensor responses are strongly correlated and have a low effective dimension; consequently, all tested linear estimators reconstruct the hidden channels accurately when five or more sensors remain. With one to three sensors the delay embedding becomes essential. The regularized projection then reaches $R^2 = 0.78$--$0.98$ and its Kalman form $0.91$--$0.99$, against $0.59$--$0.95$ for static least squares. The open-loop forecast skill decays as $R^2(h) \approx e^{-2\zeta_1 \omega_1 h}$, so the horizon over which the twin can bridge missing measurements follows from the identified damping. In the basin tests, Hankel-DMD identifies the first fore-aft and side-to-side tower modes at \(0.430\) and \(0.415\)~Hz, with mode shapes matching SSI-COV at \(\mathrm{MAC}=1.00\), and separates them from the rotor harmonics using the measured rotor speed. On unseen runs, the tower-base bending moment is reconstructed with \(R^2=0.83\)--\(0.89\) when three of the four sensors are available; with only the nacelle accelerometer, the Kalman form retains \(R^2=0.56\)--\(0.69\), within the coherence limit of the measured signals. The framework thus offers a computationally efficient and resilient structural-state layer for real-time digital twins of complex structural dynamics.
\end{abstract}

\noindent \textbf{Keywords:} Data-driven modelling, Operator-theoretic learning, Machine Learning, Dynamic Mode Decomposition, Koopman Theory, Digital Twin, System Identification, Nonlinear Dynamics

\section*{Nomenclature}
\setlength{\columnsep}{0.5cm}
\begin{multicols}{2}
\footnotesize
\noindent

\newcommand{\entry}[2]{%
  \noindent\makebox[0.25\linewidth][l]{#1}
  \parbox[t]{0.70\linewidth}{#2}\par\vspace{0.2em}
}


\entry{DMD}{Dynamic Mode Decomposition}
\entry{DOF}{Degree of Freedom}
\entry{EDMD}{Extended Dynamic Mode Decomposition}
\entry{FOWT}{Floating Offshore Wind Turbine}
\entry{OMA}{Operational Modal Analysis}
\entry{POD}{Proper Orthogonal Decomposition}
\entry{SHM}{Structural Health Monitoring}
\entry{SSI-COV}{Stochastic Subspace Identification}
\entry{SVD}{Singular Value Decomposition}
\entry{OLC}{Operational Load Cases}
\entry{FDD}{Frequency Domain Decomposition}
\entry{MAC}{Modal Assurance Criterion}
\entry{MPC}{Modal Phase Collinearity}

\entry{$\mathbf{A}$}{Full Koopman evolution operator}
\entry{$\tilde{\mathbf{A}}$}{Reduced evolution operator}
\entry{$\mathbf{b}_k$}{Modal amplitude vector at step $k$}
\entry{$\mathbf{C}$}{Damping/Coriolis matrix}
\entry{$\mathbf{C}_a$}{Output matrix of the active channels}
\entry{$\mathbf{K}_a$}{Steady-state Kalman gain}
\entry{$k_{\mathbf{C}}$}{Kruskal rank of $\mathbf{C}$}
\entry{$d$}{Time-delay embedding depth}
\entry{$\mathbf{f}$}{Nonlinear evolution map}
\entry{$\mathbf{F}_{aero}$}{Aerodynamic forcing vector}
\entry{$\mathbf{F}_{hydro}$}{Hydrodynamic forcing vector}
\entry{$g(\mathbf{x})$}{Scalar observable function}
\entry{$\mathbf{H}$}{Hankel matrix}
\entry{$\mathbf{H}_{1,2}$}{Time-shifted Hankel matrices}
\entry{$\mathcal{I}_{avail}$}{Indices of active sensors}
\entry{$\mathcal{K}$}{Infinite-dimensional Koopman operator}
\entry{$\mathbf{K}$}{Stiffness matrix}
\entry{$\mathbf{M}$}{Mass matrix}
\entry{$N$}{Total number of snapshots}
\entry{$p$}{Total sensors (Full state)}
\entry{$q$}{Active sensors ($q < p$)}
\entry{$\mathbf{q}$}{Generalized coordinates}
\entry{$r$}{Truncation rank}
\entry{$\mathbf{S}$}{Sensor selection matrix}
\entry{$\mathbf{u}$}{Generalized speeds}
\entry{$\mathbf{U}_r$}{Left singular vectors (POD)}
\entry{$\mathbf{V}_r$}{Right singular vectors}
\entry{$\mathbf{Q}, \mathbf{R}$}{Process and measurement noise covariances}
\entry{$\mathbf{V}_d$}{Vandermonde matrix of the delay structure}
\entry{$\mathbf{V}_W$}{Vandermonde matrix of the horizon $W$}
\entry{$\mathbf{w}_j$}{Eigenvectors of $\tilde{\mathbf{A}}$}
\entry{$W$}{Rolling horizon window}
\entry{$\mathbf{y}(t)$}{Sensor measurements}
\entry{$\tilde{\mathbf{y}}_k$}{Time-delay embedded state}
\entry{$\tilde{\mathbf{y}}^{obs}_k$}{Sparse observation vector}
\entry{$\hat{\mathbf{Y}}_{rec}$}{Reconstructed trajectory}

\entry{$\Delta t$}{Sampling interval}
\entry{$\zeta_j$}{Damping ratio}
\entry{$\lambda_j$}{Continuous eigenvalue}
\entry{$\mu_j$}{Discrete eigenvalue}
\entry{$\mathbf{\Lambda}$}{Eigenvalue matrix}
\entry{$\mathbf{\Sigma}_r$}{Singular value matrix}
\entry{$\mathbf{\Phi}$}{Full Koopman modes}
\entry{$\mathbf{\Phi}_{aug}$}{Augmented mode matrix}
\entry{$\mathbf{\Phi}_{phys}$}{Physical mode shapes}
\entry{$\mathbf{\Phi}_{sparse}$}{Sparse mode basis}
\entry{$\mathbf{\Psi}$}{Projection onto modal coordinates, $\mathbf{\Phi}^{\dagger}$}
\entry{$\alpha$}{Tikhonov regularization parameter}
\entry{$\kappa$}{Condition number of $\mathbf{\Phi}_{sparse}$}
\entry{$\omega_j$}{Angular frequency}
\entry{$\sigma_i$}{Standard deviation (sensor)}
\entry{$\dagger$}{Pseudoinverse operator}

\end{multicols}

\section{Introduction}

Ensuring the safety and reliability of engineering structures in harsh environments, typical of aerospace and marine applications, relies increasingly on structural health monitoring (SHM). Monitoring has traditionally been performed using networks of sensors installed at critical locations to track structural behaviour. However, under extreme conditions, physical sensors are prone to failure due to moisture ingress, signal drift and mechanical damage, potentially leaving operators with dangerous blind spots. Dense sensor networks are also often impractical to deploy. This has motivated virtual-sensing approaches, in which a limited set of measurements is combined with numerical algorithms to reconstruct unmeasured structural responses \cite{kim2024}. Developing such approaches as part of a digital-twin framework remains challenging. Modern structures can be highly flexible and are subject to nonlinear dynamics, stochastic forcing and changing operating conditions, while real-time monitoring requires computationally efficient models.

Conventional digital-twin implementations have commonly relied on model-based estimators such as the Kalman filter and its augmented variants (KF and AKF) \cite{xie2025, wei2025, ammerman2025, brijder2023}. Their performance depends on the accuracy of the underlying dynamical model and on the assumed process and measurement statistics. They can degrade as physical assets drift away from the calibrated model \cite{simpson2026}. More general probabilistic approaches, including particle filters, unscented Kalman filters and Gaussian-process latent-force models, can accommodate additional uncertainty but may become computationally demanding for high-dimensional systems \cite{ozan2025, bilbao2022}. This creates a practical trade-off between model fidelity and computational cost in real-time applications \cite{madushele2026}. Model updating can reduce discrepancies between a digital representation and measured data, but excessive tuning may also mask deficiencies in the assumed model or compensate for changes that should instead be interpreted as structural or operational effects \cite{moynihan2023}.

Data-driven identification methods provide an alternative route. Stochastic subspace identification (SSI-COV), modal expansion and related techniques have been widely used for operational modal analysis and response reconstruction. Their performance, however, can deteriorate when the measured response contains strong deterministic harmonics, narrow-band excitation or substantial coupling between structural and operational dynamics \cite{song2025, oliveira2018, weil2025, liu2023, Galvan2025}. In rotating machinery, for example, operational harmonics may be difficult to distinguish from structural resonances \cite{liu2020}. Modal expansion \cite{vettori2023}, on the other hand, normally requires a prior finite-element or modal model and sufficient spatial information to estimate the retained modal coordinates. These limitations motivate data-driven representations that exploit both the spatial and temporal structure of the measurements without requiring an explicit structural model.

Structural monitoring therefore increasingly uses spatio-temporal representations identified directly from measured response data \cite{palma2025fast}. Classical autoregressive and state-space models remain useful when a fixed linear representation is adequate \cite{liu2020}, but may become restrictive when the response changes with operating condition or exhibits strong nonlinear coupling \cite{serani2023}. Machine-learning approaches provide another alternative. Recent reviews of data-assimilation frameworks report high predictive capability for complex dynamical systems \cite{grande2026, dang2026}. Their use in SHM nevertheless introduces other challenges. Convolutional neural networks impose locality and Euclidean structure that may not naturally represent sparse and irregular sensor layouts \cite{kantamneni2024}. Graph-based and attention-based architectures may make weak or high-frequency features less visible, depending on their architecture and training \cite{duthe2025}. More generally, unconstrained black-box models may produce responses that are difficult to interpret physically \cite{shahin2025, taze2025, rajic2025}. Even physics-enhanced models remain sensitive to uncertain environmental and operational conditions \cite{Pitchforth2021}. These are the conditions in which digital-twin concepts and large monitoring datasets are expected to support damage detection and life extension \cite{Tygesen2021}. Reported gains should also be weighed against the strength of the baselines used for comparison \cite{McGreivy2024}. Related to these issues, Sanchis-Agudo et al. \cite{sanchis2025} showed that the self-attention layers of transformer architectures converge to an approximately linear state-transition operator. This suggests that operator-theoretic frameworks can match the predictive performance of modern machine-learning approaches with greater interpretability. These issues motivate representations that remain data-driven while retaining an interpretable dynamical structure.

Koopman operator theory provides such a framework by representing nonlinear dynamical systems through the linear evolution of observables. The idea originates from Koopman's operator-theoretic representation of dynamical systems \cite{koopman1931}. Mezi\'c subsequently developed the spectral formulation of the Koopman operator for nonlinear dynamical systems, model reduction and decomposition \cite{mezic2005nd}, and later related Koopman spectral objects to the geometry of the underlying state space \cite{mezic2020geometry}. Rowley et al. \cite{rowley2009} established the connection between Koopman modes and data-driven modal decompositions, while Schmid \cite{schmid2010} formulated dynamic mode decomposition (DMD) for numerical and experimental snapshot data. Williams et al. \cite{williams2015} extended the approximation through dictionaries of observables in extended DMD (EDMD). For time-delay coordinates, Arbabi and Mezi\'c \cite{arbabi2017} established the convergence of DMD applied to Hankel data matrices under ergodic assumptions. These developments provide the operator-theoretic basis for the delay-embedded representation used in the present work. Within this framework the focus shifts from the nonlinear state space to the linear evolution of observables, so that a standard eigen-decomposition characterizes the dynamics. The resulting model is data-driven and equation-free. It avoids the cost of solving explicit governing equations and retains the interpretability of classical modal analysis \cite{kutz2016, kutz2020}. In engineering applications, Koopman embeddings and DMD have been applied to fluid dynamics \cite{dai2022}, power forecasting \cite{liew2022} and rigid-body dynamics \cite{serani2023}, providing reduced-order representations that retain modal interpretability while being identified directly from data.

This paper presents an operator-theoretic identification framework for high-dimensional flexible non-linear structures. It addresses the coupled dynamics of active control systems, random vibration loads and rotating machinery. Although the framework is generalizable, we demonstrate it for floating offshore wind turbines (FOWT).

Developing a real-time digital twin for FOWT is a challenging test for any SHM approach with sparse instrumentation. These structures operate under extreme stochastic forcing and exhibit high-dimensional structural coupling. Both commonly cause traditional physics-based models to fail, which shifts their modelling towards data-driven paradigms \cite{serani2023}. Reduced-order modelling of a FOWT is also challenging because of hydrodynamic memory effects, distributed wave loading and highly coloured wind--wave spectra \cite{duarte2014, caglio2025, simpson2026}. Further challenges include coupled aero-hydro-servo-elastic dynamics and active control that induces negative damping \cite{krathe2025}. The Koopman framework and DMD have been applied to ship manoeuvring and slamming loads \cite{serani2023, palma2025fast}. However, they have never been investigated for systems with the coupled dynamics and 1P/3P rotor harmonics of FOWTs \cite{liu2020}. Existing DMD variants also lack a mechanism for chaotic divergence and often fail to forecast irregular waves outside their training region \cite{lydon2025}. At fleet scale, typically only about 10\% of the turbines in an offshore farm are equipped with SHM sensors \cite{SHM2024}. A scalable digital twin must therefore be resilient to local sensor failures \cite{kim2024} and also able to close this fleet-wide monitoring gap by using validated models for unmonitored assets.

We introduce an equation-free digital twin framework based on a rank-optimized Koopman-Hankel embedding. By learning the linear invariant subspace of the system directly from operational data, the model removes the need for \textit{a priori} mass and stiffness matrices and keeps its physical interpretability. The work departs from the input-output forward-forecasting paradigms common in the current operator-theoretic literature \cite{lydon2025, palma2025fast}. Instead, we formulate an autonomous, output-only inverse reconstruction. This output-only architecture enables virtual sensing at critical hotspots without auxiliary environmental sensors. Rather than computationally intensive repeated statistical resampling, the framework uses a deterministic spectral rank-optimization strategy. The truncation rank is selected on held-out data, and the modal amplitudes are recovered by a regularized projection, which reduces the computational cost for real-time processing. Beyond numerical accuracy, the framework links the identified model to the underlying coupled aero-hydro-servo-elastic phenomena. We show that the rank-optimized representation separates the structural resonances from the deterministic rotor harmonics (3P and its multiples) that commonly contaminate vibration measurements. The harmonics can be classified as such from their dependence on rotor speed. We also quantify the predictability horizon of the resulting estimator. Because the operator identified from output data describes the conditional-mean evolution of the response, its open-loop forecast skill decays at a rate set by the identified  structural-mode damping
 and by the unmeasured forcing \cite{majda2018}. This provides a basis for determining the data-assimilation frequency required to keep the digital twin synchronized under unseen environmental forcing.

The main contribution is a single output-only Hankel-DMD operator that provides operational modal identification, sparse-sensor reconstruction under changing sensor layouts without retraining, and a damping-based forecast horizon, together with an explicit sparse-delay identifiability condition and a stochastic Kalman realization. Classical Hankel-DMD identifies the Koopman spectrum from delay-embedded data \cite{arbabi2017, Brunton2017HAVOK} but provides neither an estimator for missing channels nor an identifiability analysis for one. Compressed and missing-data variants of DMD \cite{Brunton2015CS} address reconstruction from incomplete measurements, while gappy POD \cite{GappyPOD} reconstructs missing snapshots from a static low-rank spatial basis. Neither formulation provides the sparse-delay modal-amplitude identifiability condition developed here for changing sensor-failure patterns. Koopman-based observers \cite{Surana2016} generally rely on dictionaries of observables or on measured inputs. Modal-expansion and Kalman-filter virtual sensing for offshore wind turbines \cite{Iliopoulos2016, Maes2016, Lourens2012}, like Gaussian-process latent force models \cite{Rogers2020}, require a calibrated structural model. The contributions of this paper are: (i) a single output-only Hankel-DMD operator that identifies operational modal parameters and reconstructs unmeasured channels for any pattern of failed sensors without retraining; (ii) a sufficient identifiability condition for recovering modal amplitudes from sparse delay-embedded data, showing how time-delay information compensates for missing sensor information, together with an analysis of inversion conditioning; (iii) a stochastic-Koopman treatment of unmeasured forcing, using correlation-domain identification to reduce the damping bias of response-domain identification and, as an extension of the deterministic projection, a Kalman form of the same operator with noise covariances estimated from the same data; (iv) an assessment on spar FOWT simulations using time-blocked validation, three baseline estimators, systematic changes in sensor number and location, and robustness tests, together with experimental verification using public basin-test data from a 1:70-scale semi-submersible floating wind turbine; and (v) a predictability horizon for the virtual sensor obtained from the identified damping.

The remainder of this paper is organized as follows: \textbf{Section 2} details the OpenFAST environment and the Hankel-DMD mathematical formulation. \textbf{Section 3} describes the operational dataset, the sensor configuration and the preprocessing, and \textbf{Section 4} presents the system identification and virtual sensing results on the simulated spar FOWT. \textbf{Section 5} verifies the framework on basin-test measurements, and \textbf{Section 6} concludes with the implications for real-time SHM and fatigue life estimation. Supporting tables and sensitivity results are collected in Appendix~\ref{app:supporting}.

\section{Methodology}

\begin{figure}[htbp]
    \centering
    \includegraphics[trim={0.2cm 0.05 0 0.05}, clip, width=0.4\textwidth]{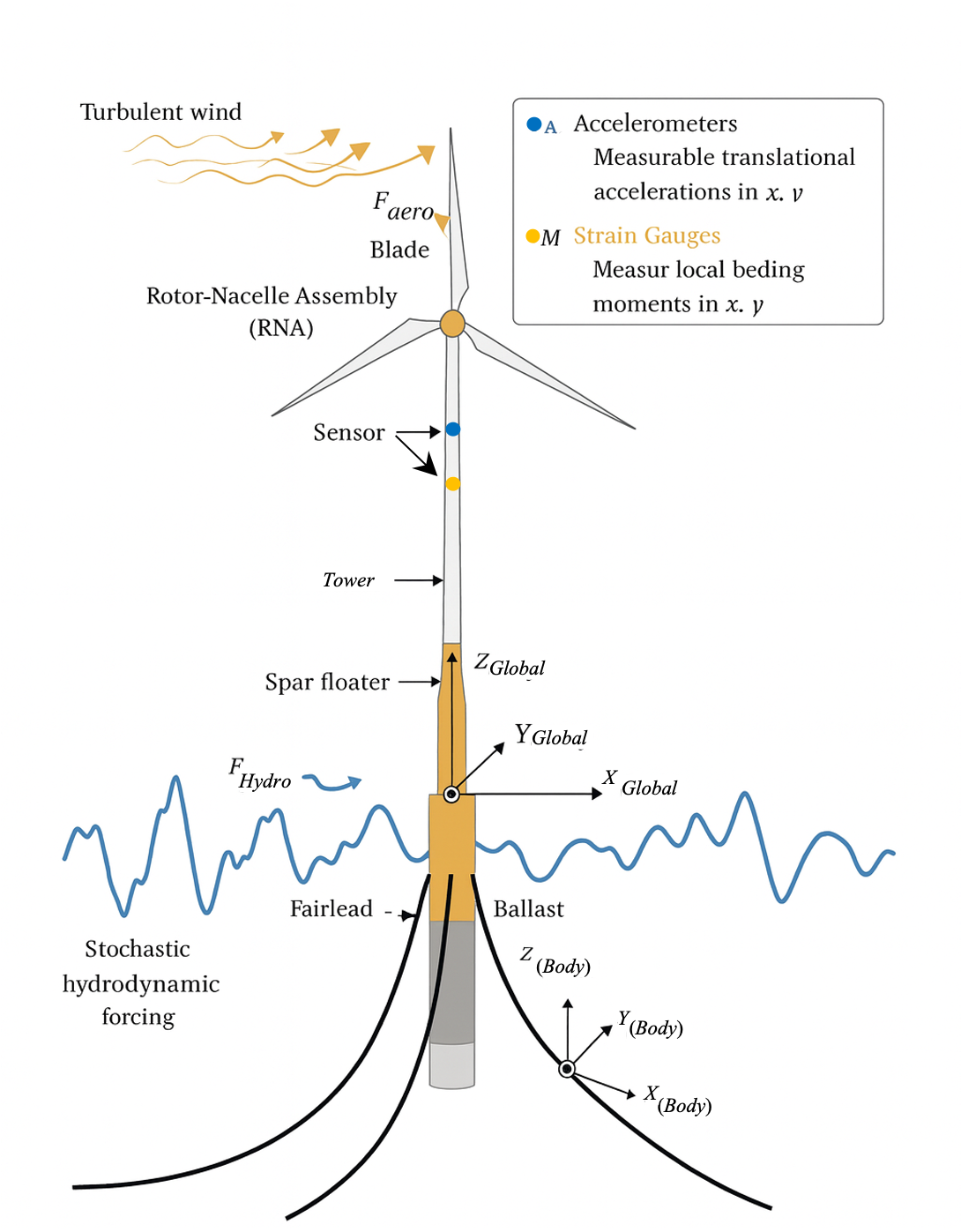}
    \caption{Schematic of the spar FOWT and the coupled aero-hydro-servo-elastic system modelled in OpenFAST.}
    \label{fig:OC3HywindSchematic}
\end{figure}

To develop the digital twin, a training dataset was generated with \textbf{OpenFAST} for the NREL 5MW wind turbine on the OC3 Hywind spar platform \cite{openfast_doc}. The simulation setup was designed to capture the irregular, coupled nature of the offshore environment, which is essential for identification. A schematic of the modelled coupled aero-hydro-servo-elastic system is presented in Figure~\ref{fig:OC3HywindSchematic}.

\subsection{Governing Dynamics }
The FOWT dynamic behaviour is governed by the multi-body equations of motion derived via Kane’s method \cite{jonkman2005fast}:

\begin{equation}
\label{eq:motion}
\mathbf{M}(\mathbf{q}, t) \dot{\mathbf{u}} + \mathbf{C}(\mathbf{q}, \mathbf{u}, t) \mathbf{u} + \mathbf{K}(\mathbf{q}, t) \mathbf{q} = \mathbf{F}_{aero} + \mathbf{F}_{hydro} + \mathbf{F}_{mooring}
\end{equation}

\noindent where $\mathbf{q}$ and $\mathbf{u}$ represent the generalized coordinates and speeds, respectively. Unlike linear formulations, the mass, damping, and stiffness matrices ($\mathbf{M}, \mathbf{C}, \mathbf{K}$) are time-varying, driven by the large-angle rigid-body motion of the platform and the state-dependent nature of the aerodynamic and hydrodynamic loads.

Equation~\eqref{eq:motion} should be read as a schematic summary of the structural equations solved in ElastoDyn rather than as the governing equation of the coupled system actually integrated. In OpenFAST the coupled problem is advanced as interconnected modules carrying their own states, $\mathbf{x} =
[\mathbf{x}_{ED},\mathbf{x}_{AD},\mathbf{x}_{HD},
\mathbf{x}_{SrvD},\mathbf{x}_{moor}]$ for the structure, unsteady aerodynamics, radiation memory, controller and mooring dynamics. The aerodynamic, hydrodynamic and mooring loads on the right-hand side of Equation~\eqref{eq:motion} are therefore state- and input-dependent operators, not equivalent constant matrices. For the present work, this distinction matters only because the data-driven model must reproduce the full coupled behaviour without access to any of these matrices. The modules, solver options, enabled degrees of freedom and structural settings used to generate the data are listed in Table~\ref{tab:openfast_config} of Appendix~\ref{app:supporting} so that the data generation can be reproduced.

To ensure that the data-driven model captures these nonlinearities, the forcing terms in Equation~\eqref{eq:motion} were selected to excite the full frequency spectrum. Full wave loading was applied via the $\mathbf{F}_{hydro}$ term using a JONSWAP spectrum ($0.05 - 0.25$ Hz). As noted by Robertson et al. \cite{Robertson2014}, the tower side-to-side (SS) mode lacks significant aerodynamic damping compared with the fore-aft direction, and it needs to be identified accurately. Consequently, wave-induced rolling is the dominant energy source required to excite and identify this mode. Neglecting it would give non-physical damping estimates. Turbulent wind fields generated by TurbSim provided broadband aerodynamic excitation via $\mathbf{F}_{aero}$. This keeps the structural modes continuously perturbed and approximates the white-noise excitation assumed for robust spectral identification.

To reconstruct the system phase space without accessing internal states, the tower was instrumented with a distributed array of virtual sensors. The state vector is constructed from bending moments, which serve as a direct proxy for stress and fatigue, and absolute accelerations, which capture the total inertial response including the coupling with the platform's rigid-body motion. This selection allows the Hankel-DMD framework to learn the operational physics directly from physical outputs, without the unmeasurable internal generalized coordinates.

\subsection{Linearization}

Nonlinear time-domain simulations do not output modal properties explicitly. A theoretical baseline is therefore established with the OpenFAST linearization module. The nonlinear dynamics are linearized about a steady-state equilibrium, which gives a linear state-space model of the evolution of perturbations $\Delta \mathbf{x}$:

\begin{equation}
\label{eq:linearization}
\Delta \dot{\mathbf{x}} = \mathbf{A} \Delta \mathbf{x} + \mathbf{B} \Delta \mathbf{u}_{in}
\end{equation}

\noindent where $\Delta \mathbf{u}_{in}$ represents the perturbation inputs from aerodynamic, hydrodynamic, and control loads. The system dynamics are governed by the state matrix $\mathbf{A}$, which has a standard block structure:

\begin{equation}
\label{eq:Amat}
    \mathbf{A} =
    \begin{bmatrix}
        \mathbf{0} & \mathbf{I} \\
        -\mathbf{M}^{-1}\mathbf{K} & -\mathbf{M}^{-1}\mathbf{C}
    \end{bmatrix}
\end{equation}

\noindent In this formulation, OpenFAST outputs the stiffness ($\mathbf{K}$) and damping ($\mathbf{C}$) pre-multiplied by the inverse mass matrix ($\mathbf{M}^{-1}$), not as separate matrices. The theoretical natural frequencies and mode shapes then follow from the eigenvalue problem:

\begin{equation}
\label{eq:eigen}
\mathbf{A} \mathbf{v} = \lambda \mathbf{v}
\end{equation}

\noindent The resulting eigenvalues ($\lambda$) and eigenvectors ($\mathbf{v}$) provide an \textbf{independent numerical reference} for this study. They are not treated as exact ground truth for the operational response. The linearization corresponds to one particular operating point, whereas the time-domain simulations contain stochastic wind and wave excitation. Comparing the data-driven Hankel-DMD predictions with this reference quantifies the ability of the digital twin to capture the fundamental physics.

\subsection{Mathematical Framework}\label{sec:framework}

The FOWT is modelled as a dynamical system evolving on a high-dimensional nonlinear manifold $\mathcal{M}$. OpenFAST solves the coupled equations of motion accurately, but the computation is expensive. The system can be viewed as a discrete-time map $\mathbf{f}: \mathcal{M} \to \mathcal{M}$, where the state $\mathbf{x}_k$ evolves as \cite{kutz2016}:

\begin{equation}
    \mathbf{x}_{k+1} = \mathbf{f}(\mathbf{x}_k)
\end{equation}

To enable real-time prediction, Koopman operator theory shifts the perspective from the state space $\mathcal{M}$ to the \textbf{space of observables}. It defines an infinite-dimensional linear operator $\mathcal{K}$ that acts on scalar observable functions $g: \mathcal{M} \to \mathbb{R}$ (such as sensor measurements), satisfying \cite{kutz2016, mezic2005}:

\begin{equation}
    \mathcal{K}g(\mathbf{x}_k) = g(\mathbf{f}(\mathbf{x}_k)) = g(\mathbf{x}_{k+1})
\end{equation}

\begin{figure}[h]
    \centering
    \includegraphics[width=0.8\textwidth]{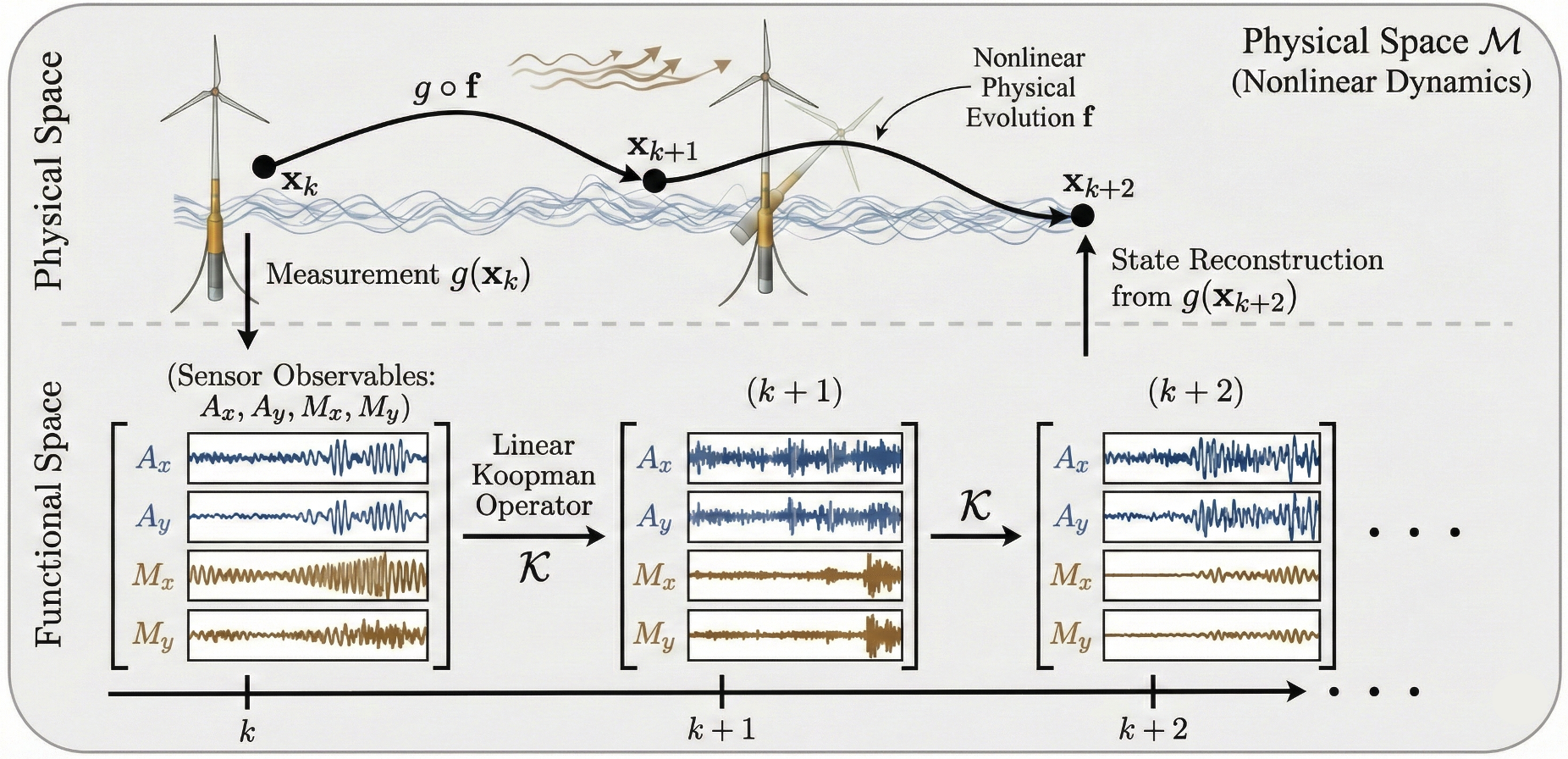}
    \caption{The evolution of states $\mathbf{x}$ in the vector space $\mathcal{M}$ (top layer) and observables $g(\mathbf{x})$ in the functional space (bottom layer). The Koopman operator $\mathcal{K}$ enables linear evolution of the observables, bypassing the nonlinear map $\mathbf{f}$.}
    \label{fig:koopman_commutative}
\end{figure}

This relationship, illustrated in Figure~\ref{fig:koopman_commutative}, establishes a commutative property: evolving the complex state nonlinearly on the low-dimensional structure is equivalent to measuring the state first and evolving the observations linearly. In the discrete formulation adopted here, the continuous-time flow of the system is replaced by the map $\mathbf{f}$, and the infinite-dimensional operator $\mathcal{K}$ is approximated by the finite matrix $\mathbf{A}$ identified by Hankel-DMD.

In practice, the full state $\mathbf{x}$ is unknown. Instead, we define a vector of available sensor measurements. The nonlinear dynamics in the state space can then be ``lifted'' to a linear evolution in the space of observables. The algorithm seeks a finite-dimensional approximation of the Koopman operator, denoted $\mathbf{A}$, which governs the time evolution of the available sensor measurements $\mathbf{y}_k \in \mathbb{R}^p$:

\begin{equation}
    \mathbf{y}_{k+1} \approx \mathbf{A} \mathbf{y}_k
\end{equation}

\paragraph{Applicability under stochastic forcing.}\label{sec:stochastic_koopman}
The FOWT is not autonomous. Its evolution is $\mathbf{x}_{k+1} = \mathbf{f}(\mathbf{x}_k, \mathbf{w}_k)$, with $\mathbf{w}_k$ the unmeasured turbulence and wave excitation, so the autonomous form adopted above requires justification. For stationary and ergodic forcing, the relevant object is the stochastic Koopman operator $\mathcal{K}g(\mathbf{x}) = \mathbb{E}\left[ g(\mathbf{f}(\mathbf{x}, \mathbf{w})) \mid \mathbf{x} \right]$ \cite{mezic2005nd, crnjaric2020}. Its spectrum can be estimated from a single long trajectory by Hankel-DMD \cite{arbabi2017, crnjaric2020}. Time-delay embedding remains valid for forced and stochastic systems under the embedding theorems of Stark et al. \cite{stark2003}. Under coloured excitation, the identified output can contain both structural poles and excitation-related poles. Deterministic rotor harmonics appear as excitation-related spectral components whose frequencies track the rotor speed; their fitted decay rates should therefore not be interpreted as structural damping. This provides the classification criterion used throughout the paper: structural poles remain approximately invariant within a given operating regime, whereas rotor-harmonic poles shift with rotor speed. Two consequences follow for the interpretation of the results. First, the operator identified from response trajectories inherits the particular realization of the forcing. Part of this realization is absorbed by the least-squares fit and biases the damping estimates low. Applying the same decomposition to the output correlation functions $\mathbf{R}_k = \mathbb{E}[\mathbf{y}_{t+k} \mathbf{y}_t^\top]$, which evolve under the conditional-mean dynamics, removes that realization (Section~\ref{sec:damping}). Second, since the identified operator propagates $\mathbb{E}[\tilde{\mathbf{y}}_{k+h} \mid \tilde{\mathbf{y}}_k]$, its open-loop predictions must lose skill as the unmeasured forcing accumulates. Reconstruction under fixed conditions and genuine input-free prediction are therefore distinguished throughout. The virtual sensor of Section~\ref{sec:virtual_sensing} re-estimates the modal amplitudes from measurements and acts as an observer, whereas the input-free forecast skill is quantified separately in Section~\ref{sec:horizon}.

\subsubsection{Time-Delay Embedding }

In complex aero-hydro-elastic systems, the instantaneous sensor vector $\mathbf{y}_k$ (dimension $p$) is often insufficient to capture the full system state due to unmeasured variables (e.g., wake memory). This violates the assumption required for linear modelling.

Motivated by delay-embedding theory for forced and stochastic systems \cite{stark2003}, the observation vector is augmented by stacking $d$ time-shifted copies of the data into a single column. This creates a ``tall'' vector that encodes the trajectory history. Let $\mathbf{y}_k \in \mathbb{R}^p$ denote the multivariate state vector at time step $k$, containing measurements from all $p$ sensors simultaneously. For a system with $p$ sensors and a delay depth of $d$, the augmented state vector $\tilde{\mathbf{y}}_k$ is constructed as:

\begin{equation}
    \tilde{\mathbf{y}}_k = \left[ \mathbf{y}_k^\top, \; \mathbf{y}_{k+1}^\top, \; \dots, \; \mathbf{y}_{k+d-1}^\top \right]^\top \in \mathbb{R}^{pd}
\end{equation}

Here $k$ indexes the start of an offline Hankel column. For online virtual sensing the same $d$-sample history is re-indexed by its final sample, $\tilde{\mathbf{y}}_k = [\mathbf{y}_{k-d+1}^\top, \dots, \mathbf{y}_k^\top]^\top$, so that the estimator uses no sample later than the instant it reports (Section~\ref{sec:rolling}).

Note that each block entry $\mathbf{y}_k$ is itself a vector of dimension $p$. Let $\Delta t$ denote the sampling interval. The discrete snapshot vector at time step $k$ is defined as $\mathbf{y}_k \equiv \mathbf{y}(t_k)$, and the subsequent snapshot at $t_{k+1} = t_k + \Delta t$ is denoted as $\mathbf{y}_{k+1}$. Rows 1 to $p$ therefore hold the current snapshot, and rows $p+1$ to $2p$ hold the next time frame $\left(t_k + \Delta t\right)$.

The full Hankel matrix $\mathbf{H}$ is constructed by arranging these augmented vectors as columns \cite{palma2025fast}. The number of columns $m$ follows from the number of snapshots and the delay depth ($m = N - d + 1$):

\begin{equation}
    \mathbf{H} = \begin{bmatrix}
    | & | & & | \\
    \tilde{\mathbf{y}}_1 & \tilde{\mathbf{y}}_2 & \dots & \tilde{\mathbf{y}}_{m} \\
    | & | & & |
    \end{bmatrix}
    = \begin{bmatrix}
    \mathbf{y}_1 & \mathbf{y}_2 & \dots & \mathbf{y}_{m} \\
    \mathbf{y}_2 & \mathbf{y}_3 & \dots & \mathbf{y}_{m+1} \\
    \vdots & \vdots & \ddots & \vdots \\
    \mathbf{y}_d & \mathbf{y}_{d+1} & \dots & \mathbf{y}_{m+d-1}
    \end{bmatrix} \in \mathbb{R}^{pd \times m}
\end{equation}

To identify the dynamics, this matrix is split into two time-shifted matrices, $\mathbf{H}_1$ (the current state) and $\mathbf{H}_2$ (the future state):

\begin{equation}
    \mathbf{H}_1 = \mathbf{H}[:, 1:m-1], \quad \mathbf{H}_2 = \mathbf{H}[:, 2:m]
\end{equation}

Here, $\mathbf{H}_1$ contains the snapshots from time $1$ to $m-1$, and $\mathbf{H}_2$ contains the snapshots from time $2$ to $m$.

The objective is to identify a linear operator $\mathbf{A}$ satisfying $\mathbf{A} \mathbf{H}_1 \approx \mathbf{H}_2$. To avoid the computational cost and overfitting risks of the high-dimensional operator, a reduced proxy $\tilde{\mathbf{A}}$ is computed by projecting the dynamics onto the Proper Orthogonal Decomposition (POD) modes of the input snapshots ($\mathbf{H}_1 \approx \mathbf{U}_r \mathbf{\Sigma}_r \mathbf{V}_r^*$). This projection restricts the operator to the reduced subspace and gives the final computational form by a similarity transformation \cite{kutz2020}:

\begin{equation} \label{eq:A_tilde}
    \tilde{\mathbf{A}} = \mathbf{U}_r^* \mathbf{A} \mathbf{U}_r \approx \mathbf{U}_r^* \left( \mathbf{H}_2 \mathbf{H}_1^{\dagger} \right) \mathbf{U}_r = \mathbf{U}_r^* \mathbf{H}_2 \mathbf{V}_r \mathbf{\Sigma}_r^{-1}
\end{equation}

\begin{figure}[h]
    \centering
    \includegraphics[width=0.72\textwidth]{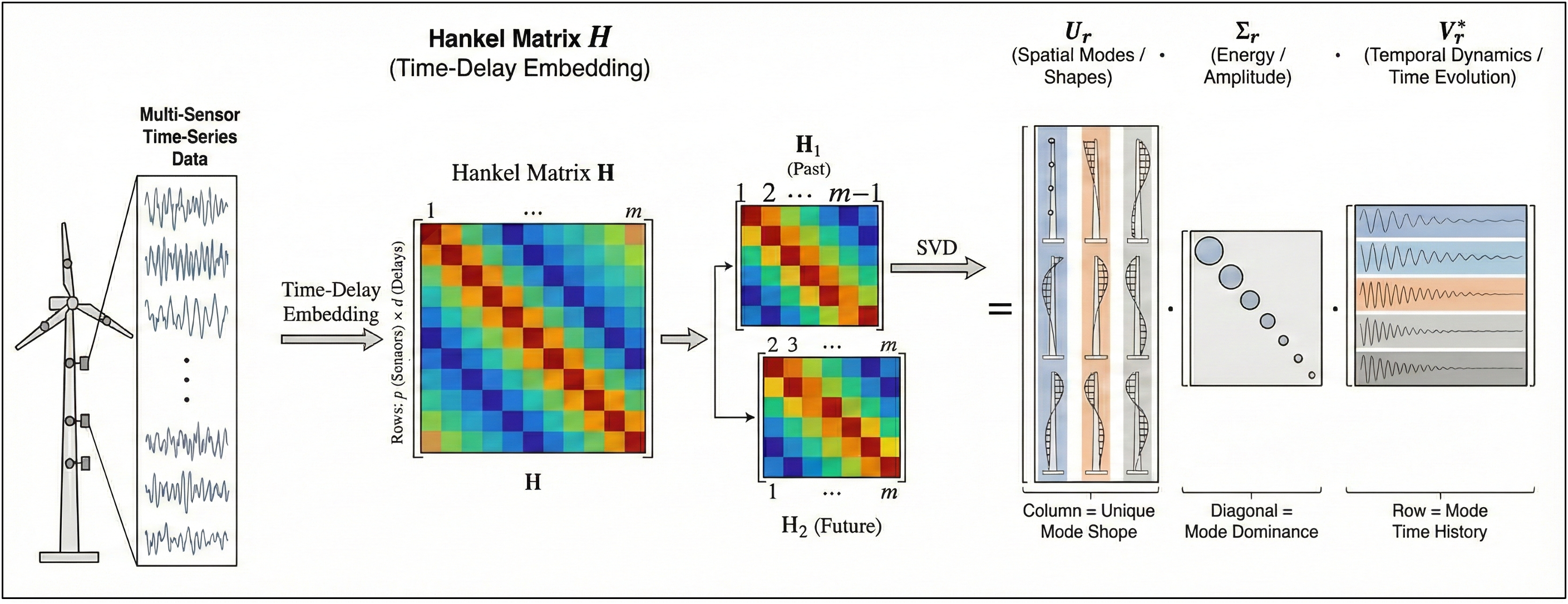}
    \caption{Extraction of spatiotemporal coherent structures from wind turbine sensor data via Hankel-matrix factorization.}
    \label{fig:Hankel-Plot}
\end{figure}

The Moore--Penrose pseudoinverse is obtained directly from the truncated SVD as $\mathbf{H}_{1}^{\dagger} \approx \mathbf{V}_r \mathbf{\Sigma}_r^{-1} \mathbf{U}_r^{*}$. This is DMD applied to time-delay coordinates (Hankel-DMD), with the POD basis providing the reduced subspace in which the underlying attractor is represented, as illustrated in Figure~\ref{fig:Hankel-Plot}.

\subsubsection{Spectral Decomposition }

The resulting matrix $\tilde{\mathbf{A}} \in \mathbb{C}^{r \times r}$ is the reduced evolution operator. Its eigen-decomposition, $\tilde{\mathbf{A}} \mathbf{w}_j = \mu_j \mathbf{w}_j$, gives the discrete eigenvalues $\mu_j$, which characterize the temporal evolution of the system per time step.

For physical interpretation, these discrete values are mapped to continuous-time eigenvalues $\lambda_j$ via the sampling interval $\Delta t$:

\begin{equation}
    \lambda_j = \frac{\ln(\mu_j)}{\Delta t} = \sigma_j + i\omega_j
    \label{eq:dmd_eigenvalue}
\end{equation}

From these, the natural frequency ($f_j = \omega_j / 2\pi$) and damping ratio ($\zeta_j = -\sigma_j / |\lambda_j|$) are derived.

The eigenvectors $\mathbf{w}_j$ obtained from the reduced operator $\tilde{\mathbf{A}}$ describe the dynamics in the low-dimensional subspace defined by the POD modes $\mathbf{U}_r$. In the exact DMD formulation, the Koopman modes ($\mathbf{\Phi}_{aug}$) are recovered in the original delay-embedded measurement space as:

\begin{equation}
    \mathbf{\Phi}_{aug} = \mathbf{H}_2 \mathbf{V}_r \mathbf{\Sigma}_r^{-1} \mathbf{W}
    \label{eq:DMD-modes}
\end{equation}

The projected alternative, $\mathbf{\Phi}_{aug} = \mathbf{U}_r \mathbf{W}$, is restricted to the POD subspace of $\mathbf{H}_1$. Both formulations share the same eigenvalues and therefore give identical frequencies and damping ratios, but they differ in the reconstructed mode shapes. The exact modes are used throughout this work, because the delay-shift structure exploited in Section~\ref{sec:virtual_sensing} is a property of the exact modes. The decomposition components play distinct roles. The POD modes ($\mathbf{U}_r$) are the principal orthogonal building blocks of the system's energy. They represent generic spatial patterns derived purely from the data variance. The reduced eigenvectors ($\mathbf{W}$) give the linear combination that forms each single-frequency dynamic mode. The temporal evolution is governed entirely by the diagonal matrix $\mathbf{\Lambda} = \text{diag}(\mu_1, \dots, \mu_r)$, which drives the signal reconstruction ($\mathbf{b}_{k+1} = \mathbf{\Lambda} \mathbf{b}_k$).

For the $j$th mode, the augmented eigenvector encodes its evolution across the delay window as a stack of time-shifted projections scaled by powers of the discrete eigenvalue $\mu_j$:
\begin{equation}
    \boldsymbol{\Phi}_{\mathrm{aug},j} =
    \big[ \boldsymbol{\phi}_{j}^\top, \; (\mu_j \boldsymbol{\phi}_{j})^\top, \dots, \; (\mu_j^{d-1} \boldsymbol{\phi}_{j})^\top \big]^\top
    \label{eq:phi_aug_stack}
\end{equation}
Since this structure contains the full history, the static \textbf{physical mode shape} is obtained by keeping the first $p$ rows of the vector:
\begin{equation}
    \mathbf{\Phi}_{phys} = \mathbf{\Phi}_{aug}[1:p, :]
\end{equation}

Online reconstruction requires the block associated with the \textit{most recent} sample of the delay window, denoted $\mathbf{\Phi}_{L}
=
\mathbf{\Phi}_{aug}[(d-1)p+1 : dp, \, :]
=
\mathbf{\Phi}_{phys}\mathbf{\Lambda}^{d-1}$,
so that for the $j$th mode $\boldsymbol{\phi}_{L,j}=\mu_j^{d-1}\boldsymbol{\phi}_j$. Because successive delay blocks of a single mode differ only by powers of $\mu_j$, either block gives the same normalized spatial mode shape. The block $\mathbf{\Phi}_{phys}$ is therefore reported as the mode-shape reference, while $\mathbf{\Phi}_{L}$ is used in the online estimator to keep causal alignment with the current measurement.


\subsubsection{Projections on Koopman Eigenfunctions and Signal Reconstruction}

With the physical mode shapes $\mathbf{\Phi}_{phys}$ and discrete eigenvalues $\mu_j$ identified, we can now formulate the signal reconstruction. In Koopman operator theory, any scalar observable $g: \mathcal{M} \to \mathbb{R}$ (e.g., a single sensor measurement) can be expanded in the basis of Koopman eigenfunctions.

In the continuous formulation, an observable $\psi(\mathbf{x})$ is projected as \cite{kutz2016}:
\begin{equation}
    \psi(\mathbf{x}) = \sum_{j=1}^{\infty} v_j \varphi_j(\mathbf{x})
\end{equation}
where $v_j$ are the scalar Koopman modes and $\varphi_j(\mathbf{x})$ are the eigenfunctions.

In our discrete, finite-dimensional data-driven framework, this projection reconstructs the state vector $\mathbf{y}_k$ at time step $k$. The infinite sum is truncated at rank $r$, and the continuous evolution $e^{\lambda_j t}$ is replaced by the discrete eigenvalue powers $\mu_j^k$. The reconstruction equation is \cite{lydon2025}:

\begin{equation}
    \mathbf{y}_k \approx \sum_{j=1}^{r} \mathbf{\phi}_j \mu_j^k b_j
    \label{eq:Reconstruction}
\end{equation}

Here, $\mathbf{\phi}_j$ is the $j$th column of the physical mode matrix $\mathbf{\Phi}_{phys}$ (representing the spatial structure), and $b_j$ is the \textbf{modal amplitude}, which corresponds to the coefficient $v_j$ in the continuous theory.

The vector $\mathbf{b} = [b_1, b_2, \dots, b_r]^T$ represents the projection of the initial system state onto the Koopman eigenbasis. It quantifies the weight or participation of each mode in the system's response at $t=0$. To construct $\mathbf{b}$, we solve the linear inverse problem at the initial snapshot $\mathbf{y}_1$. Since the mode matrix $\mathbf{\Phi}_{phys}$ is typically non-square ($p \times r$), we use the Moore--Penrose pseudoinverse ($\dagger$) to minimize the reconstruction error:

\begin{equation}
    \mathbf{b} = \mathbf{\Phi}_{phys}^{\dagger} \mathbf{y}_1
    \label{eq:b}
\end{equation}

where $\mathbf{y}_1 \in \mathbb{R}^{p}$ denotes the initial snapshot vector containing the simultaneous measurements from all $p$ sensors at time $t=0$. The resulting vector $\mathbf{b}$ contains the \textbf{initial modal participation factors}. These are the complex weights of the mode shapes that minimize the error between the reconstructed state and the observed initial condition. Once $\mathbf{b}$ is computed, the temporal evolution of the system is fully determined by the linear superposition of these modes. Future states can then be predicted without integrating the nonlinear equations of motion again.

For a finite truncation, the measured response can be written as the contribution of the retained Hankel-DMD components plus a residual term $\mathbf{r}_k$:
\begin{equation}
    \mathbf{y}_k = \underbrace{\sum_{j=1}^{r} \mathbf{\phi}_j \mu_j^k b_j}_{\text{Retained Hankel-DMD components}} + \underbrace{\mathbf{r}_k}_{\text{Unresolved residual}}
\end{equation}
Here $\mathbf{r}_k$ collects unresolved forcing, measurement noise, truncation error and any dynamics not represented by the retained modes. It should therefore not be read as a strict separation of all broadband turbulence into a continuous spectrum. The decomposition is nonetheless advantageous for SHM, because the retained components concentrate the structural signature on which the health assessment relies.

\paragraph{Correlation-domain identification.}
Direct decomposition of a continuously forced response can bias the damping estimates, because the measured trajectory carries one particular realization of the external forcing (Section~\ref{sec:stochastic_koopman}). For modal-parameter estimation, the same Hankel-DMD procedure is therefore also applied to the output correlation sequences $\mathbf{R}_{\ell} = \mathbb{E}[\mathbf{y}_{t+\ell} \mathbf{y}_{t}^{\top}]$, $\ell \ge 1$, which evolve under the conditional-mean dynamics and are free of that realization. This provides a direct comparison with SSI-COV, which is likewise built on output correlations. The two estimates are compared in Section~\ref{sec:damping}. The response-domain operator is retained for virtual sensing, where realizations rather than expectations must be reconstructed.


\subsection{Virtual Sensing }\label{sec:virtual_sensing}

Estimating missing sensor data, whether due to hardware failure or limited instrumentation, is formulated mathematically as a gappy data reconstruction problem \cite{kim2024}.

The central idea of this framework is that the high-dimensional dynamics of the FOWT evolve on a low-dimensional \textit{linear invariant subspace}. This subspace is spanned by the system's global mode shapes ($\mathbf{\Phi}$), which impose structured spatial relationships across the measured response. Consequently, surviving sensors can contain information about missing locations because their responses share common modal components.

Let the full state vector at time step $k$ be denoted by $\tilde{\mathbf{y}}_k \in \mathbb{R}^{pd}$ (the time-delay embedded history of all $p$ sensors). In a failure scenario, we only observe a subset of $q$ active sensors ($q < p$). Let $\mathbf{S}_q \in \{0,1\}^{q \times p}$ be the binary matrix that selects the active physical channels, obtained by removing from the identity matrix the rows of the failed sensors. Repeating this selection over all $d$ delays gives the selection matrix (or mask) $\mathbf{S} = \mathbf{I}_d \otimes \mathbf{S}_q \in \{0,1\}^{qd \times pd}$, where $\otimes$ denotes the Kronecker product.

The relationship between the full state and the sparse observation $\tilde{\mathbf{y}}^{obs}_k$ is given by:
\begin{equation}
    \tilde{\mathbf{y}}^{obs}_k = \mathbf{S} \tilde{\mathbf{y}}_k
    \label{eq:selection}
\end{equation}

\subsubsection{Sparse Basis Construction}
The offline training phase (Section~\ref{sec:framework}) gives the exact Hankel-DMD modes $\mathbf{\Phi}_{aug} \in \mathbb{C}^{pd \times r}$ directly. These modes are the basis functions that govern the system's trajectory. For any given time window, the time-delay embedded state vector $\tilde{\mathbf{y}}_k$ can be reconstructed by the modal expansion:

\begin{equation}
    \tilde{\mathbf{y}}_k \approx \mathbf{\Phi}_{aug} \mathbf{b}_k
\end{equation}

where $\mathbf{b}_k \in \mathbb{C}^r$ contains the \textbf{effective modal amplitudes} at time step $k$. Linear dynamics imply $\mathbf{b}_k = \mathbf{\Lambda}^k \mathbf{b}_0$, but in the virtual-sensing framework $\mathbf{b}_k$ is estimated locally from the sparse sensor data to account for nonlinearities and external disturbances.

In virtual sensing, the full vector $\tilde{\mathbf{y}}_k$ is not observed because sensors are missing. Instead, we construct a \textbf{sparse embedded vector} $\tilde{\mathbf{y}}^{obs}_k$ by stacking the histories of only the $q$ available sensors out of $p$. This corresponds to keeping only the rows of the augmented mode matrix $\mathbf{\Phi}_{aug}$ that belong to the active sensors, across all $d$ delays.

Let $\mathcal{I}_{avail}$ denote the set of row indices in the Hankel matrix corresponding to the active sensors over all $d$ delays, as selected by the selection matrix $\mathbf{S}$. We define the \textbf{sparse mode basis} $\mathbf{\Phi}_{sparse} \in \mathbb{C}^{qd \times r}$ as:
\begin{equation}
    \mathbf{\Phi}_{sparse} = \mathbf{\Phi}_{aug}[\mathcal{I}_{avail}, :] = \mathbf{S} \mathbf{\Phi}_{aug}
    \label{eq:sparse_basis}
\end{equation}

Because $\mathbf{\Phi}_{sparse}$ has only $qd$ rows for $r$ unknown amplitudes, the conditions under which $\mathbf{b}_k$ remains identifiable must be stated explicitly. The shift structure of the delay-embedded modes in Equation~\eqref{eq:phi_aug_stack} gives $\mathbf{\Phi}_{sparse}$ a Khatri--Rao form from which these conditions follow directly.

\begin{proposition}[Identifiability of the modal amplitudes]\label{prop:identifiability}
Let the delay-embedded Koopman modes have the shift structure of Equation~\eqref{eq:phi_aug_stack}, $\mathbf{\Phi}_{aug,j} = \mathbf{v}_j \otimes \mathbf{\phi}_j$ with $\mathbf{v}_j = [1, \mu_j, \dots, \mu_j^{d-1}]^\top$. For a sensor mask $\mathbf{S} = \mathbf{I}_d \otimes \mathbf{S}_q$ selecting $q$ of the $p$ channels over all $d$ delays (Equation~\eqref{eq:selection}), the sparse mode basis of Equation~\eqref{eq:sparse_basis} is the Khatri--Rao product

\begin{equation}
    \mathbf{\Phi}_{sparse}
    =
    \mathbf{V}_d \odot \mathbf{C}
    \in \mathbb{C}^{qd \times r},
    \qquad
    \mathbf{V}_d = [\mathbf{v}_1, \dots, \mathbf{v}_r] \in \mathbb{C}^{d \times r},
    \quad
    \mathbf{C} = \mathbf{S}_q \mathbf{\Phi}_{phys} \in \mathbb{C}^{q \times r}.
    \label{eq:khatri_rao}
\end{equation}

where $\odot$ denotes the column-wise Khatri--Rao product, $\mathbf{V}_d \odot \mathbf{C} = [\mathbf{v}_1 \otimes \mathbf{c}_1,\, \mathbf{v}_2 \otimes \mathbf{c}_2,\, \ldots,\, \mathbf{v}_r \otimes \mathbf{c}_r]$. If the discrete eigenvalues $\mu_j$ are distinct, then $k_{\mathbf{V}_d}=\min(d,r)$. The Khatri--Rao rank bound gives
$k_{\mathbf{V}_d\odot\mathbf{C}}\ge\min(k_{\mathbf{V}_d}+k_{\mathbf{C}}-1,r)$, so $\mathbf{\Phi}_{sparse}$ has full column rank and $\mathbf{b}_k$ is unique when $k_{\mathbf{C}}+\min(d,r)\ge r+1$. In particular, for $d\ge r$, it is sufficient that every retained mode is visible on at least one active channel.

\end{proposition}

\noindent \textit{Proof.} A Vandermonde matrix with distinct nodes has Kruskal rank $\min(d,r)$. By the Khatri--Rao rank lemma \cite{sidiropoulos2000}, $k_{\mathbf{V}_d \odot \mathbf{C}} \ge \min(k_{\mathbf{V}_d} + k_{\mathbf{C}} - 1, r)$, which equals $r$ under the stated condition. 

Proposition~\ref{prop:identifiability} shows how the delay embedding compensates for sparse instrumentation. With $q < r$ active channels the instantaneous problem is under-determined, but the $d$ delays supply the missing equations through the temporal, Vandermonde structure. The condition also shows where the construction degrades. Two eigenvalues that are close relative to the window length, $|f_i - f_j| \, d \Delta t \ll 1$, produce nearly collinear columns of $\mathbf{V}_d$, so their separation falls back on the spatial signatures in $\mathbf{C}$. The smallest singular value then bounds the noise amplification, $\| \delta \mathbf{b}_k \| \le \| \delta \tilde{\mathbf{y}}^{obs}_k \| / \sigma_{min}(\mathbf{\Phi}_{sparse})$. The singular values and condition numbers $\kappa = \sigma_{max}/\sigma_{min}$ are therefore reported for every failure scenario in Section~\ref{sec:vs_robust}. The dimensions of all matrices as implemented are summarized in Table~\ref{tab:dimensions} of Appendix~\ref{app:supporting}.

\subsubsection{Real-Time Rolling Horizon Reconstruction}\label{sec:rolling}

To keep the reconstruction robust over the full duration $t_0 \to t_n$, a \textbf{rolling horizon strategy} is used. Instead of estimating a single global amplitude vector, the algorithm updates the modal amplitudes as new data become available from the active sensors.

The scheme alternates between calibration and prediction. At time step $k$ the estimator uses the causal window of the active channels, $\tilde{\mathbf{y}}^{obs}_k = \mathbf{S}[\mathbf{y}_{k-d+1}^\top, \dots, \mathbf{y}_{k}^\top]^\top \in \mathbb{R}^{qd}$, so that no sample later than $k$ enters the estimate. This sparse measurement history is used to calibrate the local modal amplitudes $\mathbf{b}_k$ by solving the overdetermined system with the Moore--Penrose pseudoinverse:
\begin{equation}
    \mathbf{b}_k = \mathbf{\Phi}_{sparse}^{\dagger} \tilde{\mathbf{y}}^{obs}_k
    \label{Pseudo}
\end{equation}

Under extensive sensor failure the plain pseudoinverse of Equation~\eqref{Pseudo} amplifies measurement noise through the small singular values identified in Proposition~\ref{prop:identifiability}. The inversion is therefore regularized. The same normal equations are kept, with an added Tikhonov penalty scaled by the largest singular value $\sigma_{max}$ of $\mathbf{\Phi}_{sparse}$:
\begin{equation}
    \mathbf{b}_k = \arg\min_{\mathbf{b}} \; \| \mathbf{\Phi}_{sparse} \mathbf{b} - \tilde{\mathbf{y}}^{obs}_k \|_2^2 + \alpha^2 \sigma_{max}^2 \| \mathbf{b} \|_2^2
    = \mathbf{V}_s \, \text{diag}\!\left( \frac{\sigma_i}{\sigma_i^2 + \alpha^2 \sigma_{max}^2} \right) \mathbf{U}_s^{*} \, \tilde{\mathbf{y}}^{obs}_k ,
    \label{eq:tikhonov}
\end{equation}
where $\mathbf{\Phi}_{sparse} = \mathbf{U}_s \mathbf{\Sigma}_s \mathbf{V}_s^{*}$ is the SVD of the sparse mode basis and $\alpha$ is the single regularization parameter of the estimator. The limit $\alpha \to 0$ recovers Equation~\eqref{Pseudo}. The right singular vectors $\mathbf{V}_s$ are distinct from the Vandermonde matrix $\mathbf{V}_d$ in Equation~\eqref{eq:khatri_rao} and $\mathbf{V}_W$ in Equation~\eqref{eq:Y_rec}.
This step aligns the model with the instantaneous energy of the system. Next, to propagate the full state efficiently over the horizon $W$, a \textbf{Vandermonde matrix} $\mathbf{V}_W \in \mathbb{C}^{r \times W}$ is constructed from the powers of the discrete eigenvalues $\mu_j$:
\begin{equation}
    \mathbf{V}_W =
    \begin{bmatrix}
    1 & \mu_1 & \dots & \mu_1^{W-1} \\
    \vdots & \vdots & \ddots & \vdots \\
    1 & \mu_r & \dots & \mu_r^{W-1}
    \end{bmatrix}
\end{equation}
The reconstructed trajectory $\hat{\mathbf{Y}}_{rec} \in \mathbb{R}^{p \times W}$ is then computed in a single vectorized matrix operation:
    \begin{equation}
        \hat{\mathbf{Y}}_{rec} = \text{Re} \left( \underbrace{\mathbf{\Phi}_{L}}_{\text{Mode Shapes}} \cdot \underbrace{\text{diag}(\mathbf{b}_k)}_{\text{Mode Energy}} \cdot \underbrace{\mathbf{V}_W}_{\text{Time Evolution}} \right)
        \label{eq:Y_rec}
    \end{equation}
Here, $\mathbf{\Phi}_{L}$ is the spatial mode block aligned with the current end of the delay window (\textit{where} the structure moves), $\text{diag}(\mathbf{b}_k)$ scales the energy (\textit{how much the structure moves}), and $\mathbf{V}_W$ governs the temporal dynamics (\textit{when the structure moves}). The block $\mathbf{\Phi}_{L}$ is obtained from the training model and is an invariant parameter. This formulation replaces iterative time stepping with efficient linear algebra and acts as a filter that continuously prevents drift. The complete computational procedure is summarized in Algorithm~\ref{alg:virtual_sensing}.

Two settings of the horizon $W$ are used in what follows, and the distinction drawn in Section~\ref{sec:stochastic_koopman} applies to them. Reconstruction of hidden channels is performed causally with $W = 1$, so that the window ending at step $k$ uses no sample later than $k$ and the modal amplitudes are re-estimated at the measurement rate. A horizon $W > 1$ corresponds to an input-free forecast over $W$ steps and is used only to quantify how long an estimate remains valid without new measurements (Section~\ref{sec:horizon}).

\subsubsection{Stochastic Extension: Kalman}\label{sec:kalman}

The regularized projection of Equation~\eqref{eq:tikhonov} is the deterministic estimator of the framework. It re-fits the modal amplitudes in every window, so the unmeasured excitation and the measurement noise are absorbed by the fit. Since the unmeasured forcing is the dominant source of model error under stochastic excitation, the same identified operator can also be written as a linear stochastic state-space model in the modal coordinates $\mathbf{b}_k = \mathbf{\Psi} \tilde{\mathbf{y}}_k$ with $\mathbf{\Psi} = \mathbf{\Phi}^{\dagger}$,
\begin{equation}
    \mathbf{b}_{k+1} = \mathbf{\Lambda} \mathbf{b}_k + \mathbf{w}_k, \qquad
    \mathbf{y}_k = \mathbf{\Phi}_{L} \mathbf{b}_k + \mathbf{v}_k ,
    \label{eq:koopman_ss}
\end{equation}
in which $\mathbf{w}_k$ carries the evolution driven by the unmeasured forcing, that is, the residual of the stochastic Koopman operator of Section~\ref{sec:stochastic_koopman}, and $\mathbf{v}_k$ the model-reduction residual error. No additional data are required to characterize them. Their covariances follow from the same training correlations that Hankel-DMD already forms, $\mathbf{\Sigma}_0 = \mathbb{E}[\tilde{\mathbf{y}}_k \tilde{\mathbf{y}}_k^\top]$ and $\mathbf{\Sigma}_1 = \mathbb{E}[\tilde{\mathbf{y}}_{k+1} \tilde{\mathbf{y}}_k^\top]$,
\begin{equation}
    \mathbf{Q} = \text{Cov}\!\left( \mathbf{\Psi} \tilde{\mathbf{y}}_{k+1} - \mathbf{\Lambda} \mathbf{\Psi} \tilde{\mathbf{y}}_k \right), \qquad
    \mathbf{R} = \mathbf{E} \mathbf{\Sigma}_0 \mathbf{E}^{\top}, \quad
    \mathbf{E} = \mathbf{M}_L - \mathbf{\Phi}_{L} \mathbf{\Psi} ,
    \label{eq:QR}
\end{equation}
Here, $\mathbf{Q}$ is the covariance of the modal prediction error, while $\mathbf{R}$ is the covariance of the output reconstruction error, with $\mathbf{M}_L=[\mathbf{0},\ldots,\mathbf{0},\mathbf{I}_p]$, so that $\mathbf{M}_L\tilde{\mathbf{y}}_k=\mathbf{y}_k$ and $\mathbf{\Phi}_L\mathbf{\Psi}\tilde{\mathbf{y}}_k=\hat{\mathbf{y}}_k$. For the active channels, with $\mathbf{C}_a = \mathbf{S}_q \mathbf{\Phi}_{L}$ and the corresponding noise block $\mathbf{R}_{aa}$, the steady-state Kalman gain $\mathbf{K}_a$ follows from the discrete algebraic Riccati equation \cite{AndersonMoore1979} and the modal amplitudes are updated at every sample,
\begin{equation}
    \hat{\mathbf{b}}_k = (\mathbf{I} - \mathbf{K}_a \mathbf{C}_a) \mathbf{\Lambda} \hat{\mathbf{b}}_{k-1} + \mathbf{K}_a \mathbf{y}^{obs}_k , \qquad
    \hat{\mathbf{y}}_k = \text{Re}\!\left( \mathbf{\Phi}_{L} \hat{\mathbf{b}}_k \right).
    \label{eq:kf}
\end{equation}
The pair $(\mathbf{\Lambda}, \mathbf{\Phi}_{aug})$ is identical to that of the projection form. A change of the sensor failure pattern alters only $\mathbf{C}_a$ and $\mathbf{R}_{aa}$, so the rank $r$ remains the only tuned parameter. The filter is used here purely as an estimator of the hidden channels. The white-noise assumptions on $\mathbf{w}_k$ and $\mathbf{v}_k$ are approximations, and the state covariance is not interpreted as a quantification of uncertainty. Both realizations are inexpensive online. The projection requires $\mathcal{O}(qdr)$ and the Kalman form $\mathcal{O}(r^2)$ operations per sample. This is a few tens of microseconds for the ranks considered here, far below the sampling intervals of interest.

\begin{algorithm}[hbt!]
\caption{Data-Driven Virtual Sensing via Rolling Horizon Hankel-DMD}
\label{alg:virtual_sensing}
\footnotesize
\begin{algorithmic}[1]

\State \textbf{Input:} Raw Sensor Data $\mathbf{Y}_{raw} \in \mathbb{R}^{p \times N}$, Delay $d$, Rank $r$, Horizon $W$, Regularization $\alpha$.
\State \textbf{Output:} Reconstructed Full State $\hat{\mathbf{Y}}_{rec}$.

\Statex \textbf{--- Phase I: Offline Training (System Identification) ---}
\State \textbf{Preprocessing:} Apply zero-phase bandpass filter and Z-score normalization.
\State \textbf{Embedding:} Construct Hankel matrix $\mathbf{H}$ from training snapshots.
\State \textbf{Decomposition:} Compute Exact Hankel-DMD:
    \State \quad $\mathbf{H}_1 = \mathbf{H}[:, :-1], \quad \mathbf{H}_2 = \mathbf{H}[:, 1:]$ \Comment{Time-shifted matrices}
    \State \quad $\mathbf{U}, \mathbf{\Sigma}, \mathbf{V}^* \leftarrow \text{SVD}(\mathbf{H}_1, \text{rank}=r)$
    \State \quad $\tilde{\mathbf{A}} \leftarrow \mathbf{U}^* \mathbf{H}_2 \mathbf{V} \mathbf{\Sigma}^{-1}$ \Comment{Reduced Operator}
    \State \quad $\tilde{\mathbf{A}} \mathbf{W} = \mathbf{W} \mathbf{\Lambda}$; exact modes $\mathbf{\Phi}_{aug} \leftarrow \mathbf{H}_2 \mathbf{V} \mathbf{\Sigma}^{-1} \mathbf{W}$ \Comment{Eq.~\eqref{eq:DMD-modes}}
\State \textbf{Partitioning:} Isolate the mode-shape reference, the current block and the active rows:
    \State \quad $\mathbf{\Phi}_{phys} \leftarrow \mathbf{\Phi}_{aug}[1:p, :]$ \Comment{Mode-shape reference}
    \State \quad $\mathbf{\Phi}_{L} \leftarrow \mathbf{\Phi}_{aug}[(d-1)p+1 : dp, :]$ \Comment{Most recent delay block}
    \State \quad $\mathbf{\Phi}_{sparse} \leftarrow (\mathbf{I}_d \otimes \mathbf{S}_q) \mathbf{\Phi}_{aug}$ \Comment{Active rows, all $d$ delays}

\Statex \textbf{--- Phase II: Online Virtual Sensing (Rolling Horizon) ---}
\State Initialize $k \leftarrow d-1$
\While{$k < N$}
    \State \textbf{1. Observation:}
    \State \quad Extract the causal trajectory window $\tilde{\mathbf{y}}^{obs}_k$ from the active sensors.

    \State \textbf{2. Calibration (Inverse Problem):}
    \State \quad Estimate local modal amplitudes via regularized inversion, Eq.~\eqref{eq:tikhonov}:
    \State \quad $\mathbf{b}_k \leftarrow \mathbf{V}_s \text{diag}\!\left( \sigma_i / (\sigma_i^2 + \alpha^2 \sigma_{max}^2) \right) \mathbf{U}_s^{*} \tilde{\mathbf{y}}^{obs}_k$
    \State \quad \textit{(Optional, stochastic extension)} $\hat{\mathbf{b}}_k \leftarrow$ Eq.~\eqref{eq:kf} with $\mathbf{K}_a$ from $\mathbf{Q}, \mathbf{R}$ of Eq.~\eqref{eq:QR}

    \State \textbf{3. Prediction (Matrix Propagation):}
    \State \quad Construct Vandermonde matrix $\mathbf{V}_W$ from eigenvalues $\boldsymbol{\mu}$.
    \State \quad Compute the full-state trajectory using $\mathbf{b}_k$ for the projection:
    \State \quad $\hat{\mathbf{Y}}_{win} \leftarrow \text{Re}(\mathbf{\Phi}_{L} \cdot \text{diag}(\mathbf{b}_k) \cdot \mathbf{V}_W)$
    \State \quad For the Kalman form, use the updated state directly:
    \State \quad $\hat{\mathbf{y}}_k \leftarrow \text{Re}(\mathbf{\Phi}_{L}\hat{\mathbf{b}}_k)$
    \State \textbf{4. Update:}
    \State \quad Store the first column of $\hat{\mathbf{Y}}_{win}$ in $\hat{\mathbf{Y}}_{rec}$ as the current estimate;
    \State \quad retain the remaining columns only when an open-loop forecast is required.
    \State \quad $k \leftarrow k + 1$ \Comment{Advance the causal window by one sample}
\EndWhile

\State \textbf{Return} $\hat{\mathbf{Y}}_{rec}$
\end{algorithmic}
\end{algorithm}

\section{Results and Validation}

To validate the proposed framework, the Hankel-DMD algorithm was applied to a high-fidelity dataset of the NREL 5MW spar-buoy under realistic operational conditions. The analysis evaluates two capabilities of the digital twin. The first is \textbf{system identification}, which extracts accurate structural parameters (frequencies, damping and mode shapes) from stochastic environmental noise. The second is \textbf{real-time virtual sensing}, which reconstructs missing sensor fields under unseen operational regimes.

\subsection{Operational Data and Sensor Configuration}
The dataset was generated with the OpenFAST aero-hydro-elastic simulation tool. To capture the full operational variance of the FOWT, simulations were run for 12 operational load cases (OLCs). Table~\ref{tab:openfast_config} of Appendix~\ref{app:supporting} lists the complete solver configuration: modules, options, enabled degrees of freedom, structural settings and time steps.

\subsubsection{Linearization vs. Operational Simulation }
The theoretical baseline and the operational training data are kept distinct, so that the effect of physical nonlinearities can be quantified:

\begin{itemize}
    \item \textbf{Theoretical Baseline (Linear):} The reference model was derived with the OpenFAST linearization module. To ensure convergence to a steady equilibrium, this configuration uses simplified physics: quasi-static mooring (MAP++), steady uniform wind ($8$ m/s) and still water. At the linearization instant the rotor was nearly stationary ($0.18$ rpm). The model therefore gives a snapshot of the stiffness and damping at that operating point.
    \item \textbf{Operational Training Data (Nonlinear):} In contrast, the Hankel-DMD training data use the full nonlinear capabilities of OpenFAST. These simulations include turbulent wind fields (TurbSim), irregular wave spectra (JONSWAP), hydrodynamic loads (BEM model) and, in particular, dynamic mooring loads (FEAMooring). This setup captures the time-varying, stochastic nature of the floating system, including hysteresis and platform--tower coupling. The digital twin therefore learns from the complex reality of operation rather than from an idealized steady state.
\end{itemize}

\subsubsection{Operational Load Cases }
The wave parameters were varied systematically to cover four sea states ($H_s \in [1, 4]$ m) and three incidence angles ($0^\circ, 45^\circ, 90^\circ$), so that coupled motions in several DOFs are excited (Table~\ref{tab:load_cases}). The wind speed was held constant at the rated value ($12$ m/s) to keep the aerodynamic thrust at its highest (rotor speed $12.1$ rpm).

Virtual sensing requires the digital twin to generalize to unseen conditions. Training and validation data are therefore kept disjoint in both the environmental and the temporal dimension. The twelve OLCs vary the sea state under a single turbulent inflow realization, so a held-out OLC would otherwise still share its wind history with the training set. The validation therefore follows a time-blocked leave-one-OLC-out design, described in Section~\ref{sec:vs_protocol}, in which neither the sea state nor the time interval of a test record is seen during training or during the selection of hyper-parameters.

\begin{table}[htbp]
    \centering
    \caption{Simulated operational load cases.}
    \label{tab:load_cases}
    \footnotesize
    \renewcommand{\arraystretch}{1.2}
    \begin{tabular}{c c c c c}
        \toprule
        \textbf{Case ID} & \textbf{$H_s$ (m)} & \textbf{$T_p$ (s)} & \textbf{Dir ($^\circ$)} & \textbf{Wind (m/s)} \\
        \midrule
        1 & 1.0 & 6.0 & 0 & 12 \\
        2 & 1.0 & 9.0 & 45 & 12 \\
        3 & 1.0 & 12.0 & 90 & 12 \\
        \midrule
        4 & 2.0 & 8.0 & 0 & 12 \\
        5 & 2.0 & 10.0 & 45 & 12 \\
        6 & 2.0 & 14.0 & 90 & 12 \\
        \midrule
        7 & 3.0 & 10.0 & 0 & 12 \\
        8 & 3.0 & 12.0 & 45 & 12 \\
        9 & 3.0 & 16.0 & 90 & 12 \\
        \midrule
        10 & 4.0 & 9.0 & 0 & 12 \\
        11 & 4.0 & 12.0 & 45 & 12 \\
        12 & 4.0 & 20.0 & 90 & 12 \\
        \bottomrule
    \end{tabular}
    \vspace{1ex}
    \parbox{\linewidth}{\scriptsize \textit{Note:} Wave direction is defined relative to the wind vector: $0^\circ$ is aligned with the wind (fore-aft dominant) and $90^\circ$ is perpendicular to it (side-to-side dominant).}
\end{table}
\subsubsection{Hybrid State Vector Formulation}

To provide complementary sensitivity to the dominant structural dynamics, the state vector combines two physical quantities: absolute acceleration and bending moment (derived from strain). For a planar analysis (e.g., the fore-aft direction), this gives an 18-dimensional state vector $\mathbf{x}_k$ at each time step $t_k$:

\begin{equation}
    \mathbf{x}_k = \begin{bmatrix} \mathbf{a}_{1:9}(t_k) \\ \mathbf{M}_{1:9}(t_k) \end{bmatrix} \in \mathbb{R}^{18}
\end{equation}

This hybrid formulation offers a distinct advantage over single-modality approaches, in line with the Takens embedding theorem. Strain gauges give high signal-to-noise ratios for elastic deformation modes, particularly the fundamental bending modes, where curvature is high but acceleration may be low. Conversely, accelerometers capture the rigid-body platform dynamics and higher-frequency structural modes that induce small strains but generate significant inertial forces. Combining the two measurement types therefore improves sensitivity to both rigid-body and elastic structural dynamics and provides a richer output space for identification than either measurement type alone.

\subsubsection{Sensor Scope and Layout}

The study focuses on the tower substructure, instrumented with nine virtual gauge locations corresponding to ElastoDyn tower nodes 1, 3, 5, 7, 10, 13, 15, 17 and 20 of the 20-node tower discretization, spanning $11.9$--$85.7$~m. The gauges are therefore distributed non-uniformly along the tower height. This work targets the tower dynamics, but the framework is general. The state vector can be expanded to include blade roots, mooring lines or nacelle quantities without changing the underlying Hankel-DMD algorithm. The data-driven architecture is also not restricted to FOWT applications. It applies to any high-dimensional dynamical system with coherent spatiotemporal structures.

Table~\ref{tab:openfast_outputs} lists the OpenFAST outputs used for the fore-aft (FA) and side-to-side (SS) analyses.

\begin{table}[htbp]
    \centering
    \caption{OpenFAST output channels used in the Hankel-DMD analysis.}
    \label{tab:openfast_outputs}
    \scriptsize  
    \renewcommand{\arraystretch}{1.2} 

    \begin{tabular}{l p{0.25\linewidth} p{0.45\linewidth}} 
        \toprule
        \textbf{Channel Name} & \textbf{Description} & \textbf{Purpose in DMD Framework} \\
        \midrule
        \texttt{TwHt[1-9]MLxt} & Tower local side-to-side moment & Primary source for SS strain mode shapes (captures curvature). \\
        \texttt{TwHt[1-9]MLyt} & Tower local fore-aft moment & Primary source for FA strain mode shapes (captures curvature). \\
        \texttt{TwHt[1-9]ALxt} & Tower absolute acceleration (local $x$, fore-aft) & Captures FA bending modes coupled with platform surge and pitch kinematics. \\
        \texttt{TwHt[1-9]ALyt} & Tower absolute acceleration (local $y$, side-to-side) & Captures SS bending modes coupled with platform sway and roll kinematics. \\
        \bottomrule
    \end{tabular}
\end{table}

\begin{figure}[htbp]
    \centering
    \includegraphics[width=0.7\textwidth]{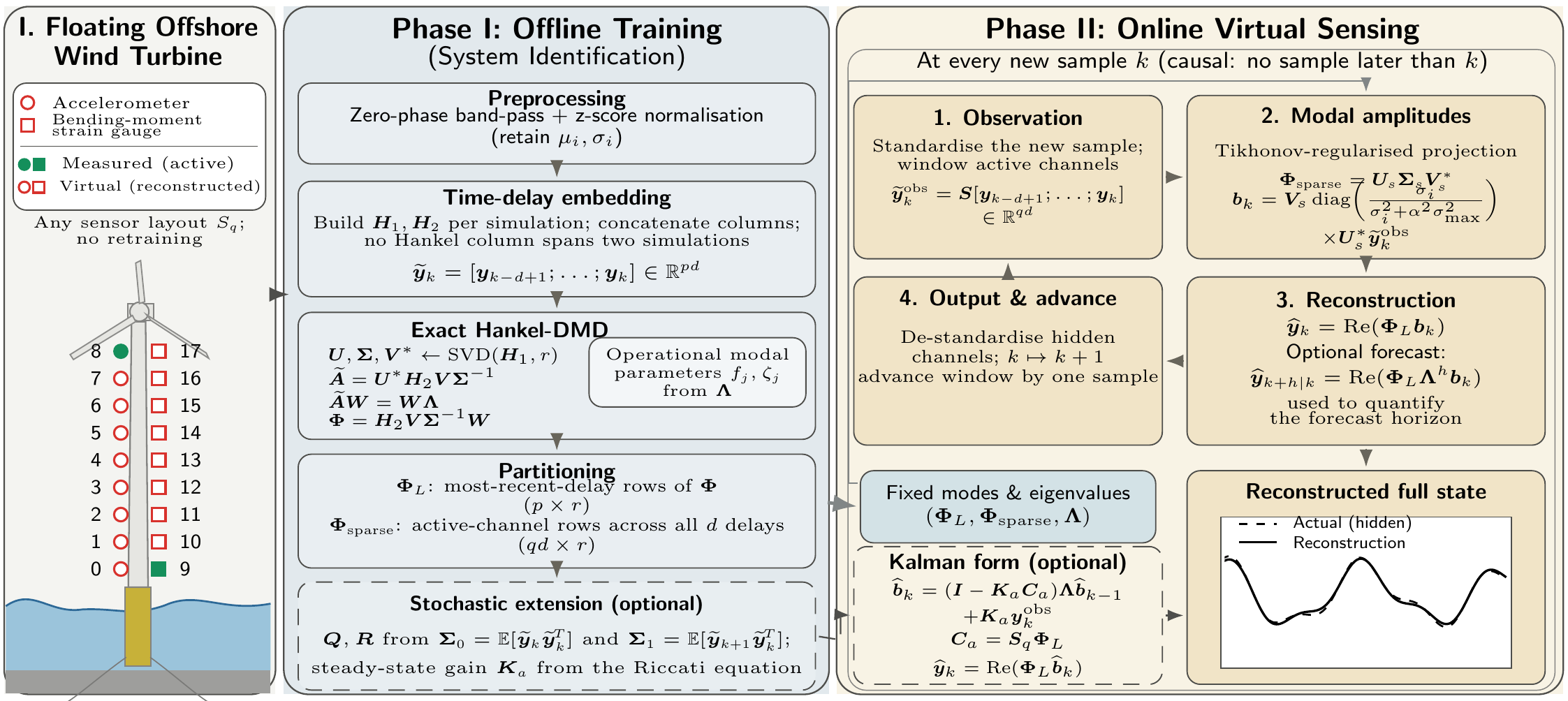}
    \caption{Schematic of the data-driven virtual-sensing framework: data flow from the physical FOWT through offline Hankel-DMD training to the online rolling horizon loop for real-time full-state reconstruction.}
    \label{fig:digital_twin_arch}
\end{figure}

Figure~\ref{fig:digital_twin_arch} shows the overall architecture of the digital twin, from physical data acquisition to the virtual-sensing module. Algorithm~\ref{alg:virtual_sensing} summarizes the computational procedure for the rolling horizon updates.

\subsection{Data Preprocessing}\label{sec:preprocessing}

Before identification, the raw sensor data $\mathbf{Y}_{raw} \in \mathbb{R}^{p \times N}$ pass through a preprocessing pipeline that isolates the structural dynamics and standardizes the signal magnitudes. All channels are processed at the native OpenFAST output interval ($\Delta t = 0.05$~s, 20~Hz) and the first 300~s of each simulation are discarded as start-up transient. First, high-frequency noise and low-frequency drift are removed using a 4th-order Butterworth bandpass filter ($0.25 - 5.0$ Hz). A \textbf{zero-phase filtering} scheme (forward--backward) is used to preserve the temporal alignment of peaks across sensors, which is a prerequisite for accurate mode-shape identification:
\begin{equation}
    \mathbf{y}_{filt}(t) = \mathcal{F}_{backward}(\mathcal{F}_{forward}(\mathbf{y}_{raw}(t)))
\end{equation}
\textbf{Z-score standardization} is then applied, so that the singular value decomposition (SVD) is not biased numerically by the difference in magnitude between accelerations ($m/s^2$) and bending moments ($kN \cdot m$). Each sensor channel $i$ is normalized independently to zero mean ($\mu_i$) and unit variance ($\sigma_i$), so that all sensors contribute equally to the Hankel energy spectrum:
\begin{equation}
    \mathbf{y}_{norm}^{(i)}(t) = \frac{\mathbf{y}_{filt}^{(i)}(t) - \mu_i}{\sigma_i}
\end{equation}

Two properties of this pipeline matter for the validation protocol of Section~\ref{sec:vs_protocol}. First, for virtual-sensing validation the normalization parameters $\mu_i$ and $\sigma_i$ are determined from the training data alone and applied unchanged to the corresponding held-out data, so that no information from a scored record enters its own preprocessing. Second, the forward--backward filter is non-causal by construction and is applied offline to the complete training and test records alike. The causality of the estimator described in Section~\ref{sec:rolling} therefore refers to the estimator window: no sample later than the reporting instant enters the reconstruction. A deployed implementation would use a causal filter at this preprocessing stage.

\section{System Identification Results}\label{sec:sysid}

To assess the physical fidelity of the Hankel-DMD model, the extracted modal parameters were compared with three established baselines: the theoretical linearization provided by OpenFAST, the standard covariance-driven stochastic subspace identification (SSI-COV) algorithm \cite{SSICOV} and frequency domain decomposition (FDD) \cite{Brincker2001FDD}. The correlation between the identified operational modes ($\boldsymbol{\phi}_{op}$) and the reference linear modes ($\boldsymbol{\phi}_{ref}$) was quantified with the modal assurance criterion (MAC), defined as:

\begin{equation}
    \text{MAC}(\boldsymbol{\phi}_{op}, \boldsymbol{\phi}_{ref}) = \frac{\left| \boldsymbol{\phi}_{op}^H \boldsymbol{\phi}_{ref} \right|^2}{(\boldsymbol{\phi}_{op}^H \boldsymbol{\phi}_{op})(\boldsymbol{\phi}_{ref}^H \boldsymbol{\phi}_{ref})}
    \label{eq:mac_definition}
\end{equation}

where $(\cdot)^H$ denotes the Hermitian transpose. A MAC value approaching unity indicates a high degree of linear dependence between the mode shape vectors.

To avoid tuning the Hankel-DMD parameters to a single favourable setting, the delay depth was varied over $d \in \{30, 60, 90\}$ samples ($1.5$--$4.5$~s, i.e.\ $0.75$--$2.2$ periods of the first tower mode) and the rank over $r \in [10, 50]$. Within each frequency band, the pole with the largest modal energy was retained, and the median and inter-quartile range (IQR) over all $(d, r)$ are reported. Each OLC was treated as a separate segment, so that no Hankel column spans two simulations. SSI-COV used $i = 60$ block rows over model orders 20--80. At each order, the admissible in-band pole with the largest mode-shape norm was retained, and the median over orders is reported. FDD used the peak of the first singular value of the cross-spectral density matrix. Mode shapes were de-standardized, that is, multiplied back by the channel standard deviations $\sigma_i$, before any comparison. The MAC was evaluated separately for the acceleration and bending-moment components. Modal phase collinearity (MPC), which measures the phase complexity of the identified mode shapes, was also computed; values near unity indicate an essentially real mode. Following Section~\ref{sec:stochastic_koopman}, the decomposition was applied both to the response data and to the output correlation functions $\mathbf{R}_k$.

\subsection{Comparison with Linear Theory and Standard OMA}
Table~\ref{tab:sys_id_comparison} summarizes the results. All four methods identify the same four tower modes: the first fore-aft (FA) and side-to-side (SS) modes at $0.50$ and $0.51$~Hz, and the second FA and SS modes at $3.71$ and $3.60$~Hz. The three output-only methods agree within $1.2\%$ in frequency, which is comparable with the $0.01$~Hz resolution of FDD. The corresponding OpenFAST linearized modes show larger operating-condition-dependent differences, discussed below. The MAC between the Hankel-DMD and SSI-COV shapes is $\ge 0.99$ for every tower mode, and the modes are essentially real (MPC $\ge 0.998$). Tracking the modes separately in each OLC gives frequency standard deviations below $0.001$~Hz for SSI-COV and $0.007$~Hz for response-domain Hankel-DMD on the first modes, with shape correlations MAC $\ge 0.94$ against the global estimates. The structural poles are therefore invariant to the sea state, as expected for structural rather than excitation poles (Section~\ref{sec:stochastic_koopman}).

The deterministic rotor forcing appears as a distinct object rather than as a contaminated structural mode. The 3P blade-passing harmonic is returned by all methods as a separate pole at $0.57$--$0.61$~Hz, consistent with $3\Omega_{rotor}/2\pi \approx 0.61$~Hz at the rated rotor speed ($\Omega \approx 12.1$~rpm) and clearly distinct from the first tower modes at $0.50$--$0.51$~Hz. Its large apparent damping ($6$--$8\%$) is the signature of a harmonic whose frequency follows the rotor speed, which varies under turbulent inflow, rather than of structural dissipation. This is the criterion of Section~\ref{sec:stochastic_koopman}, and it is verified directly against a measured rotor speed in Section~\ref{sec:focal_sysid}. The Hankel embedding makes the separation possible at this frequency spacing. For $d\Delta t=3.0$ and $4.5$~s, the delay window spans more than one period of the first tower mode, helping distinguish the $0.50$~Hz structural mode from the $0.61$~Hz rotor harmonic through the Vandermonde structure of Proposition~\ref{prop:identifiability}.

As set out in Section~\ref{sec:framework}, the exact and projected DMD formulations share the same reduced eigenvalues, so this choice does not affect the identified frequencies and damping ratios. The exact DMD modes are used here for the reported spatial mode shapes.

The same four tower modes are found in the OpenFAST linearization. Its modes were paired with the identified ones by their fore-aft and side-to-side energy share and by MAC, not by frequency order. The linear first FA and SS modes lie at $0.477$ and $0.494$~Hz and the second FA and SS modes at $3.711$ and $3.811$~Hz, with MAC $\ge 0.97$ for the acceleration and $\ge 0.99$ for the bending-moment components. The remaining linear modes below $4$~Hz, notably those at $1.54$ and $2.00$~Hz, are dominated by blade flapwise and edgewise motion and are not tower modes at all. Matching the identified second tower modes against them by frequency proximity gives a spurious mode-shape mismatch. The linearized baseline represents a nearly stationary rotor ($0.18$~rpm) in $8$~m/s steady inflow, still water and quasi-static mooring, whereas the identification refers to the operating rotor at $12$~m/s with dynamic mooring. In operation the first modes have $4$--$5\%$ higher frequencies and, in the fore-aft direction, $2.0$--$2.3$ percentage points more damping than the linear ones (Section~\ref{sec:damping}). The second FA mode coincides with the linear one, and the second SS mode is $5\%$ lower.

\begin{table}[htbp]
\centering
\caption{System identification from the 12 OLCs (native output interval $\Delta t = 0.05$~s,
$0.25$--$5$~Hz) and comparison with the OpenFAST linearization. Hankel-DMD: medians [inter-quartile range] over
$d \in \{30, 60, 90\}$ and $r \in [10, 50]$, applied to the output correlation functions (corr.,
Section~\ref{sec:stochastic_koopman}) or to the response data (resp.). SSI-COV: medians over model orders 20--80
($i = 60$), selecting at each order the admissible in-band pole with the largest mode-shape norm. FDD: peak of the first singular value. Linear: eigen-analysis of the linearized model (nearly stationary rotor, $0.18$~rpm,
$8$~m/s steady inflow, still water, MAP++), modes tracked by their fore-aft / side-to-side energy share and MAC.
MAC: acceleration / bending-moment components of the de-standardized shapes.}
\label{tab:sys_id_comparison}
\footnotesize
\resizebox{\textwidth}{!}{%
\begin{tabular}{l c cc cc cc cc}
\toprule
\textbf{Mode} & \textbf{FDD} & \multicolumn{2}{c}{\textbf{SSI-COV}} & \multicolumn{2}{c}{\textbf{Hankel-DMD (corr.)}} &
\multicolumn{2}{c}{\textbf{Hankel-DMD (resp.)}} & \multicolumn{2}{c}{\textbf{Linear (OpenFAST)}} \\
 & $f$ (Hz) & $f$ (Hz) & $\zeta$ (\%) & $f$ (Hz) & $\zeta$ (\%) & $f$ (Hz) & $\zeta$ (\%) & $f$ (Hz) & $\zeta$ (\%) \\
\midrule
1st FA tower & 0.500 & 0.500 & 3.54 & \textbf{0.498} [0.497--0.499] & \textbf{3.26} [3.02--3.41] & 0.502 [0.496--0.519] & 0.80 [0.37--1.40] & 0.477 & 1.22 \\
1st SS tower & 0.510 & 0.514 & 1.19 & \textbf{0.514} [0.513--0.514] & \textbf{1.57} [1.49--1.73] & 0.516 [0.511--0.518] & 0.30 [0.14--0.72] & 0.494 & 0.70 \\
2nd FA tower & 3.720 & 3.714 & 1.31 & \textbf{3.712} [3.711--3.714] & \textbf{1.36} [1.29--1.45] & 3.698 [3.670--3.732] & 0.41 [0.27--0.61] & 3.711 & 1.09 \\
2nd SS tower & 3.600 & 3.596 & 2.17 & \textbf{3.598} [3.597--3.600] & \textbf{2.15} [2.08--2.30] & 3.610 [3.577--3.645] & 0.53 [0.35--0.68] & 3.811 & 1.87 \\
\midrule
3P harmonic (FA / SS) & 0.58 / 0.57 & 0.583 / 0.589 & 7.5 / 7.9 & 0.582 / 0.602 & 6.9 / 6.1 & 0.605 / 0.601 & 1.1 / 1.8 & -- & -- \\
\bottomrule
\end{tabular}}
\vspace{1ex}
\parbox{\linewidth}{\scriptsize \textit{Note:} MAC between the Hankel-DMD (corr.) and SSI-COV shapes is
$0.99$--$1.00$ for all four tower modes, and between Hankel-DMD and the linear reference $0.97$--$1.00$
(acceleration / bending-moment components).}
\end{table}

\subsubsection{Topological Accuracy: Platform Participation in the First Tower Modes}
The most visible feature of the first identified tower modes is an acceleration node at about $51$--$55$~m, with the tower base and top moving in opposite directions. This shape cannot be explained by a fixed-base cantilever mode alone. We therefore checked it against the ElastoDyn kinematics. At height $z$, the tower acceleration contains contributions from platform translation, platform rotation, and the two assumed tower modes $\phi_1(z)$ and $\phi_2(z)$. Fitting the identified first-mode shapes using only the two tower modes explains about $75$--$78\%$ of the shape. When platform translation and rotation are also included, the fit increases to $R^2>0.9999$. About $30\%$ of the fitted first-mode shape comes from platform rigid-body motion acting in the opposite direction to the tower deflection, which creates the observed node. This also explains the relatively low MAC of $0.55$--$0.67$ with the first fixed-base assumed mode. The first tower mode is therefore a coupled tower--platform mode rather than a fixed-base cantilever mode. In contrast, the second modes are already well represented by the assumed tower modes alone ($R^2=0.99$, MAC $=1.00$), with only about $1\%$ platform contribution. Their bending moment changes sign near $38$~m. SSI-COV gives the same topology, confirming that it is a property of the response rather than the identification method.

\subsubsection{Capturing Operational Physics: Frequency and Damping Shifts}\label{sec:damping}
Beyond mode topology, the analysis identifies distinct shifts in the eigenvalues. The first-mode frequencies are $4$--$5\%$ higher in operation than in the linearized model ($0.477 \rightarrow 0.498$~Hz in FA, $0.494 \rightarrow 0.514$~Hz in SS). Several mechanisms can produce such a shift. The rotor speed, the thrust-induced platform offset and the dynamic rather than quasi-static mooring model all differ between the two configurations. Separating their individual contributions would require linearizations at the operating point. We therefore report the difference without attributing it to a single mechanism.

The damping estimates are more informative. They expose a bias that follows directly from the stochastic-Koopman argument of Section~\ref{sec:stochastic_koopman}. Hankel-DMD applied to the response data returns systematically low damping ratios ($0.3$--$0.8\%$ for the first modes), whereas SSI-COV and correlation-domain Hankel-DMD return $1.2$--$3.5\%$. The response-domain estimate for the first SS mode ($0.30\%$) is below the $1\%$ structural damping prescribed for the tower modes in the simulation model. This identifies it as biased rather than merely different. Response trajectories contain the particular realization of the unmeasured coloured excitation, part of which is absorbed by the least-squares operator. The correlation functions describe the average response dynamics and reduce the influence of the particular realization of the random forcing. Correlation-domain Hankel-DMD and SSI-COV agree within $0.4$ percentage points and are insensitive to the embedding parameters (frequency IQR $\approx 0.001$~Hz for the first modes and $<0.004$~Hz for all tower modes, Figure~\ref{fig:sensitivity}). The response-domain estimates, in contrast, vary by up to $0.06$~Hz in frequency and by a factor of several in damping as the rank admits or removes neighbouring components such as 3P.

The physically meaningful result is the difference between the fore-aft and side-to-side directions. The first fore-aft mode is markedly more damped ($3.3$--$3.5\%$) than the first side-to-side mode ($1.2$--$1.6\%$). This is consistent with the aerodynamic damping that an operating rotor provides in the fore-aft direction, where the rotor thrust varies with the relative inflow velocity at the tower top. Side-to-side motion is dissipated mainly by the structure ($1\%$ prescribed) and the platform \cite{Burton2011}. The ordering is reproduced on a physical model in Section~\ref{sec:focal_sysid}, which is the closest independent evidence available here. Modal damping is therefore estimated in the correlation domain throughout. The response-domain operator is retained for virtual sensing, where realizations rather than expectations must be reconstructed (Section~\ref{sec:vs_results}).

\subsubsection{Mode Shape Sensitivity and Physical Coupling}
\begin{figure}[htbp]
    \centering
    \begin{subfigure}[b]{0.48\textwidth}
        \centering
        \includegraphics[width=\textwidth]{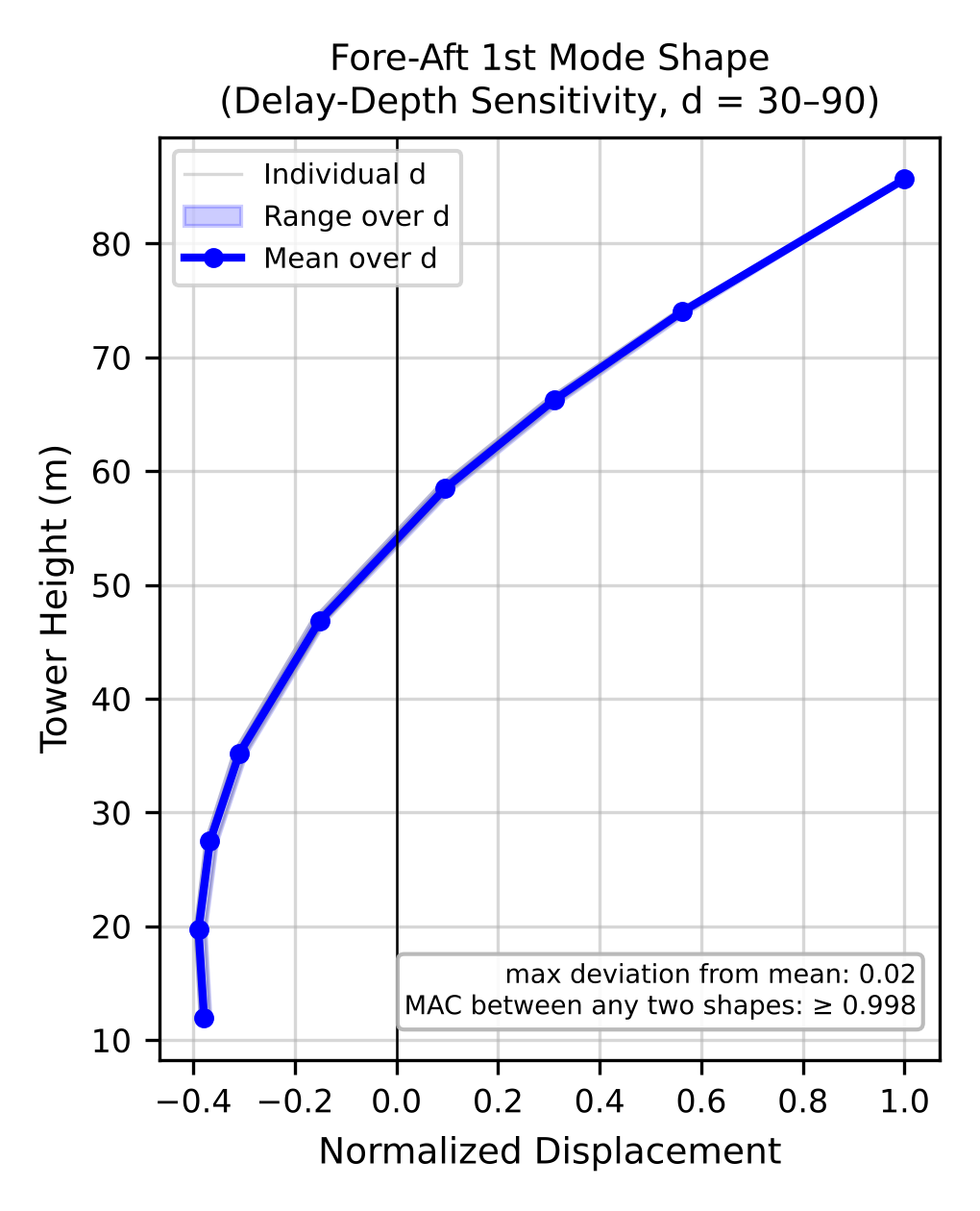}
        \caption{Fore-aft (FA) mode shape}
        \label{fig:modes_fa}  
    \end{subfigure}
    \hfill
    \begin{subfigure}[b]{0.48\textwidth}
        \centering
        \includegraphics[width=\textwidth]{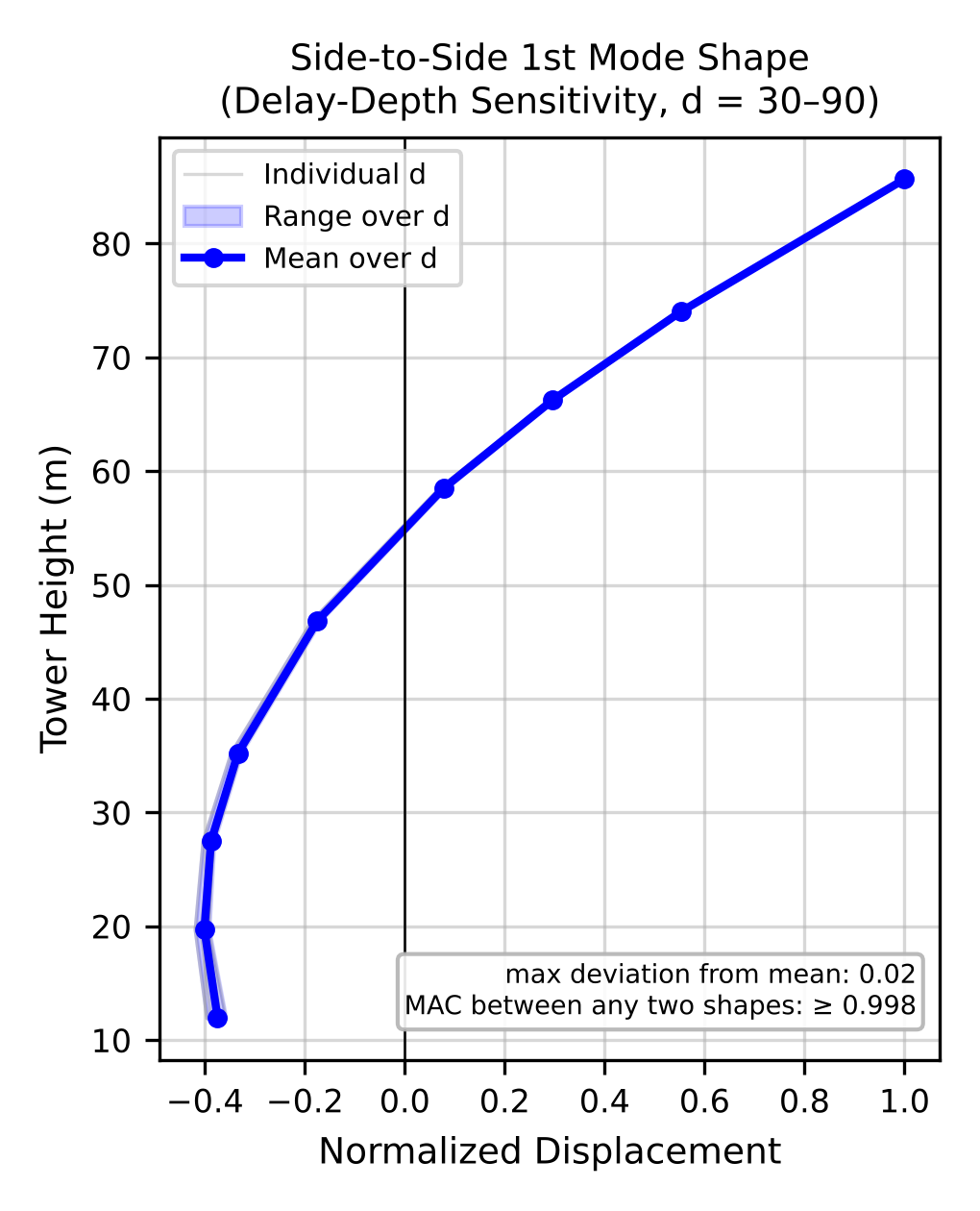}
        \caption{Side-to-side (SS) mode shape}
        \label{fig:modes_ss}  
    \end{subfigure}

    \caption{First tower-mode shapes identified by Hankel-DMD on the response data of all twelve OLCs ($r = 35$) and their sensitivity to the Hankel delay depth: thin grey lines, shapes for $d = 30, 40, \dots, 90$ samples ($1.5$--$4.5$~s); blue line, mean over $d$; shaded band, range over $d$. The shapes are normalized to unit amplitude at the top sensor. They are essentially independent of $d$: no shape deviates from the mean by more than $0.02$, and the MAC between any two shapes is at least $0.998$ in both directions.}
    \label{fig:modes_ss_fa} 
\end{figure}

A sensitivity analysis over the delay depth was performed to ensure that the identified mode shapes are not numerical artefacts of the embedding. The de-standardized first-mode shapes are presented in Figure~\ref{fig:modes_fa} and Figure~\ref{fig:modes_ss}. The thin grey lines are the shapes obtained for $d = 30, 40, \dots, 90$ samples, and the shaded band is the range over $d$, which is the same delay range as used for Table~\ref{tab:sys_id_comparison}. The shapes are essentially independent of the delay depth. No shape deviates from the mean by more than $0.02$ of the tower-top amplitude, and the MAC between any two shapes is at least $0.998$ in both directions. The embedding depth is therefore not adjusted to improve the results. The truncation rank has more influence. At $r = 20$ and $r = 50$, the fore-aft spread over the same delay range widens to $0.10$ (MAC $\ge 0.978$).

The shapes themselves show the coupling discussed above. The node of the first-mode acceleration shape lies at $51$--$55$~m in both directions, and the anti-phase motion of base and top reflects the participation of the platform rigid-body degrees of freedom quantified in Table~\ref{tab:shape_check}. The amplitude at the tower base is governed by the shifting inertial boundary condition imposed by platform surge and pitch, and at the nacelle by the mass and rotational inertia of the rotor-nacelle assembly. Both are reproduced by the kinematic fit, so no additional mechanism is needed to explain the shape. This confirms that Hankel-DMD gives a shape consistent with the kinematics of the underlying model and remains robust to the embedding parameters.

\subsection{Full-Field Reconstruction}
With the spectral fidelity of the identified modes established, the assessment now turns to the time-domain predictive capability of the digital twin. The main objective of the framework is to act as a real-time \textbf{virtual sensor}. It infers the dynamic state at unobservable or difficult-to-monitor locations of the FOWT from only a sparse subset of accessible sensors.

\subsubsection{Validation Protocol and Metrics}\label{sec:vs_protocol}
To test robustness, the trained Hankel-DMD model was evaluated with a time-blocked leave-one-OLC-out protocol over the operational load cases of Table~\ref{tab:load_cases}. After discarding the first 300~s, each record was divided into two halves of 1650~s. In every fold the model was identified from one half of the eleven other OLCs and tested on the other half of the held-out OLC, so that the test sea state and the test time interval are both excluded from training. This gives 24 folds per direction, and every OLC is reconstructed over its full length by models that have seen neither its sea state nor its time interval. The first 60~s of each test half were not scored, to exclude the start-up of the estimators. Within the test record, specific sensor channels were removed from the input vector, while the stochastic wind and wave loading remained active but unmeasured. The reconstruction performance therefore reflects the ability of the model to generalize spatial correlations and recover missing physics rather than to memorize training data.

The delay was fixed at $d = 60$ samples (3~s, about $1.5$ periods of the first tower modes). The rank $r \in \{24, 34, 50, 70, 90, 120\}$ and the Tikhonov parameter $\alpha \in \{0.001, 0.003, 0.01, 0.03\}$ of Equation~\eqref{eq:tikhonov} were cross-fitted. Folds whose test OLC lies in OLC~1--6 use the setting selected on the folds of OLC~7--12, and vice versa, so no held-out OLC contributes to the choice of its own hyper-parameters. Reconstruction is causal throughout, with the window advancing one sample at a time ($W = 1$ in Equation~\eqref{eq:Y_rec}), so no estimate uses a sample later than the instant it reports.

Three baseline estimators were evaluated under the same protocol. Each uses the same training covariance and hence no additional data, so the comparison isolates the contribution of the delay-embedded Koopman model. The baselines are (i) \textbf{gappy POD} \cite{GappyPOD}, a static estimator fitting $\min(q, 5)$ POD modes of the instantaneous covariance to the active channels, which like Hankel-DMD uses one model for all layouts; (ii) \textbf{static least squares}, the best instantaneous linear map from the active to the hidden channels, computed for each layout; and (iii) a \textbf{Wiener filter}, that is, delay ridge regression mapping the same $d$-sample window of the active channels to the hidden channels, computed for each layout with a cross-fitted ridge parameter. The Wiener filter is the minimum-variance linear estimator for a given window and layout and therefore serves as an upper reference for all window-based linear estimators.

The reconstruction accuracy is quantified by comparing the prediction of the digital twin ($\hat{\mathbf{x}}$) with the OpenFAST simulation ground truth ($\mathbf{x}_{true}$) at every node along the tower. Two complementary metrics are used: the normalized root mean square error (NRMSE) for the error magnitude and the coefficient of determination ($R^2$) for the captured variance:

\begin{equation}
    \text{NRMSE} = \frac{\sqrt{\frac{1}{N} \sum ( \mathbf{x}_{true} - \hat{\mathbf{x}} )^2}}{\max(\mathbf{x}_{true}) - \min(\mathbf{x}_{true})}, \quad
    R^2 = 1 - \frac{\sum (y_i - \hat{y}_i)^2}{\sum (y_i - \bar{y})^2}
    \label{eq:metrics}
\end{equation}

where $\mathbf{x}_{true}$ ($y_i$) is the ground truth, $\hat{\mathbf{x}}$ ($\hat{y}_i$) is the model prediction, and $\bar{y}$ is the mean of the observed data. An NRMSE below $5\%$ is generally considered sufficient for high-fidelity digital-twin applications and fatigue load estimation \cite{mehlan2023, branlard2024}. An $R^2$ close to $1.0$ shows that the model captures the dynamic variance of the signal under stochastic excitation.

\subsection{Temporal Evolution and Modal Decoupling: From Energy to Physics}

The temporal evolution of the identified Koopman modes is governed by the discrete-time linear recurrence $\mathbf{b}_{k+1} = \boldsymbol{\Lambda} \mathbf{b}_k$ as described in Equation~\eqref{eq:Reconstruction}. In continuous time, these dynamics are represented by the complex exponential expansion:
\begin{equation}
    \label{eq:dmd_continuous}
    \mathbf{Dynamics}_j(t) = b_j e^{(\sigma_j + i\omega_j)t}
\end{equation}
where $b_j$ is the initial modal amplitude (from the projection of Equation~\eqref{eq:b}, $\mathbf{b} = \mathbf{\Phi}^{\dagger} \mathbf{x}_0$) and $\omega_j$ is the oscillation frequency from Equation~\eqref{eq:dmd_eigenvalue}.

These physically isolated trajectories differ from the standard POD temporal modes. The POD temporal modes, $\mathbf{\Psi}_{POD}(t) = \mathbf{\Sigma}_r \mathbf{V}_r^*$, follow from Equation~\eqref{eq:A_tilde} and are visualized in Figure~\ref{fig:Hankel-Plot}. POD modes are orthogonal energy structures that often mix several frequencies in a single vector. The DMD formulation separates these components. The relationship follows from equating the two reconstructions of the data:
\begin{equation}
    \mathbf{H}_1 \approx \mathbf{U}_r \mathbf{\Sigma}_r \mathbf{V}_r^* \quad \text{(SVD Representation )}
\end{equation}
\begin{equation}
    \mathbf{H}_1 \approx \mathbf{\Phi} \cdot \mathbf{Dynamics}(t) \quad \text{(Spectral Representation)}
\end{equation}
For the projected modes, $\mathbf{\Phi}_{proj} = \mathbf{U}_r \mathbf{W}$ (Section~\ref{sec:framework}), substituting into the spectral representation and using $\mathbf{U}_r^* \mathbf{U}_r = \mathbf{I}$ gives the transformation that decouples the mixed energy modes into pure spectral dynamics:
\begin{equation}
    \label{eq:unmixing}
    \mathbf{Dynamics}_{proj}(t) \approx \mathbf{W}^{-1} \left( \mathbf{\Sigma}_r \mathbf{V}_r^* \right)
\end{equation}
Equation~\eqref{eq:unmixing} shows that the inverse eigenvector matrix $\mathbf{W}^{-1}$ acts as a spectral filter. It transforms the generic orthogonal basis ($\mathbf{V}_r^*$) into the mono-frequency evolution of the structural eigenmodes.

The exact modes used in this work, $\mathbf{\Phi}_{aug} = \mathbf{H}_2 \mathbf{V}_r \mathbf{\Sigma}_r^{-1} \mathbf{W}$ (Equation~\eqref{eq:DMD-modes}), are not restricted to the POD subspace. From Equation~\eqref{eq:A_tilde} and $\tilde{\mathbf{A}} \mathbf{W} = \mathbf{W} \mathbf{\Lambda}$, their coordinates in the POD basis are $\mathbf{U}_r^* \mathbf{\Phi}_{aug} = \tilde{\mathbf{A}} \mathbf{W} = \mathbf{W} \mathbf{\Lambda}$. Projecting $\mathbf{H}_1 \approx \mathbf{\Phi}_{aug} \cdot \mathbf{Dynamics}(t)$ onto the POD basis in the same way gives, for non-zero $\mu_j$,
\begin{equation*}
    \mathbf{Dynamics}(t) \approx \mathbf{\Lambda}^{-1} \mathbf{W}^{-1} \left( \mathbf{\Sigma}_r \mathbf{V}_r^* \right) = \mathbf{\Lambda}^{-1} \, \mathbf{Dynamics}_{proj}(t).
\end{equation*}
The exact-mode dynamics are therefore the projected dynamics with row $j$ scaled by $\mu_j^{-1}$. For a mono-frequency evolution, this scaling is a shift of one time step, so both formulations give the same frequencies and decay rates. In practice, the amplitudes $\mathbf{b}$ of the exact modes are obtained by projecting the data onto the modes, as in Equation~\eqref{eq:b}, and they evolve with the powers $\mu_j^k$. In continuous time, this is Equation~\eqref{eq:dmd_continuous}.

\begin{figure}[htbp]
    \centering
    \begin{subfigure}[b]{0.8\textwidth}
        \centering
        \includegraphics[width=\textwidth]{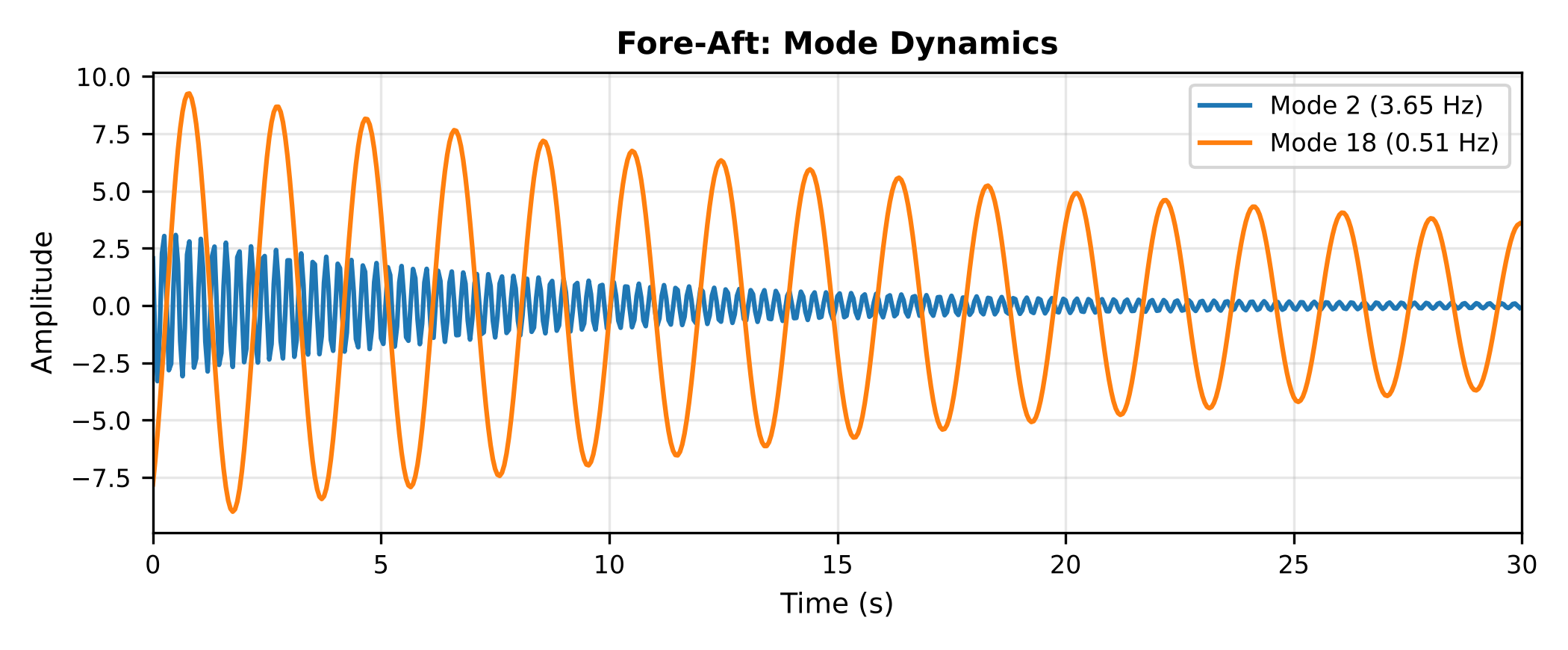}
        \caption{Fore-aft mode dynamics: separation of the first ($0.50$ Hz) and second ($3.71$ Hz) tower modes.}
        \label{fig:dynamics_fa}
    \end{subfigure}

    \vspace{0.5cm}

    \begin{subfigure}[b]{0.8\textwidth}
        \centering
        \includegraphics[width=\textwidth]{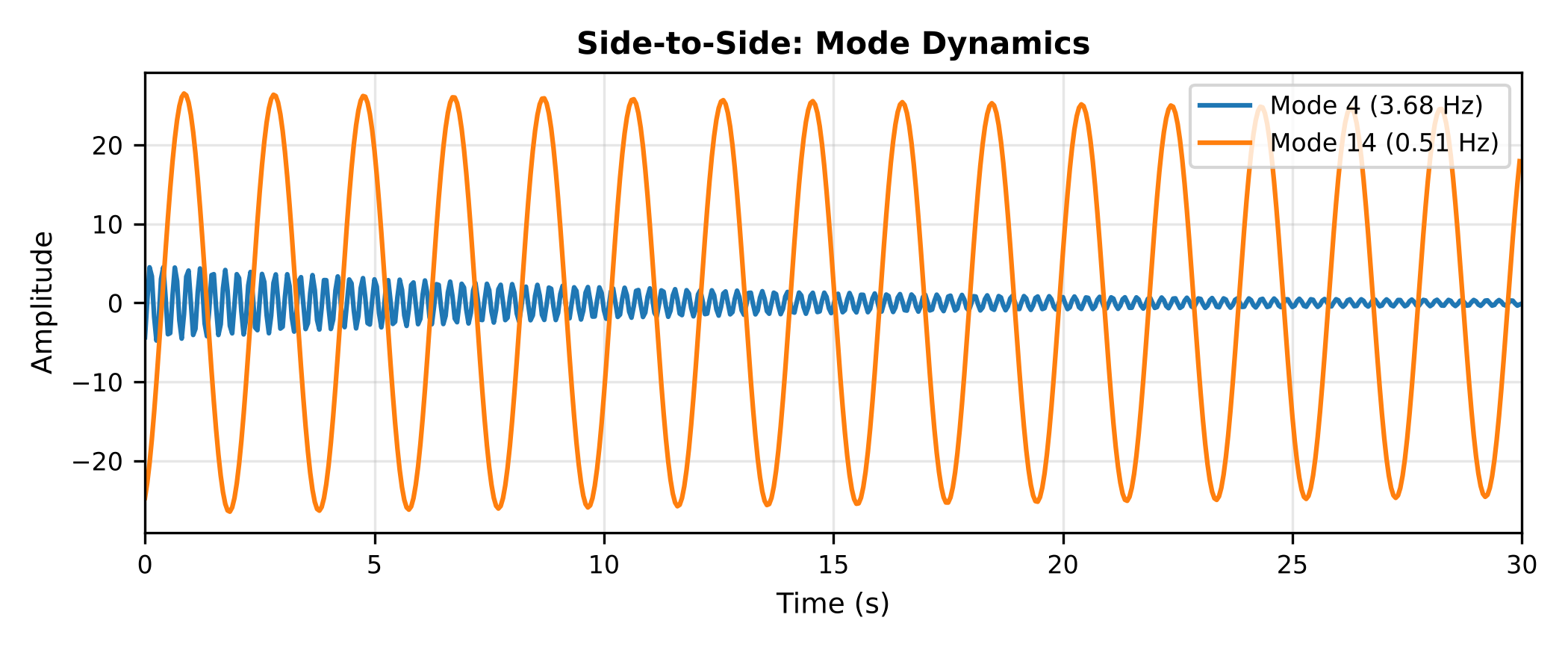}
        \caption{Side-to-side mode dynamics: separation of the first ($0.51$ Hz) and second ($3.60$ Hz) tower modes.}
        \label{fig:dynamics_ss}
    \end{subfigure}

    \caption{Reconstructed time-domain evolution of the identified structural modes: (a) fore-aft and (b) side-to-side. The trajectories are the decoupled mono-frequency dynamics of Equation~\eqref{eq:dmd_continuous}, which isolate the exponential decay of each eigenmode from orthogonal energy mixing.}
    \label{fig:temporal_dynamics}
\end{figure}

Figure~\ref{fig:temporal_dynamics} presents these decoupled trajectories for the fore-aft and side-to-side directions, obtained with $d = 60$ and $r = 24$. The time histories show a clear separation of scales in the modal amplitudes. The first tower modes ($0.50$~Hz in FA, Figure~\ref{fig:temporal_dynamics}(a), and $0.51$~Hz in SS, Figure~\ref{fig:temporal_dynamics}(b)) dominate the system energy, with normalized amplitudes several times larger than those of the second tower modes. Despite this difference in energy, the spectral separation of Hankel-DMD resolves the low-energy second bending modes at $3.71$~Hz (FA) and $3.60$~Hz (SS) without their being masked by the dominant signal.

The extracted time histories show a smooth, mono-frequency sinusoidal decay, without the beating typical of harmonic interference or raw sensor data. This visually confirms the spectral selectivity of the method. Unlike the irregular superposition of waves and rotor vibrations in the input data, the reconstructed dynamics represent the isolated structural response.

\subsubsection{Spectral Stability and Mode Selection}
\begin{figure}[htbp]
    \centering
    \begin{subfigure}[b]{0.48\textwidth}
        \centering
        \includegraphics[width=\textwidth]{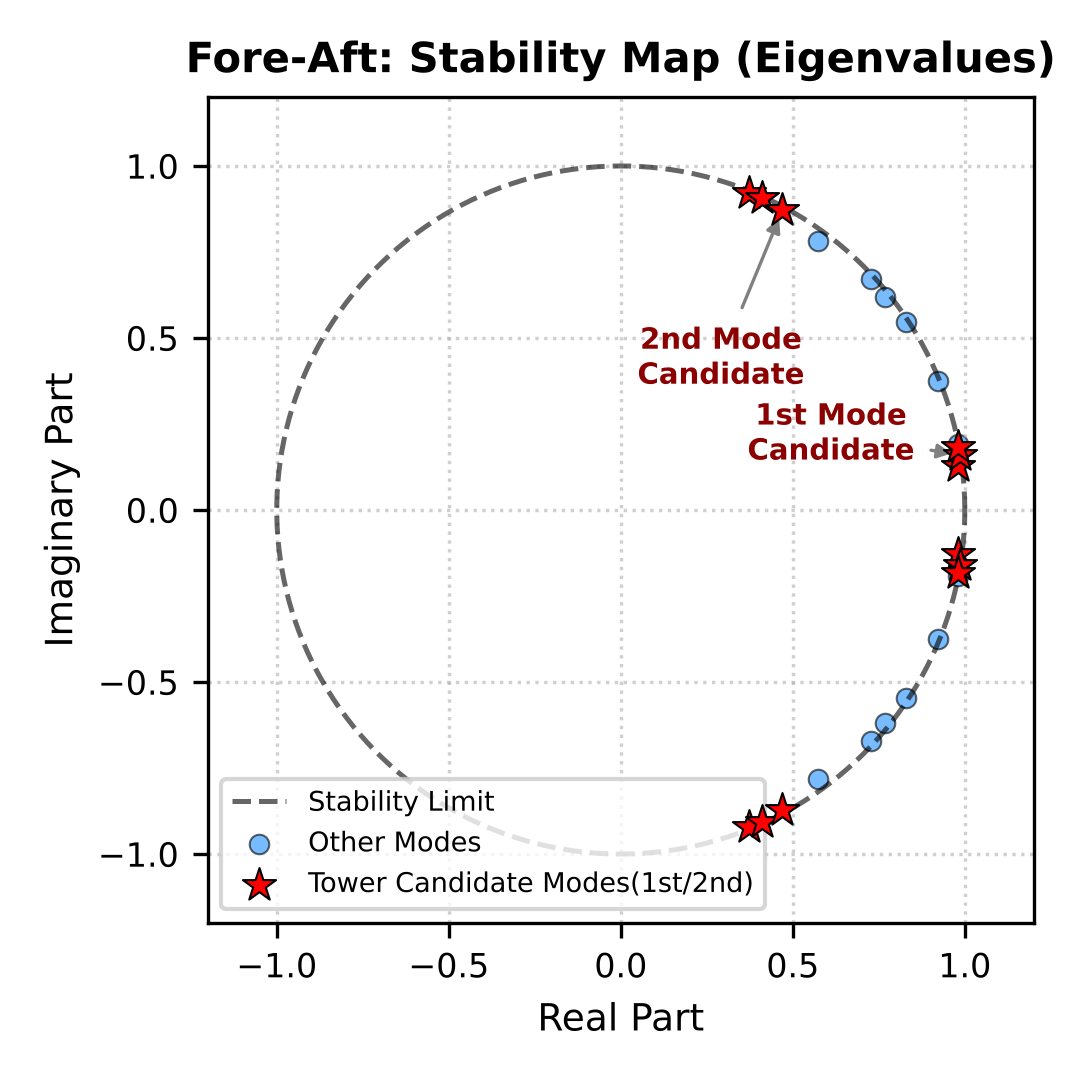}
        \caption{Fore-aft stability map}
        \label{fig:stab_fa}
    \end{subfigure}
    \hfill
    \begin{subfigure}[b]{0.48\textwidth}
        \centering
        \includegraphics[width=\textwidth]{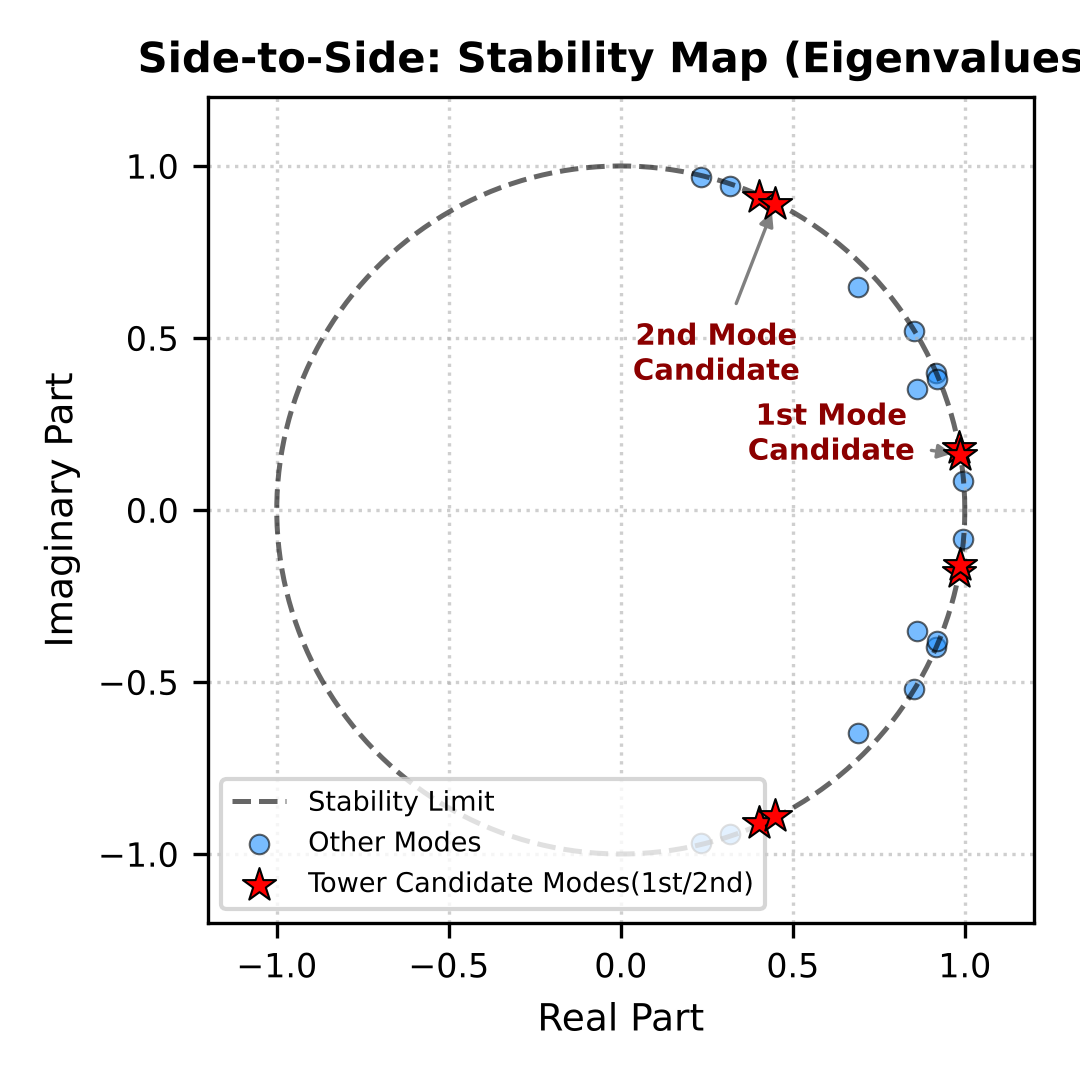}
        \caption{Side-to-side stability map}
        \label{fig:stab_ss}
    \end{subfigure}

    \caption{Discrete eigenvalue spectra of the identified system. The structural modes (red stars) lie within the unit circle, confirming numerical stability and positive physical damping. The position of each pole sets the frequency (angle) and decay rate (radius) seen in the time histories.}
    \label{fig:stability_maps}
\end{figure}

The discrete eigenvalue spectrum in Figure~\ref{fig:stability_maps} gives a further check of the physical validity of the identified dynamics. The eigenvalues $\mu_j$ govern the temporal evolution of Equation~\eqref{eq:dmd_continuous} through the relation $\mu_j = e^{(\sigma_j + i\omega_j)\Delta t}$.

The selected tower modes (red stars) lie strictly inside the unit circle ($|\mu_j|<1$), consistent with stable, positively damped structural dynamics. The remaining poles (blue circles) are not interpreted as structural modes; they may include low-energy, forcing-related or numerical components. Structural interpretation is therefore restricted to the stable and repeatable poles selected by their spectral and mode-shape characteristics.

In typical CFD applications (e.g., wake flow analysis), eigenvalues lie well inside the unit circle, reflecting rapid viscous dissipation of transient flow structures. In contrast, the eigenvalues identified for the FOWT lie close to the unit circle ($|\mu| \approx 1$). This is physically consistent with steady turbine operation. Continuous energy input from the stochastic wind excitation balances the structural damping, so the oscillatory response is sustained rather than decaying.

\subsection{Reconstruction Accuracy}

To evaluate the digital twin as a virtual sensor, the framework was tested on a graded set of blind test scenarios. The first three represent realistic operational failures, from the loss of specific tower nodes to a massive data loss in which nearly 50\% of the sensor network is unavailable. The dependence of performance on the number and placement of sensors cannot be assessed from three fixed patterns, so these are complemented by a systematic variation of the active set.

The test cases are detailed in Table~\ref{tab:virtual_sensing_cases} and were all evaluated with the time-blocked protocol of Section~\ref{sec:vs_protocol}. Sensors 0--8 are translational accelerations (Acc.) and sensors 9--17 are bending moments (Moment). Cases A--C are the three failure scenarios above. Two practical layouts are also considered. L1 uses a nacelle accelerometer only, while L2 combines a nacelle accelerometer with a tower-base strain gauge. For each number of active channels, \(q \in \{1,2,3,4,6,9,14\}\), six additional layouts are selected randomly using a fixed seed. The random layouts show how accuracy changes with sensor sparsity, while L1 and L2 represent practical turbine instrumentation.

\begin{table}[htbp]
    \centering
    \caption{Virtual-sensing test scenarios. Sensors 0--8: accelerations, 9--17: bending moments
    (tower base to nacelle). All scenarios use the time-blocked leave-one-OLC-out protocol of
    Section~\ref{sec:vs_protocol}.}
    \label{tab:virtual_sensing_cases}
    \footnotesize
    \renewcommand{\arraystretch}{1.2}
    \begin{tabular}{c c l p{0.40\textwidth}}
        \toprule
        \textbf{Scenario} & \textbf{Test OLCs}& \textbf{Hidden Sensors (Indices)} & \textbf{Failure Description} \\
        \midrule
        Case A & 5 & 2, 3, 5, 7, 11, 12, 14, 16 & \textbf{Massive Failure:} 8 sensors lost (distributed along the tower). \\
        Case B & 7 & 0, 8, 9, 17 & \textbf{Boundary Loss:} Tower base (0) and top/nacelle (17) lost. \\
        Case C & 2 & 0, 8, 9, 17 & \textbf{Low-Wave Condition:} Same layout as Case B.\\
        \midrule
        L1 & 1--12 & all but 8 & \textbf{Nacelle accelerometer only} ($q = 1$). \\
        L2 & 1--12 & all but 8, 9 & \textbf{Nacelle accelerometer and tower-base moment} ($q = 2$). \\
        Random & 1--12 & $18 - q$ channels & Six random layouts per $q \in \{1,2,3,4,6,9,14\}$ (fixed seed). \\
        \bottomrule
    \end{tabular}
\end{table}

\subsection{Reconstruction Fidelity and Sensor Sparsity}\label{sec:vs_results}

Before the estimators are compared, the effective dimension of the simulated response must be established, because it determines which sensor-loss scenarios can distinguish between the estimators. A principal component analysis (PCA) of the 18 standardized channels shows that, within the analysis band, five principal components explain 99.997\% of the variance in the fore-aft direction, while six explain 99.9999\% in the side-to-side direction. In every individual OLC, five components explain at least 99.99\% (FA) and 99.94\% (SS). This low instantaneous dimension reflects the reduced-order structural model, in which tower deflection in each direction is represented by two assumed modes, with the remaining response governed by a small number of platform and rotor degrees of freedom (Table~\ref{tab:shape_check}). Consequently, when five or six suitably placed channels remain and span the dominant response subspace, the hidden channels can be reconstructed approximately as a static linear function of the active ones.

Table~\ref{tab:reconstruction_metrics} confirms this directly. In Cases A--C, which retain $q = 10$ and $q = 14$ active channels, every estimator, including static least squares, reaches a mean $R^2 \ge 0.999$ with a minimum over the hidden sensors of at least $0.997$. The NRMSE of the two Hankel-DMD realizations stays below $0.2\%$ for every hidden sensor, including the mid-tower accelerometer (sensor 5) and the nacelle moment (sensor 17). These scenarios therefore confirm that the identified model is consistent with the data, but they do not discriminate between estimators. The informative regime for a virtual sensor is $q < 5$, where the instantaneous problem is under-determined and the time-delay structure of Proposition~\ref{prop:identifiability} is required.

\begin{figure}[htbp]
    \centering
    \includegraphics[width=\textwidth]{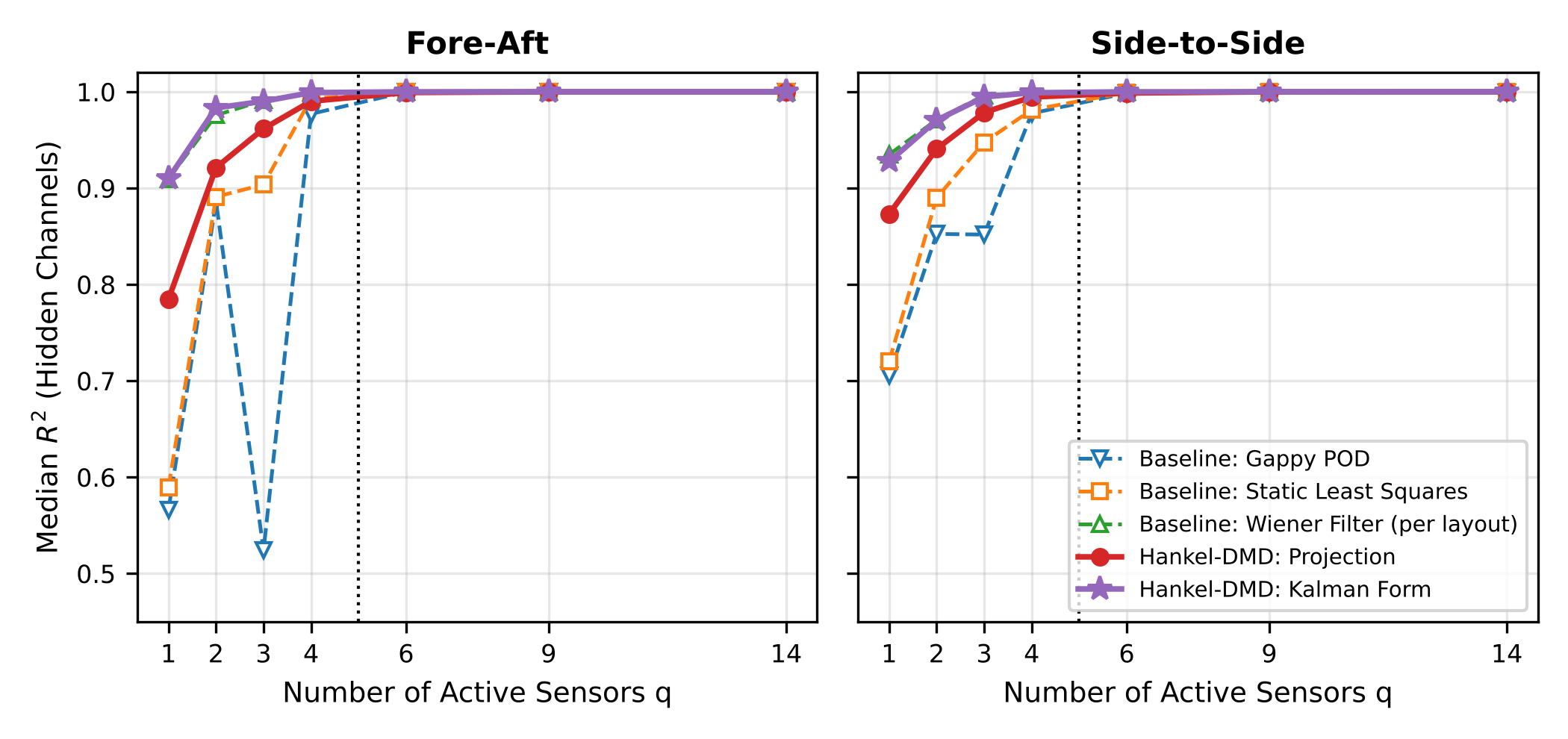}
    \caption{Median $R^2$ of the hidden channels versus the number of active sensors (six random layouts per $q$,
    time-blocked leave-one-OLC-out): baseline estimators (dashed lines, open markers) and the two realizations of the
    Hankel-DMD model (solid lines, filled markers). The dotted line marks the effective instantaneous dimension of the
    simulated response, below which the delay embedding becomes necessary.}
    \label{fig:vs_sensors}
\end{figure}

Figure~\ref{fig:vs_sensors} reports the accuracy over the random layouts and isolates that regime. For $q \ge 6$ all estimators reconstruct the hidden channels with median $R^2 \ge 0.999$. Below that, the static estimators lose accuracy and become sensitive to placement (static least squares: $R^2 = 0.59$ and $0.72$ at $q = 1$ for FA and SS; gappy POD: $0.52$ at $q = 3$ in FA). The estimators that use the delay window remain accurate. The regularized projection of Equation~\eqref{eq:tikhonov} reaches $0.78/0.87$ at $q = 1$ (FA/SS), $0.92/0.94$ at $q = 2$ and $0.96/0.98$ at $q = 3$, that is, $0.03$--$0.20$ above static least squares. This is the behaviour predicted by the Vandermonde structure of Proposition~\ref{prop:identifiability}. The layout-specific Wiener filter is more accurate still ($0.91/0.93$ at $q = 1$), as it must be, since it is the minimum-variance linear estimator for that window and layout. However, a separate Wiener map must be fitted for each sensor layout, whereas the Hankel-DMD operator remains unchanged when the active sensors change. The stochastic extension of Section~\ref{sec:kalman} closes that gap. The Kalman form reaches the Wiener accuracy to within $0.01$ for every $q$, while using a single model for all layouts. It propagates the modal state between samples instead of re-fitting the amplitudes in each window, so the unmeasured forcing is absorbed by the process noise rather than by the fit. Its accuracy saturates with rank (less than $0.01$ change from $r = 120$ to $r = 200$), whereas the projection has an optimum at $r = 50$--$90$.

The two practical layouts are reported at the foot of Table~\ref{tab:reconstruction_metrics} and illustrated in Figure~\ref{fig:vs_sparse}. With the nacelle accelerometer alone (L1), the projection reconstructs the hidden channels with mean $R^2 = 0.78$ (FA) and $0.88$ (SS) and the Kalman form with $0.90$ and $0.93$. The tower-base bending moment, the channel most relevant for fatigue, is dominated by the first mode and is already recovered by static least squares ($R^2 = 0.91$ and $0.94$). The projection gives $0.89$ and $0.96$, and the Kalman form $0.97$ and $0.99$. Adding a tower-base strain gauge (L2) raises the mean accuracy to $0.87/0.92$ for the projection and $0.96/0.97$ for the Kalman form. A single nacelle accelerometer therefore suffices to track the fatigue-driving base moment. This is the practically relevant finding for sparsely instrumented turbines.

\begin{table}[htbp]
\centering
\caption{Virtual-sensing accuracy (mean $R^2$ over hidden channels; minimum in brackets) for the scenarios of Table~\ref{tab:virtual_sensing_cases}, using $d=60$ and cross-fitted ranks. Time-blocked leave-one-OLC-out excludes the test OLC and test interval from training and hyper-parameter selection. Cases A--C use the OLC specified in Table~\ref{tab:virtual_sensing_cases}; L1 and L2 report the median over 12 held-out OLCs. $\kappa$ is the condition number of $\mathbf{\Phi}_{sparse}$. Per-sensor results are given in Table~\ref{tab:vs_per_sensor}.}
\label{tab:reconstruction_metrics}
\footnotesize
\resizebox{\textwidth}{!}{%
\begin{tabular}{ll c c ccccc}
\toprule
 & & & & \multicolumn{3}{c}{\textit{Baselines}} & \multicolumn{2}{c}{\textit{Hankel-DMD}}\\
\cmidrule(lr){5-7}\cmidrule(lr){8-9}
Dir. & Scenario & $q$ & $\kappa$ & \textbf{Gappy POD} & \textbf{Static LS} & \textbf{Wiener} & \textbf{Projection} & \textbf{Kalman form}\\
\midrule
FA & Case A & 10 & $1\times10^{3}$ & 1.000 [1.000] & 1.000 [1.000] & 1.000 [1.000] & \textbf{1.000} [1.000] & \textbf{1.000} [1.000]\\
 & Case B & 14 & $2\times10^{3}$ & 1.000 [0.999] & 1.000 [1.000] & 1.000 [1.000] & \textbf{1.000} [1.000] & \textbf{1.000} [1.000]\\
 & Case C & 14 & $2\times10^{3}$ & 1.000 [1.000] & 1.000 [1.000] & 1.000 [1.000] & \textbf{1.000} [1.000] & \textbf{1.000} [1.000]\\
 & L1 & 1 & $3\times10^{14}$ & 0.562 [$-0.040$] & 0.586 [$-0.009$] & 0.894 [0.223] & \textbf{0.784} [$-0.165$] & \textbf{0.898} [0.269]\\
 & L2 & 2 & $2\times10^{9}$ & 0.017 [$-2.163$] & 0.698 [0.034] & 0.962 [0.641] & \textbf{0.870} [0.077] & \textbf{0.963} [0.691]\\
\midrule
SS & Case A & 10 & $1\times10^{3}$ & 0.999 [0.997] & 1.000 [1.000] & 1.000 [1.000] & \textbf{1.000} [1.000] & \textbf{1.000} [1.000]\\
 & Case B & 14 & $2\times10^{3}$ & 0.999 [0.998] & 1.000 [1.000] & 1.000 [1.000] & \textbf{1.000} [0.999] & \textbf{1.000} [1.000]\\
 & Case C & 14 & $2\times10^{3}$ & 0.999 [0.997] & 1.000 [1.000] & 1.000 [1.000] & \textbf{1.000} [1.000] & \textbf{1.000} [1.000]\\
 & L1 & 1 & $3\times10^{14}$ & 0.731 [0.091] & 0.737 [0.109] & 0.936 [0.735] & \textbf{0.883} [0.643] & \textbf{0.927} [0.733]\\
 & L2 & 2 & $3\times10^{9}$ & $-0.043$ [$-3.645$] & 0.788 [0.238] & 0.978 [0.922] & \textbf{0.921} [0.762] & \textbf{0.968} [0.912]\\
\bottomrule
\end{tabular}}
\end{table}

\begin{figure}[htbp]
    \centering
    \includegraphics[width=\textwidth]{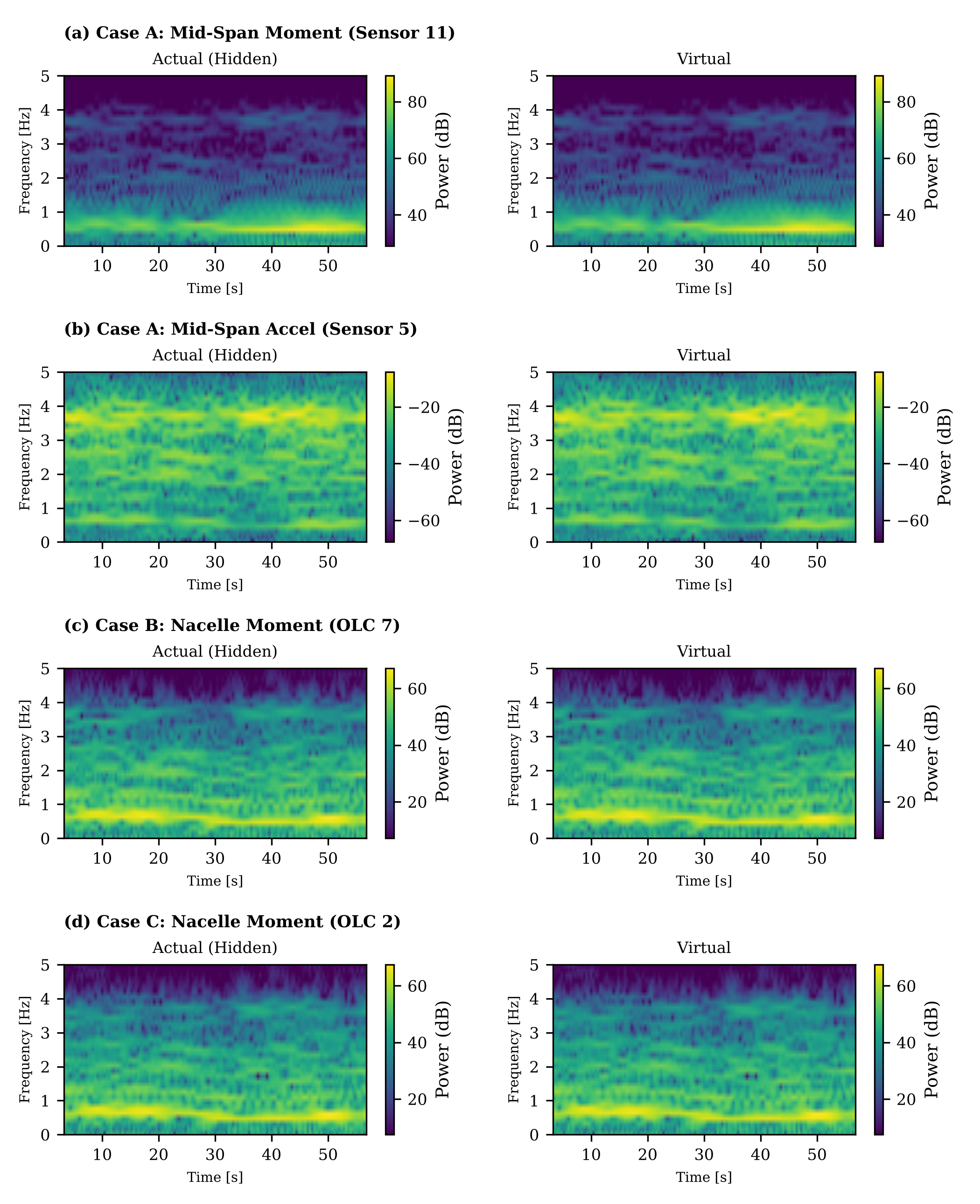}
    \caption{Time--frequency comparison of the hidden (actual) and reconstructed (virtual) channels for Cases A--C on
    unseen data (time-blocked protocol, Hankel-DMD projection). Rows pair each hidden channel with its reconstruction,
    showing that the reconstruction reproduces both the dominant structural frequency bands and their temporal intermittency.}
    \label{fig:reconstruction_results}
\end{figure}

\subsection{Physical Interpretation of Reconstruction Challenges}

\begin{figure}[htbp]
    \centering
    \includegraphics[width=\textwidth]{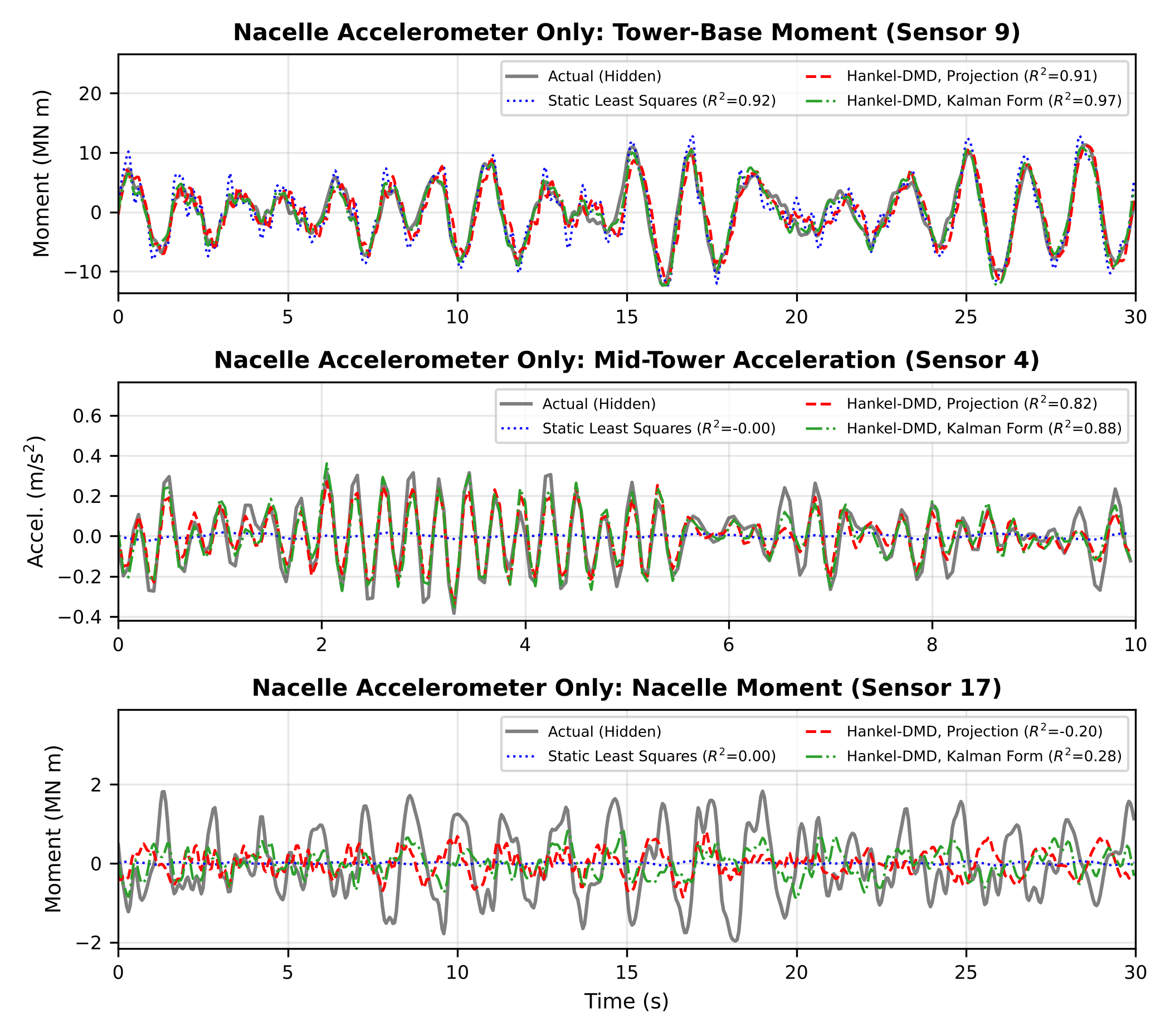}
    \caption{Reconstruction from the nacelle accelerometer alone (layout L1, OLC 7, unseen data) by the Hankel-DMD
    projection and its Kalman form, compared with static least squares: tower-base moment, mid-tower acceleration near
    the node of the first coupled mode, and nacelle moment.}
    \label{fig:vs_sparse}
\end{figure}

Where the reconstruction is difficult, the sparse layouts show why. They expose the information that the active channels do not carry (Table~\ref{tab:vs_l1_channels}, Figure~\ref{fig:vs_sparse}). The limitation is location-dependent and is set by the modal content of each channel rather than by the estimator.

\subsubsection{Modal Nodal Interaction (Mid-Tower / Sensors 3--6)}\label{sec:vs_physics}
The accelerometers near the node of the first coupled mode are the most difficult channels for a static estimator. At these locations, the identified acceleration mode shape satisfies $a_1(z_{\rm node})\approx 0$, as described by the kinematic decomposition in Table~\ref{tab:shape_check}, so the first coupled-mode contribution is small and the relative influence of the other modal components becomes more important. A single instantaneous measurement cannot reliably separate these contributions. With only the nacelle accelerometer active, static least squares decreases to $R^2=0.01$--$0.32$ in FA and $0.11$--$0.61$ in SS at sensors 3--6. The delay-based estimators use the different modal frequencies ($0.50$ and $3.71$~Hz in FA) to separate these contributions through the temporal Vandermonde structure of Proposition~\ref{prop:identifiability}. They recover $R^2=0.82$--$0.86$ (FA) and $0.83$--$0.92$ (SS) with the projection, and $0.88$--$0.91$ and $0.86$--$0.94$ with the Kalman form. This provides the clearest evidence in the simulation study that the time-delay embedding adds information beyond a static modal basis.

\subsubsection{Boundary Loads and Observability (Sensor 17)}\label{sec:vs_boundary}
A distinct conflict between inertial response and direct forcing is observed at the boundaries. The fore-aft moment immediately below the yaw bearing (sensor 17) is the least observable channel in the model. With the nacelle accelerometer alone, every estimator remains below $R^2 = 0.3$ in the fore-aft direction, while in the side-to-side direction the projection and the Kalman form reach $0.65$ and $0.86$ (Table~\ref{tab:vs_l1_channels}). This is consistent with the fore-aft moment at that location containing a substantial contribution from loads transferred from the rotor, and turbulence-driven thrust variations, that is not fully represented by the tower acceleration. Reconstructing this component would require either an additional load-sensitive measurement or a model of the rotor forcing, which lies outside the output-only setting adopted here. This therefore represents a limit of the information available to the present output-only estimator and is reported as such rather than absorbed by raising the rank.

\subsubsection{Conditioning and Robustness}\label{sec:vs_robust}
The numerical stability of the inversion under extensive sensor failure is quantified by the condition number of $\mathbf{\Phi}_{sparse}$ reported in Table~\ref{tab:reconstruction_metrics}. It remains $\mathcal{O}(10^{2})$--$\mathcal{O}(10^{3})$ for $q \ge 6$ and exceeds $10^{6}$ for $q \le 2$ and for $q = 3$--$4$ at the larger selected ranks. For \(q=1\) and \(d=60\), \(\mathbf{\Phi}_{\mathrm{sparse}}\in\mathbb{C}^{60\times70}\), so \(\operatorname{rank}(\mathbf{\Phi}_{\mathrm{sparse}})\le60<70=r\), and the 70 modal coefficients in \(\mathbf b_k\) are \textit{not} uniquely identifiable in the sense of Proposition~\ref{prop:identifiability}.  The reported reconstruction there reflects the regularized estimator of Equation~\eqref{eq:tikhonov} and should not be read as unique recovery of the modal coefficients. We state this explicitly because the reconstruction accuracy at $q = 1$ is nonetheless high.

Three pre-declared tests were run for $q \le 3$ and layout L2, with all hyper-parameters fixed to the values selected above (Table~\ref{tab:vs_robust}). These are (T1) training on 600~s of a single OLC; (T2) training on the calm sea states ($H_s = 1$--$2$~m) and testing on the rough ones ($H_s = 3$--$4$~m), and vice versa; and (T3) adding $5\%$ white noise, relative to the standard deviation of each channel, to all channels in training and testing, with the reconstruction scored against the noise-free signals. T1 and T2 change the median $R^2$ by at most $0.04$ for every estimator and every $q \le 3$. This is consistent with the invariance of the structural poles across sea states established in Section~\ref{sec:sysid}. Within the tested range of sea states, the reconstruction is insensitive to these changes in the excitation statistics of the training data. T3 separates the two realizations sharply. The projection, whose Tikhonov parameter was selected on noise-free data, fails at $q = 1$ ($R^2 = 0.14$ in FA and $-0.25$ in SS) and degrades at $q = 2$ ($0.79/0.82$). The noise is amplified through the ill-conditioned sparse mode matrix, as the bound in Section~\ref{sec:virtual_sensing} predicts. The Kalman form, which weights the measurements through the noise covariance $\mathbf{R}$ identified from the same noisy training data, retains $R^2 = 0.86/0.87$ at $q = 1$ and $0.96/0.94$ at $q = 2$, equalling the Wiener filter. With noisy measurements, the projection regularization parameter \(\alpha\) should therefore be selected at the corresponding noise level; otherwise, the stochastic extension, which accounts for measurement noise through \(\mathbf{R}\), should be used. Both are applied to the measured data of Section~\ref{sec:focal}.

The time--frequency analysis of Figure~\ref{fig:reconstruction_results} gives insight into the aero-hydro-elastic coupling of the system. For the nacelle moment (sensor 17), the dominant frequency bands vary with time, with short periods of high energy rather than continuously strong response.

These time-varying spectral bands demonstrate that the digital twin can represent non-stationary excitation. Unlike the tower-base response, which is dominated by low-frequency platform surge and sway, the nacelle moment contains strong rotor harmonics ($1P/3P$) whose frequencies vary with rotor speed. As the variable-speed controller adjusts the rotor speed under turbulent wind, these harmonics shift relative to the structural eigenfrequencies, producing temporary strengthening or weakening of the response. The spectrogram captures these periods of frequency alignment and separation: bright regions indicate stronger spectral response when excitation approaches a structural mode, whereas weaker or disappearing bands indicate movement away from the modal frequency or reduced excitation intensity. The broader high-frequency energy in the mid-span acceleration is expected because acceleration scales with \(\omega^2\), thereby amplifying higher-frequency modal and harmonic contributions compared with the bending-moment response, which depends mainly on modal curvature at that location. The virtual reconstruction follows these rapid spectral changes closely. This confirms that the Hankel-DMD model captures the variable-speed and strongly damped response characteristics of the floating wind turbine.

\begin{figure}[!t]
    \centering
    \includegraphics[
        width=0.92\textwidth,
        height=0.58\textheight,
        keepaspectratio
    ]{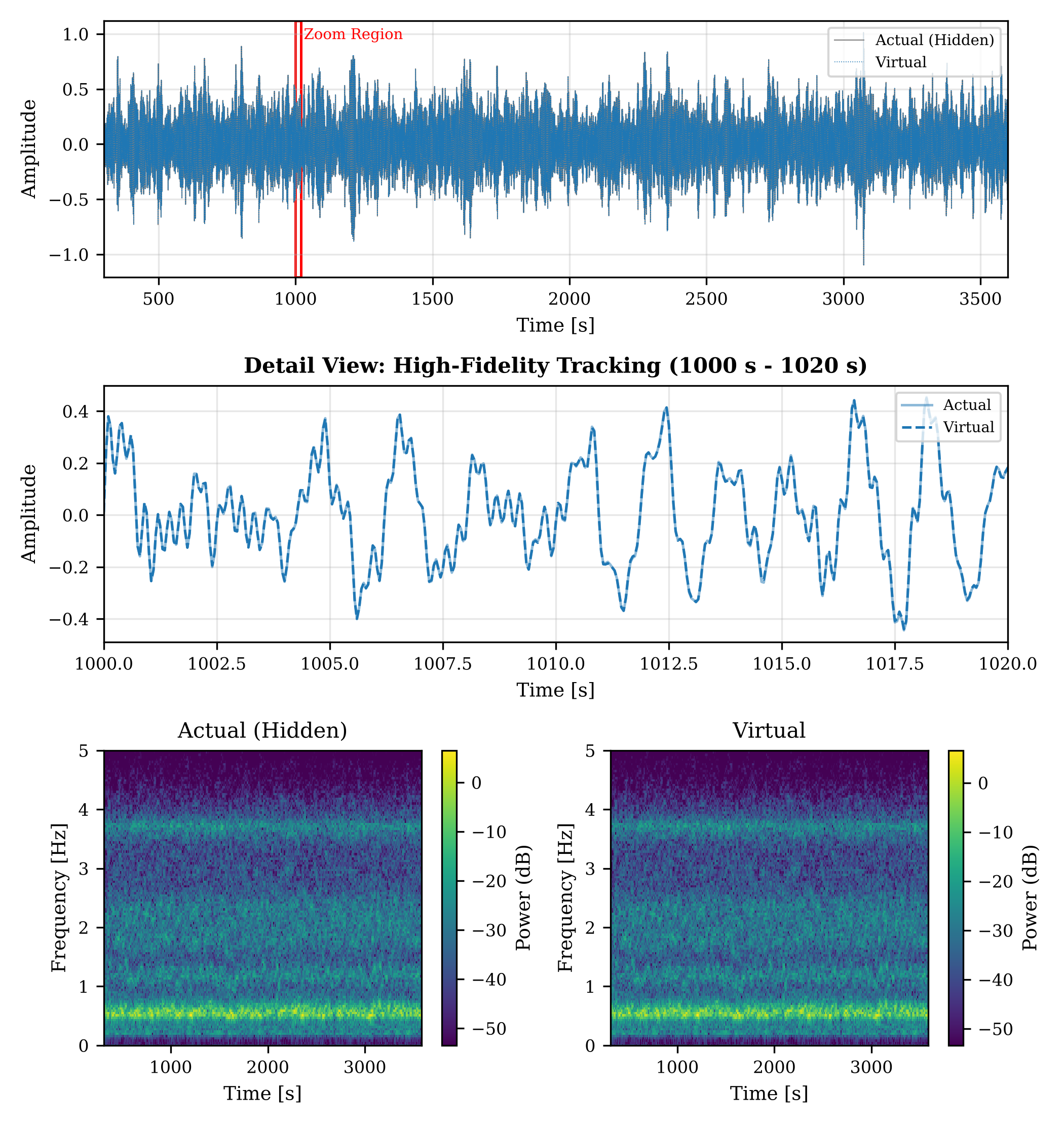}
    \caption{Long-term reconstruction of sensor 8 (nacelle acceleration) over $3600$~s of OLC 7 with the Case B layout.}
    \label{fig:sensor8_longterm}
\end{figure}

The one-hour real-time reconstruction in Figure~\ref{fig:sensor8_longterm} demonstrates the long-duration stability of the virtual-sensing model. The virtual response closely follows the hidden measurement over the full record, including the short-time fluctuations shown in the enlarged interval. The corresponding spectrograms also show close agreement in the dominant low- and high-frequency energy bands, indicating that the Hankel-DMD reconstruction preserves the principal time-varying spectral content over extended operation.

\subsection{Spectral Energy Analysis and Rank Optimization}

\begin{figure}[htbp]
    \centering

    \includegraphics[width=0.9\textwidth]{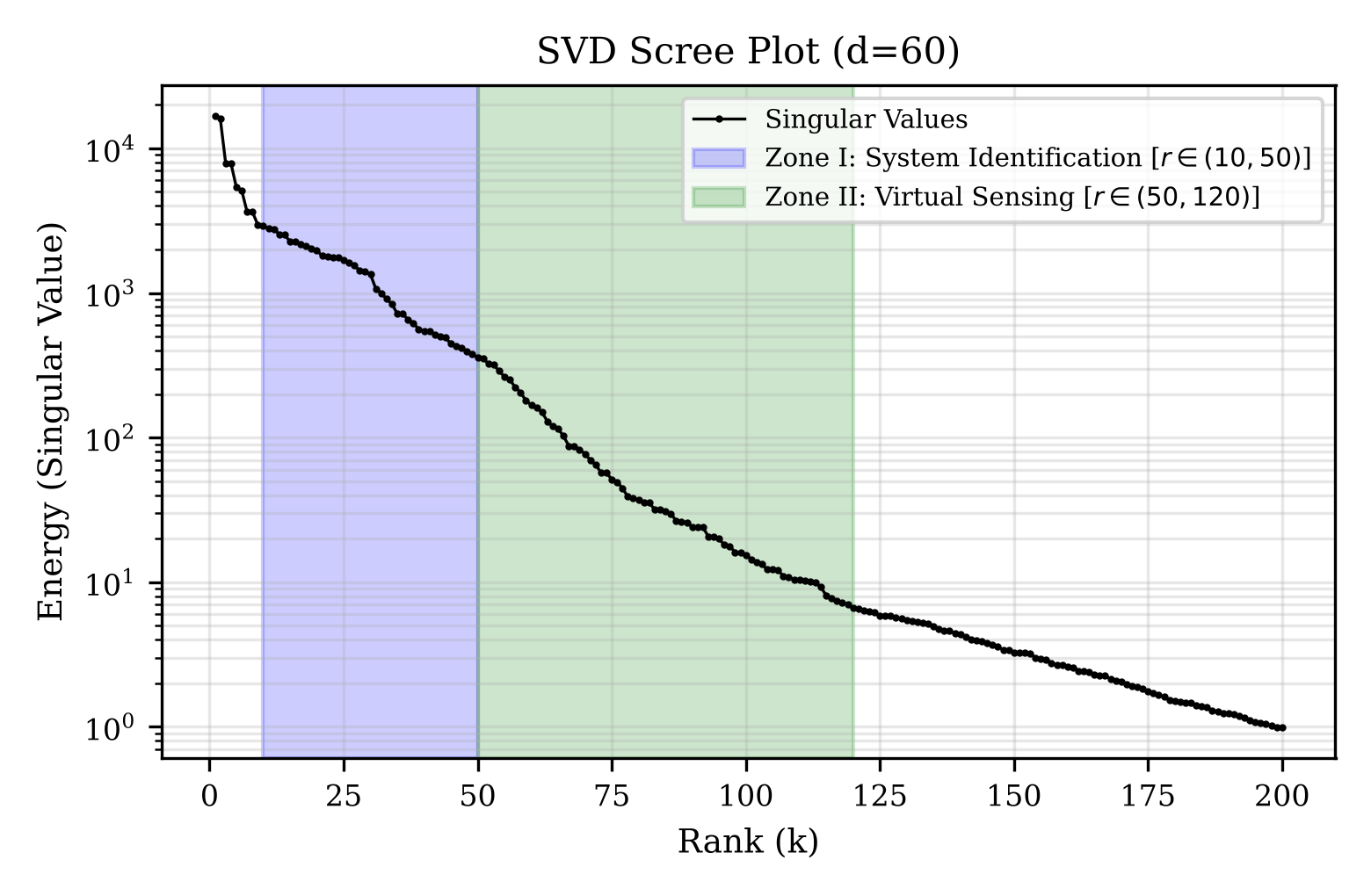}
    \caption{Singular value spectrum ($\sigma_k$) of the Hankel matrix ($d = 60$). The spectrum decays smoothly, with no noise floor, as expected for noise-free simulated data. \textbf{Zone I} is the high-energy subspace of the global structural modes, used for operational modal analysis. \textbf{Zone II} is the sub-dominant, low-energy subspace selected by cross-validation for virtual sensing. Zone II is needed to resolve local boundary loads (e.g., sensor 17) and nodal-point dynamics (e.g., sensor 5) that are energetically distinct from the global dynamics.}
    \label{fig:scree_plot}
\end{figure}

The accuracy at critical nodes depends on the SVD truncation rank, so we state explicitly how that rank was chosen. The SVD of the Hankel matrix shows the energy distribution of the system (Figure~\ref{fig:scree_plot}). The singular values decrease smoothly over about four orders of magnitude and do not show a distinct noise floor, consistent with the absence of measurement noise in the simulated data. The rank cannot be read from the spectrum alone, so two operating ranges are used instead:
\begin{itemize}
    \item \textbf{Zone I ($10 \le k \le 50$):} The dominant subspace capturing the global tower and platform modes. Ranks in this range give stable modal parameters (Section~\ref{sec:damping}) and are used for system identification.
    \item \textbf{Zone II ($50 \le k \le 120$):} A heavy tail containing sub-dominant, low-energy components. Ranks in this range are selected by cross-validation for virtual sensing.
\end{itemize}

A moderate rank ($r \approx 24$--$35$) is sufficient for system identification, whereas the cross-fitted ranks for virtual sensing are $50$--$120$. The estimate of a hidden channel benefits from low-energy components that are not needed to identify the dominant modes. These components carry the fine-scale spatial correlations that are energetically overshadowed by the fundamental sway but are physically essential at boundaries and nodal points. The rank is thus selected by held-out accuracy rather than by an energy criterion, which also avoids attributing a physical meaning to the truncation point.

\subsection{Forecast Horizon under Stochastic Forcing}\label{sec:horizon}

\begin{figure}[htbp]
    \centering
    \includegraphics[width=\textwidth]{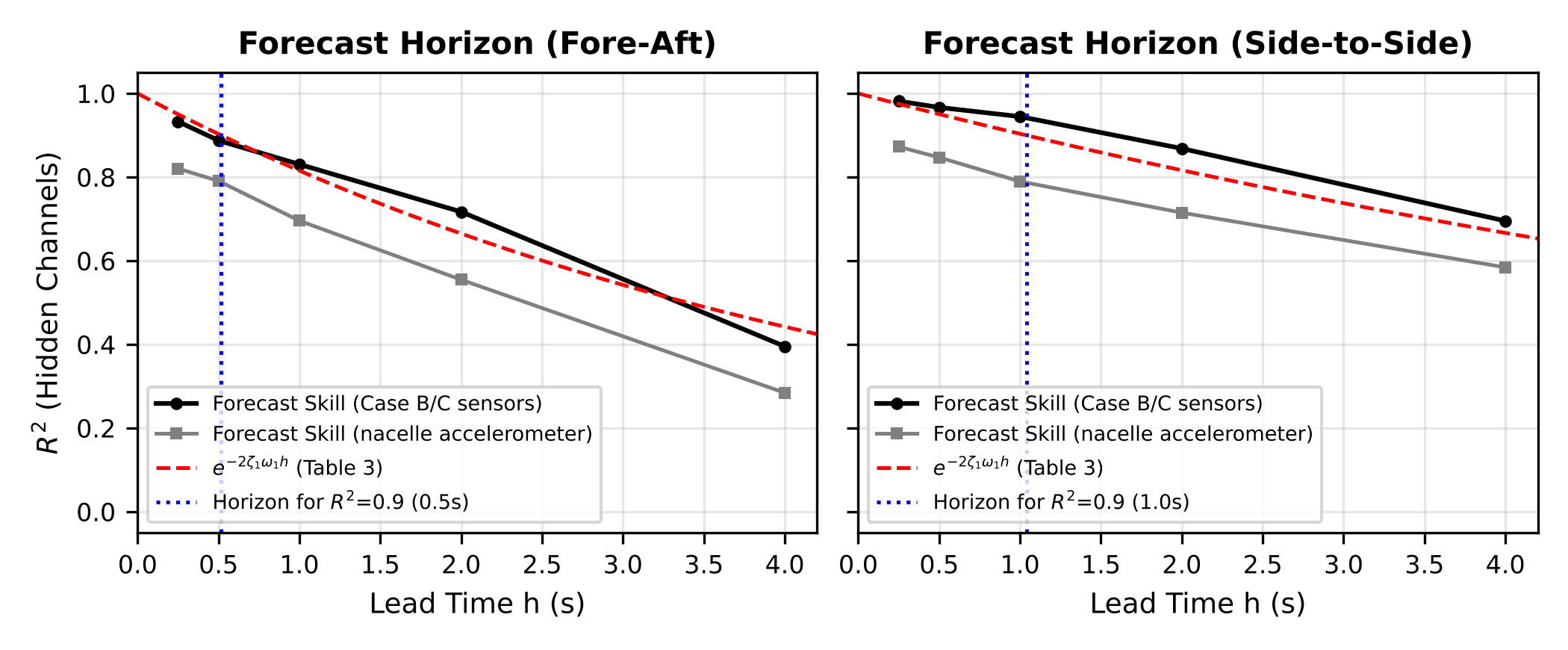}
    \caption{Open-loop forecast skill of the hidden channels versus lead time $h$ (median over the 12 held-out OLCs; Kalman form of the Hankel-DMD model, the projection giving practically the same skill for the Case B/C layout), compared with $e^{-2\zeta_1\omega_1 h}$ evaluated from the identified first-mode damping of Table~\ref{tab:sys_id_comparison}. The dotted line marks the lead time at which $R^2 = 0.9$.}
    \label{fig:lyapunov_limit}
\end{figure}

An important property of the digital twin is how long an estimate remains useful when no new measurements arrive, for example, when sensor data transmission is temporarily lost or when a known measurement delay must be addressed. This is the input-free prediction problem distinguished from reconstruction in Section~\ref{sec:stochastic_koopman}, and it is quantified separately here. The open-loop forecast of Equation~\eqref{eq:Y_rec} with $W = h > 1$ was evaluated at lead times $h = 0.25$--$4$~s under the same time-blocked protocol and cross-fitted hyper-parameters (Figure~\ref{fig:lyapunov_limit}). For the four hidden channels of Cases B and C the forecast skill decreases from $R^2 = 0.93$ at $h = 0.25$~s to $0.83$ at $h = 1$~s and $0.40$ at $h = 4$~s in the fore-aft direction, and from $0.98$ through $0.95$ to $0.70$ in the side-to-side direction. The projection and its Kalman form give practically the same values. With the nacelle accelerometer as the only input, the skill at $h = 1$~s is $0.55$ (FA) and $0.68$ (SS) for the projection and $0.70$ and $0.79$ for the Kalman form. The Kalman form filters the modal state over the whole measurement history rather than fitting it to a single window.

The forecast horizon follows directly from the identified modal frequency and damping. For a stable mode, the $h$-step conditional mean evolves as $\mathbf{\Lambda}^h\mathbf{b}_k$. Because the broadband excitation entering during the forecast interval is unmeasured, only the response determined by the current modal amplitudes remains predictable. For a lightly damped mode under approximately white excitation, this predictable fraction therefore decays at a rate governed by $\zeta_j\omega_j$.

\begin{equation}
    R_j^2(h)
    \approx
    \left(\frac{A_j(h)}{A_j(0)}\right)^2
    =
    \left(e^{-\zeta_j\omega_j h}\right)^2
    =
    e^{-2\zeta_j\omega_j h},
    \label{eq:horizon}
\end{equation}

\noindent and the horizon corresponding to a required accuracy $R^2_{req}$ is $h_{req} = \ln(1/R^2_{req}) / (2 \zeta_j \omega_j)$. With the first-mode parameters of Table~\ref{tab:sys_id_comparison} from correlation-domain Hankel-DMD, Equation~\eqref{eq:horizon} gives $0.82$ and $0.90$ at $h = 1$~s and $0.44$ and $0.67$ at $h = 4$~s for FA and SS, close to the measured skill of $0.83/0.95$ and $0.40/0.70$. The somewhat faster decay at the shortest lead times reflects the more strongly damped high-frequency content of the second tower modes, for which $\zeta_2 \omega_2 \approx 0.3$--$0.5~\mathrm{s}^{-1}$.

The predictability horizon of the virtual sensor is thus set by the damping of the dominant mode and by the unmeasured forcing, both of which the identified model quantifies. The longer horizon in the side-to-side direction is a direct consequence of its lower damping. For an accuracy of $R^2 \ge 0.9$, Equation~\eqref{eq:horizon} gives $h_{req} \approx 0.5$~s in the fore-aft and $1.0$~s in the side-to-side direction. This is the interval over which the twin can bridge missing measurements. In normal operation, the modal state is updated at the measurement rate. Modern offshore wind farms support this rate, since fibre-optic SCADA networks stream high-frequency logs at $1$--$50$~Hz \cite{Gonzalez2019}. The proposed framework therefore fits well within the bandwidth and telemetry constraints of existing operational infrastructure.

\section{Experimental Verification on Basin-Test Data}\label{sec:focal}

The results of Sections~\ref{sec:sysid} and \ref{sec:vs_results} rest entirely on simulated data. A reduced-order
structural model cannot by itself establish that the framework works on a physical floating turbine. The
identification and virtual-sensing procedures were therefore applied, without any modification, to measured data from
a scaled physical model.

\subsection{Data and Processing}\label{sec:focal_data}
The measurements come from Campaign~IV of the FOCAL programme \cite{Fowler2023ISOPE, Fowler2023FOCAL, FOCALdata}. They were taken on a
$1$:$70$ Froude-scaled model of the IEA Wind 15-MW turbine on the VolturnUS-S semi-submersible, with a flexible tower
and the ROSCO controller, tested in combined wind and waves at the University of Maine. The platform, rotor, tower and
controller all differ from the simulated spar, so the test is one of transferability rather than of calibration. Four
runs were used in which the same wind input (mean speed about $12$~m/s, mean rotor speed $7.3$~rpm) was combined with
four irregular wave realizations ($H_s \approx 3.2$~m, $T_p \approx 9$~s). The calibrated wind time series is identical
across the four runs, and the wave elevations are uncorrelated.This provides multiple statistically independent wave realizations under the same wind and nominal sea-state condition, allowing the same time-blocked validation design to be applied without treating the runs as different sea states.

All quantities are reported at full scale, where the model-scale sampling rate of $200$~Hz corresponds to $23.9$~Hz. Four channels are available per direction: tower-top and tower-base bending moments measured by six-component load cells, which provide \((F_x,F_y,F_z,M_x,M_y,M_z)\), and translational accelerations measured by accelerometers at the tower top (nacelle) and tower base (platform), using the \(x\)-component for fore-aft and the \(y\)-component for side-to-side motion. The fore-aft axis was identified as the mean rotor-thrust direction, based on a mean surge displacement of approximately \(14\,\mathrm{m}\) and a mean tower-base bending moment of approximately \(0.61\,\mathrm{GN\,m}\) measured along that axis. The tower-base accelerometer is mounted on the platform rather than on the flexible tower. Its measured signal agrees closely with the rigid-body platform acceleration predicted from the measured surge and pitch motions: the kinematic fit over $0.03$--$0.25$~Hz gives an effective lever arm of about $5$--$10$~m from the motion-capture reference point, while the mean coherence over $0.05$--$0.20$~Hz is $0.95$. Applying the same check to the tower-top accelerometer gives an effective lever arm of about 160 m, consistent with its much higher nacelle-level location. It is
therefore referred to below as the platform accelerometer. After discarding the first $600$~s of ramp-up, $8400$~s of
stationary operation per run were processed exactly as in Section~\ref{sec:preprocessing}: band-pass filtering
($0.25$--$5$~Hz for the tower modes, $0.01$--$0.2$~Hz for the platform modes), standardization, and each run treated as
a separate segment. The delay windows were scaled to the first-mode period estimated from the FDD peak (about
$2.4$~s), so that they span the same number of periods as in the simulation study. This gives $d \in \{43, 86, 129\}$ for
identification, $d = 86$ for virtual sensing and $i = 86$ for SSI-COV.

\subsection{Modal Identification}\label{sec:focal_sysid}
\begin{figure}[htbp]
    \centering
    \includegraphics[width=\textwidth]{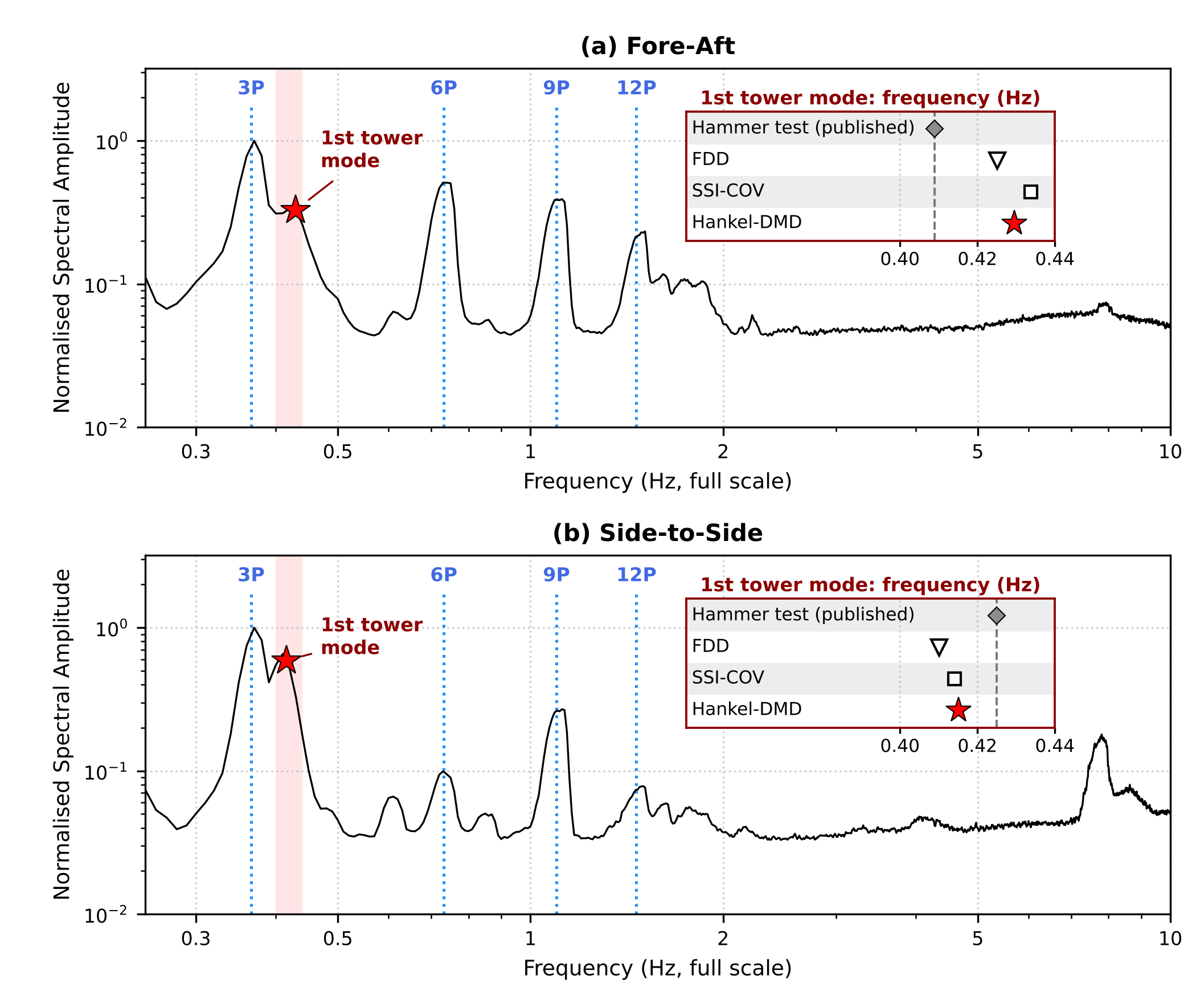}
    \caption{FOCAL Campaign~IV basin tests (full scale): measured spectra and first tower modes. (a) Fore-aft and (b) side-to-side normalized spectral amplitude over $0.25$--$10$~Hz. Dotted lines denote the rotor harmonics 3P, 6P, 9P and 12P; stars denote the first tower modes identified by Hankel-DMD. Insets compare the first tower-mode frequencies from the published hammer tests \cite{Fowler2023ISOPE}, FDD, SSI-COV and Hankel-DMD; values are listed in Table~\ref{tab:focal_sysid}. Second tower modes were not identified consistently and are therefore not marked.}
    \label{fig:focal_id}
\end{figure}

Figure~\ref{fig:focal_id} shows the measured spectra over $0.25$--$10$~Hz together with the rotor harmonics and the
identified first tower mode. Table~\ref{tab:focal_sysid} compares the methods. The measured response is dominated
by the rotor harmonics 3P, 6P, 9P and 12P. Their locations are fixed independently by the measured rotor speed, so the
separation of structural from excitation poles can be tested directly rather than inferred. Between the 3P and 6P
harmonics, Hankel-DMD identifies the first fore-aft and side-to-side tower modes at $0.430$ and $0.415$~Hz with
$3.9\%$ and $2.2\%$ damping. SSI-COV and FDD return the same modes, with frequencies within $0.01$~Hz and MAC $= 1.00$.
The frequencies identified separately for the four wave realizations differ by less than $0.004$~Hz.

\begin{figure}[htbp]
    \centering
    \includegraphics[width=\textwidth]{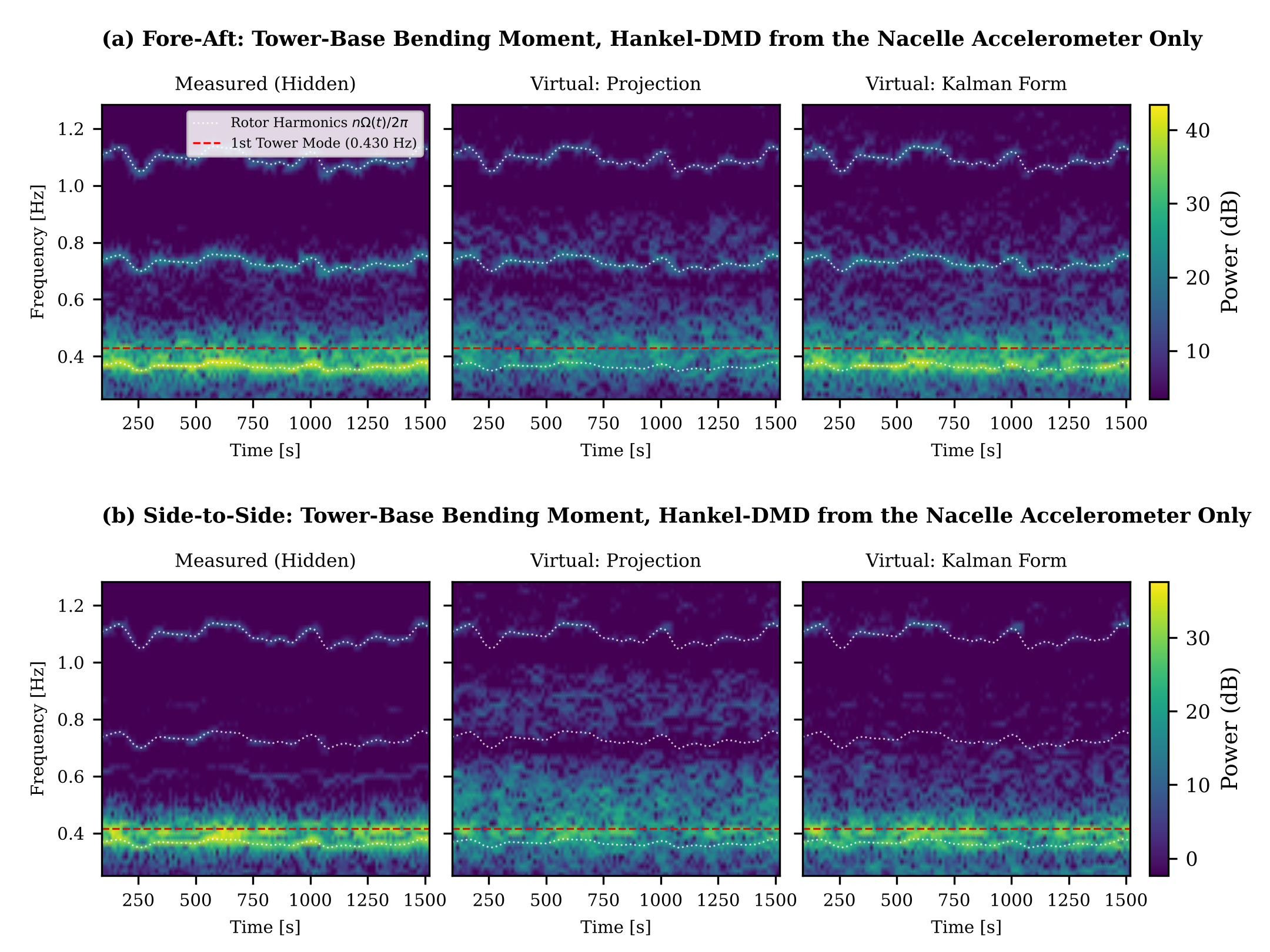}
    \caption{FOCAL Campaign~IV basin tests: time--frequency representation of the measured tower-base bending moment
    and of its reconstructions from the nacelle accelerometer alone by the Hankel-DMD projection and its Kalman form
    (run E11, second half, not used for training). Dotted lines: rotor harmonics 3P, 6P and 9P computed from the
    measured rotor speed; dashed line: identified first tower mode.}
    \label{fig:focal_tf}
\end{figure}

The time--frequency representation of Figure~\ref{fig:focal_tf} makes the distinction visible. The harmonic bands
follow the measured rotor speed, $n\Omega(t)/2\pi$, whereas the tower-mode band holds its frequency over the whole
record. Grouping the 200-s blocks by rotor speed confirms this distinction: between the lowest and highest rotor-speed groups, the mean rotor speed changes by $4.3\%$ and the 3P pole shifts by $3.8$--$4.1\%$, whereas the tower poles change by at most $0.5\%$. The strong band near $0.35$--$0.38$~Hz therefore represents the rotor-driven 3P response rather than the first tower mode near $0.43$~Hz. This is a direct experimental confirmation of the classification criterion of
Section~\ref{sec:stochastic_koopman} and of the 3P interpretation given for the simulated data in
Section~\ref{sec:sysid}. The damping bias of response-domain identification is also reproduced. Applied to the
response data, Hankel-DMD returns the same frequencies but much lower damping ($0.4\%$ and $1.8\%$ for the
side-to-side and fore-aft modes). In contrast, the correlation-domain estimate agrees with SSI-COV within $0.5$ percentage
points for the side-to-side mode. The fore-aft mode is more strongly damped than the side-to-side mode by every
estimate. This reproduces on a physical model the anisotropy attributed to aerodynamic damping in
Section~\ref{sec:damping}. The fore-aft damping is the less certain of the two (SSI-COV $4.8$--$7.8\%$ across runs), because the
mode is heavily damped and lies between the 3P and 4P harmonics.

The second tower modes could not be identified consistently from these measurements by FDD, SSI-COV or Hankel-DMD and
are therefore not reported. Above $1$~Hz the stable poles of Hankel-DMD and SSI-COV either follow the rotor speed
(6P--18P and a rotor-locked line near $7.9$~Hz corresponding to a rotor harmonic, \(f_{nP}=n\Omega/(2\pi)\)) or are not repeatable. Candidates near $1.4$, $1.6$--$1.8$, $4.0$, $8.7$ and $9.6$~Hz shift by up to $9\%$ between the lowest and highest rotor-speed groups, vary between runs, and give MAC $0.17$--$0.80$ between data subsets. The limitation lies in the instrumentation rather than in the method. Above about $2$~Hz, the acceleration spectra are dominated by broadband content with very low coherence between the two accelerometer signals, while the bending-moment spectra contain less than $1.5\%$ of their total $0.25$--$5$~Hz variance in this range. The second tower modes are therefore not reliably observable from the four channels located at the tower ends.

The identified modes were finally compared with independent measurements published for the same model
(Table~\ref{tab:focal_sysid}). The platform roll period identified from the tower channels, $31.1$~s, agrees with the
free-decay value of $31.0$--$31.1$~s \cite{Fowler2023ISOPE, Fowler2023FOCAL}. The roll mode, which is expected to be the
least affected by rotor thrust, is reproduced by all three methods with MAC $= 1.00$. It is therefore the most direct
independent check available. The first tower-mode frequencies lie within $6\%$ of the hammer-test values
\cite{Fowler2023ISOPE}. They are $4$--$6\%$ higher in the fore-aft and $2$--$4\%$ lower in the side-to-side direction, so
their order is reversed relative to the published pair. The platform pitch mode is strongly damped in operation
($17.5$--$18.8\%$) and has a longer period ($35.3$--$35.6$~s) than in free decay ($30.8$--$31.4$~s). The reference
values were obtained in dedicated tests of the moored model, for which the rotor operating state and the state of the hull tuned-mass dampers are not fully documented. The identification refers to near-rated operation with a mean platform pitch of $8^\circ$ and
a mean surge offset of $14$~m. These differences are therefore reported as they stand and are not attributed to a specific
mechanism. More generally, operational and test-based modal parameters of a floating turbine should be
compared only for documented conditions.

\begin{table}[htbp]
\centering
\caption{Modal identification from the FOCAL Campaign~IV basin tests (full scale; four runs with the same wind input and
different wave realizations; four channels per direction; 0.25--5~Hz, platform modes 0.01--0.2~Hz) and comparison
with published measurements on the same model. Hankel-DMD: medians [inter-quartile range] over $d\in\{43,86,129\}$
and $r\in[10,50]$, the same delay windows in units of the first-mode period as in Table~\ref{tab:sys_id_comparison},
applied to the output correlation functions (corr.) or to the response data (resp.). SSI-COV: medians over stabilized
orders ($i=86$). FDD: peak of the first singular value. MAC between the de-standardized Hankel-DMD (corr.) and SSI-COV
shapes.}
\label{tab:focal_sysid}
\footnotesize
\resizebox{\textwidth}{!}{%
\begin{tabular}{l c cc cc cc c c}
\toprule
\textbf{Mode} & \textbf{FDD} & \multicolumn{2}{c}{\textbf{SSI-COV}} & \multicolumn{2}{c}{\textbf{Hankel-DMD (corr.)}} &
\multicolumn{2}{c}{\textbf{Hankel-DMD (resp.)}} & \textbf{Published} & \textbf{MAC}\\
 & $f$ (Hz) & $f$ (Hz) & $\zeta$ (\%) & $f$ (Hz) & $\zeta$ (\%) & $f$ (Hz) & $\zeta$ (\%) & $f$ (Hz) & DMD--SSI\\
\midrule
1st FA tower & 0.425 & 0.434 & 6.6 & 0.430 [0.428--0.431] & 3.9 [3.6--4.8] & 0.430 [0.425--0.449] & 1.8 [0.6--2.5] & 0.409$^{a}$ & 1.00\\
1st SS tower & 0.410 & 0.414 & 1.8 & 0.415 [0.415--0.416] & 2.2 [1.7--2.3] & 0.414 [0.413--0.416] & 0.4 [0.2--0.5] & 0.425$^{a}$ & 1.00\\
2nd FA / SS tower & \multicolumn{9}{c}{not identifiable from the available channels (Section~\ref{sec:focal_sysid})}\\
\midrule
3P (FA / SS) & 0.375 / 0.375 & 0.366 / 0.368 & 1.5 / 1.2 & 0.369 / 0.368 & 1.6 / 1.4 & 0.371 / 0.371 & 0.2 / 0.2 & 0.366$^{b}$ & 1.00 / 1.00\\
Platform roll & 0.0320 & 0.0322 & 1.9 & 0.0321 [0.0321--0.0321] & 2.1 [2.1--2.1] & -- & -- & 0.0322--0.0323$^{c}$ & 1.00\\
Platform pitch & 0.0290 & 0.0283 & 18.8 & 0.0281 [0.0263--0.0292] & 17.5 [15.1--21.0] & -- & -- & 0.0318--0.0325$^{c}$ & 1.00\\
\bottomrule
\end{tabular}}
\par\smallskip\raggedright\scriptsize
$^{a}$ Hammer tests of the floating system \cite{Fowler2023ISOPE}. $^{b}$ Three times the mean rotor frequency,
$3\bar{\Omega}/2\pi$. $^{c}$ Free-decay tests, periods of 31.0--31.1~s (roll) and 30.8--31.4~s (pitch)
\cite{Fowler2023ISOPE,Fowler2023FOCAL}; identified periods: roll 31.1~s, pitch 35.3--35.6~s.
\end{table}

\subsection{Virtual Sensing}\label{sec:focal_vs}
The time-blocked protocol of Section~\ref{sec:vs_protocol} was applied with the four runs in place of the OLCs. Each
run was divided into two halves of $4200$~s, and each half was reconstructed by models identified from the other half
of the other three runs. This gives eight folds per direction. Hyper-parameters were cross-fitted between the run pairs
E11/E13 and E14/E15, and the first $60$~s of each test half were not scored. Because the regularization selected for
the projection and the rank selected for the Kalman form fell at the edge of grids designed for noise-free
simulations, the grids were extended to $\alpha \le 1$ and $r \le 250$. The projection then selects interior values for
the one- and two-sensor layouts ($\alpha = 0.1$--$0.3$, $r = 24$--$34$). The Kalman form saturates and changes by less
than $0.02$ between $r = 200$ and $r = 250$.

Table~\ref{tab:focal_vs} reports the $R^2$ values computed against the measured, noisy hidden signals. The information available in the active sensors is quantified by the coherence bound $\mathcal{B}_{h|\mathbf{a}}$
\begin{equation}
\mathcal{B}_{h|\mathbf{a}}
=
\frac{
\displaystyle \int_{f_1}^{f_2}
\gamma^2_{h|\mathbf{a}}(f)\,S_{hh}(f)\,df
}{
\displaystyle \int_{f_1}^{f_2}
S_{hh}(f)\,df
},
\qquad
\gamma^2_{h|\mathbf{a}}(f)
=
\frac{
\mathbf{S}_{h\mathbf{a}}(f)\,
\mathbf{S}_{\mathbf{aa}}^{-1}(f)\,
\mathbf{S}_{\mathbf{a}h}(f)
}{
S_{hh}(f)
}.
\label{eq:coherence_bound}
\end{equation}
where $\gamma^2_{h|\mathbf{a}}(f)$ is the squared multiple coherence between the hidden channel $h$ and the active sensor set $\mathbf{a}$, and $S_{hh}(f)$ is the power spectral density of the hidden channel. The bound is the fraction of hidden-channel variance that is linearly predictable from the active sensors over the analysis band, here $f_1=0.25$~Hz and $f_2=5$~Hz. In this linear, stationary setting, it is an upper limit on the achievable reconstruction accuracy. It therefore separates the performance of the estimator from the limits set by the available measurements.

\begin{figure}[htbp]
    \centering
    \includegraphics[width=\textwidth]{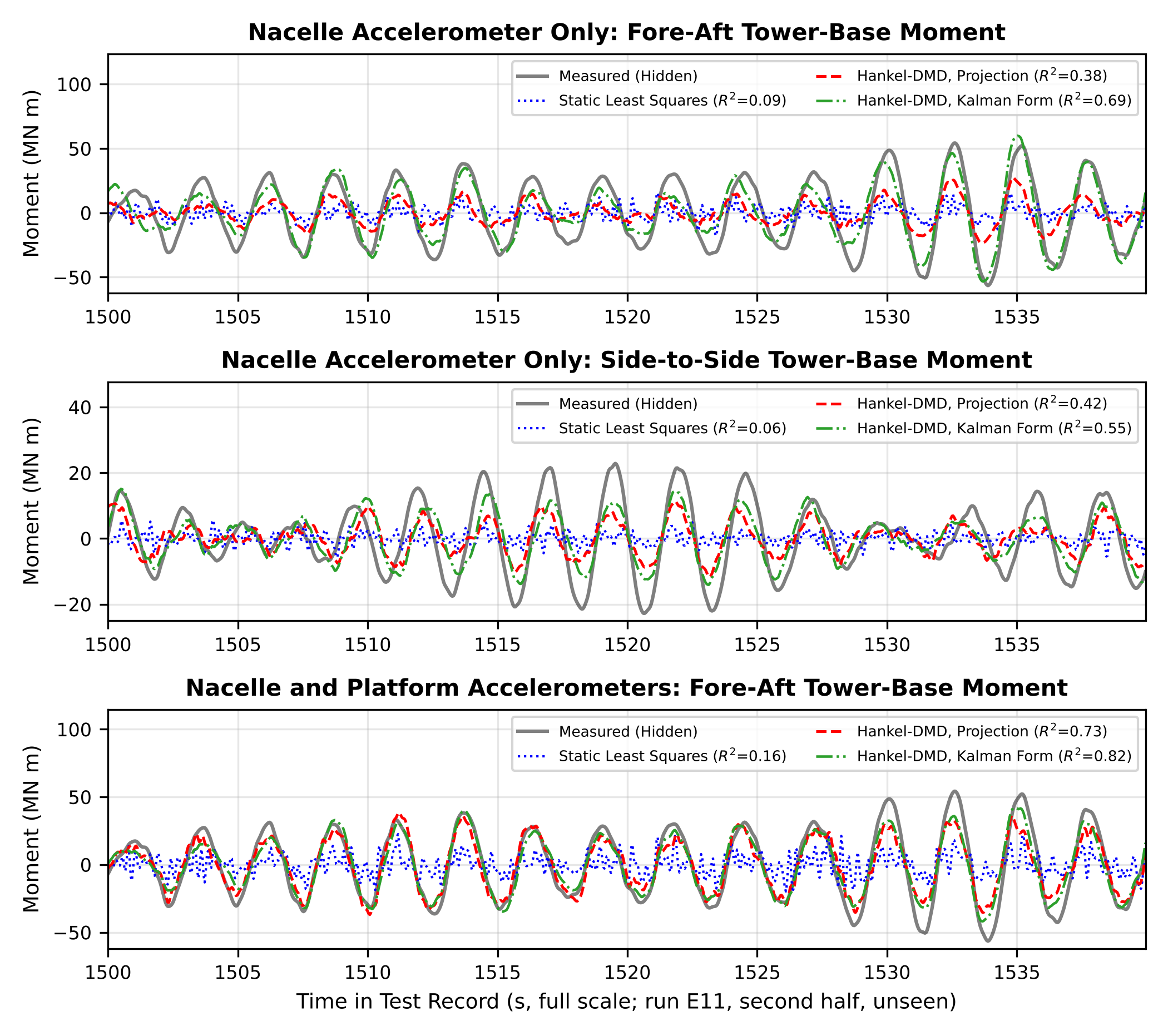}
    \caption{FOCAL Campaign~IV basin tests: reconstruction of the tower-base bending moment on a run and time interval
    not used for training (run E11, second half; model identified from the first half of the other three runs), from
    the nacelle accelerometer alone and from the nacelle and platform accelerometers. $R^2$ over the whole test half
    ($0.25$--$5$~Hz); the $R^2$ of the four held-out runs differ by at most $0.05$.}
    \label{fig:focal_vs}
\end{figure}

The measured response is not instantaneously low-dimensional. This removes the regime in which the static baselines
succeeded on simulated data. For the four corresponding channels, three principal components (PCs) explain $99.5\%$ (FA) and
$99.4\%$ (SS) of the variance in the simulation but only $86.9\%$ and $93.3\%$ in the basin test. The static baselines
accordingly fail. Static least squares reaches $R^2 \le 0.20$ except for the side-to-side moments with three sensors.
Gappy POD, which assumes a low-dimensional instantaneous basis, gives negative $R^2$ in almost all cases. Both
realizations of the Hankel-DMD model use the delay structure and recover considerably more. From the nacelle
accelerometer alone, the tower-base bending moment is reconstructed with $R^2 = 0.40$ (FA) and $0.44$ (SS) by the
projection and $0.69$ and $0.56$ by the Kalman form, against $0.10$ and $0.06$ for static least squares
(Figure~\ref{fig:focal_vs}). With three of the four sensors, the two realizations reach $0.83$--$0.84$ and
$0.88$--$0.89$. The values of the four held-out runs differ by at most $0.05$. The projection is the less accurate of
the two on these data because the measurements are noisy, as in the robustness test T3 of
Section~\ref{sec:vs_robust}. The Kalman form weights the measurements through a noise covariance
identified from the same data. It is as accurate as a Wiener filter fitted separately for each layout ($-0.02$ to
$+0.11$ in $R^2$), while the proposed framework uses one model for all layouts.

The absolute reconstruction accuracies are lower than in the simulation, but the coherence bound $\mathcal{B}_{h|\mathbf{a}}$ shows that this is largely due to limited information in the measured sensor set. Between $31\%$ and $60\%$ of the accelerometer variance within $0.25$--$5$~Hz lies above $2$~Hz, where the signals have very low inter-sensor coherence. Consequently, the platform acceleration has a low linear predictability bound, $\mathcal{B}_{h|\mathbf{a}}=0.07$--$0.53$, so no linear estimator can reconstruct more than this fraction of its variance from the available active sensors. The bending moments are instead dominated by rotor harmonics, which account for $66$--$81\%$ of their in-band variance and are only partly coherent with the tower accelerations. Using the nacelle accelerometer alone, the maximum linearly predictable variance is about $80\%$ (FA) and $74\%$ (SS) for the tower-base moment, and $69\%$ and $61\%$ for the tower-top moment. Relative to these bounds, the projection recovers $46$--$90\%$ and the Kalman estimator $69$--$95\%$ of the predictable variance, measured by $R^2/\mathcal{B}_{h|\mathbf{a}}$. Thus, the virtual-sensing accuracy in the experiment is limited mainly by the information content of the available sensors rather than by the estimator itself.

\begin{table}[htbp]
\centering
\caption{FOCAL Campaign~IV virtual sensing: median held-out-run $R^2$ over $0.25$--$5$~Hz.
N and P denote the nacelle and platform accelerometers; a$_N$, a$_P$ are accelerations and
M$_T$, M$_B$ are tower-top and tower-base bending moments.
$\mathcal{B}_{h|\mathbf{a}}$ is the linear coherence bound.}
\label{tab:focal_vs}
\footnotesize
\resizebox{\textwidth}{!}{%
\begin{tabular}{ll cc cc cccc}
\toprule
 & & \multicolumn{2}{c}{N} & \multicolumn{2}{c}{N + P} & \multicolumn{4}{c}{Three of four sensors}\\
\cmidrule(lr){3-4}\cmidrule(lr){5-6}\cmidrule(lr){7-10}
Dir. & Estimator & M$_T$ & M$_B$ & M$_T$ & M$_B$ & a$_N$ & a$_P$ & M$_T$ & M$_B$\\
\midrule
FA & \textit{Baselines} & & & & & & & & \\
 & \quad Gappy POD & $<\!-1$ & $<\!-1$ & $-0.25$ & $-0.32$ & $-0.42$ & $<\!-1$ & $<\!-1$ & $-0.26$\\
 & \quad Static least squares & 0.03 & 0.10 & 0.15 & 0.17 & 0.17 & 0.16 & 0.19 & 0.20\\
 & \quad Wiener filter (per layout) & \textbf{0.48} & 0.58 & \textbf{0.62} & 0.80 & \textbf{0.30} & \textbf{0.53} & \textbf{0.82} & 0.87\\
 & \textit{Hankel-DMD} & & & & & & & & \\
 & \quad Projection & 0.39 & 0.40 & 0.48 & 0.75 & 0.22 & 0.46 & 0.58 & 0.84\\
 & \quad Kalman form (stochastic extension) & \textbf{0.48} & \textbf{0.69} & \textbf{0.60} & \textbf{0.83} & \textbf{0.30} & \textbf{0.52} & \textbf{0.80} & \textbf{0.88}\\
 & \textit{Coherence bound} $\mathcal{B}_{h|\mathbf{a}}$ & \textit{0.69} & \textit{0.80} & \textit{0.79} & \textit{0.91} & \textit{0.33} & \textit{0.53} & \textit{0.92} & \textit{0.94}\\
\midrule
SS & \textit{Baselines} & & & & & & & & \\
 & \quad Gappy POD & $<\!-1$ & $<\!-1$ & $<\!-1$ & $<\!-1$ & $<\!-1$ & $<\!-1$ & 0.40 & 0.45\\
 & \quad Static least squares & 0.00 & 0.06 & 0.04 & 0.08 & 0.13 & 0.04 & 0.50 & 0.53\\
 & \quad Wiener filter (per layout) & 0.34 & 0.49 & 0.42 & 0.55 & \textbf{0.21} & \textbf{0.13} & \textbf{0.82} & 0.86\\
 & \textit{Hankel-DMD} & & & & & & & & \\
 & \quad Projection & 0.28 & 0.44 & 0.34 & 0.49 & 0.15 & 0.06 & 0.67 & 0.83\\
 & \quad Kalman form (stochastic extension) & \textbf{0.43} & \textbf{0.56} & \textbf{0.50} & \textbf{0.62} & \textbf{0.21} & \textbf{0.13} & \textbf{0.83} & \textbf{0.89}\\
 & \textit{Coherence bound} $\mathcal{B}_{h|\mathbf{a}}$ & \textit{0.61} & \textit{0.74} & \textit{0.68} & \textit{0.78} & \textit{0.25} & \textit{0.15} & \textit{0.90} & \textit{0.94}\\
\bottomrule
\end{tabular}}
\end{table}

\subsection{Scope of the Digital Twin and Limitations}\label{sec:scope}
Taken together with the simulation study, the basin tests define what the framework establishes. Following the usual
definitions \cite{Wright2020DT, Kapteyn2021DT, NASEM2024DT}, a structural digital twin comprises a computational model
of the asset, a mechanism that keeps the model synchronized with the measured state, and a feedback path to decisions.
The present framework provides the first two components for the tower dynamics: a reduced-order dynamical model
identified from operational data (Section~\ref{sec:sysid}) and a state estimator that synchronizes it with the
available measurements and reconstructs the unmeasured channels (Section~\ref{sec:vs_results}). The operator is
identified offline and held fixed during operation. Online updating of the operator, for instance by recursive or
windowed DMD, propagation of the uncertainty of the identified parameters, and the connection of the estimates to
fatigue assessment or control are not demonstrated here. The term digital twin is therefore used in this restricted
sense throughout: an equation-free model--estimator pair that can serve as the structural-state layer of a digital
twin.

Three limitations arise from the simulation setting. First, in the OpenFAST/ElastoDyn model used here, tower deformation is represented by only two assumed bending modes in each direction. The simulated response is therefore strongly low-dimensional (Section~\ref{sec:vs_results}), which explains why static estimators already perform well with five or more sensors. A physical tower contains higher modes, local dynamics and measurement noise, increasing its effective dimension. Delay embedding can therefore remain useful at larger sensor counts, although the achievable accuracy is lower. The basin tests show both effects. Second, all 18 channels were available during training. This represents a commissioning stage with temporary dense instrumentation, followed by operation with a smaller permanent sensor set. Third, the wind speed was fixed at the rated value, so changes in aerodynamic damping with operating condition were not examined. The structural properties were also unchanged across all load cases; sensitivity of the identified parameters to structural change, required for damage detection, remains to be demonstrated.

The experimental verification has similar limits. Sensors are available only at the tower top and base, so the full along-tower reconstruction demonstrated in simulation cannot be verified experimentally. Only one near-rated wind condition was considered. The physical model follows Froude scaling to preserve the gravity--inertia balance of the floating system; because Reynolds similarity cannot be maintained simultaneously, the rotor is performance-matched rather than Reynolds-scaled \cite{Fowler2023FOCAL}. Nevertheless, the basin tests show that the identification, rotor-harmonic classification, damping-bias correction, and the benefit of delay embedding under sparse sensing all transfer from simulation to a physical floating turbine with different geometry.

\section{Conclusion}
This study shows that the Hankel dynamic mode decomposition (Hankel-DMD) framework provides a robust, computationally efficient basis for digital twins of floating offshore wind turbines. Grey-box approaches incorporate predefined physical equations or state-space structure, which can become restrictive when the coupled dynamics are uncertain or incompletely represented. This work instead validates an equation-free, operator-theoretic framework. The method is purely data-driven, yet it retains the spectral interpretability of physics-based models. It works as a “glass-box” approach and avoids the limitations of both less interpretable neural networks and rigid physical assumptions. One Hankel-DMD approximation of the Koopman operator supplies both the operational modal parameters and the virtual sensing of the tower. The framework was assessed on OpenFAST simulations of the NREL 5MW OC3-Hywind spar under twelve wind--wave load cases and verified on basin-test data of a $1$:$70$-scale IEA 15MW semi-submersible floating turbine.

By shifting the modelling perspective from nonlinear state spaces to linear evolution in observable coordinates, the framework retains the computational efficiency of a reduced-order linear representation while preserving the spectral interpretability of modal analysis. The main findings are as follows.

\begin{itemize}
    \item The first and second fore-aft and side-to-side tower modes ($0.50/0.51$~Hz and $3.71/3.60$~Hz) are identified consistently by Hankel-DMD, SSI-COV and FDD (MAC $\ge 0.99$). The deterministic 3P rotor harmonic is returned as a separate pole with large apparent damping. The structural poles are invariant across sea states, as expected of a stochastically forced system. The mode shapes are reproduced by the kinematics of the simulation model, with about $30\%$ platform rigid-body participation in the first tower modes.
    \item Damping identified from response trajectories is biased low. Identifying the operator from the output correlation functions, which is the stochastic-Koopman form of the method, removes the bias and agrees with SSI-COV within $0.4$ percentage points. The fore-aft modes are consistently more damped than the side-to-side modes, in line with the aerodynamic damping of an operating rotor.
    \item Because the simulated response is instantaneously low-dimensional, any linear estimator reconstructs the hidden channels once five or more sensors remain. With one to three sensors, the time-delay embedding becomes essential. The regularized projection reaches $R^2 = 0.78$--$0.98$, that is, $0.03$--$0.20$ above static least squares. Its Kalman form reaches $R^2 = 0.91$--$0.99$, within $0.01$ of a Wiener filter fitted separately for each layout. Both use a single model for every failure pattern. A sufficient identifiability condition (Proposition~\ref{prop:identifiability}) and the reported condition numbers make the dependence on the sensor layout explicit, including where the modal amplitudes cease to be uniquely identifiable.
    \item The open-loop forecast skill decays as $e^{-2\zeta_1\omega_1 h}$ with the identified first-mode damping. The horizon over which the twin can bridge missing measurements therefore follows from the identified parameters rather than from a chaotic divergence rate. For $R^2 \ge 0.9$, it is about $0.5$~s in the fore-aft and $1.0$~s in the side-to-side direction.
    \item Applied without modification to the FOCAL basin tests, the framework identified the first tower modes ($0.430$ and $0.415$~Hz) consistently with SSI-COV and FDD. It separated them from the rotor harmonics, which the measured rotor speed confirmed directly. It also reproduced the free-decay roll period of the model ($31.1$~s) and the damping bias of response-domain identification. On unseen runs, the tower-base bending moment was reconstructed from the nacelle accelerometer alone with $R^2 = 0.40$ (FA) and $0.44$ (SS) by the projection and $0.69$ and $0.56$ by the Kalman form, against $0.10$ and $0.06$ for static least squares. From three of the four sensors, it was reconstructed with $R^2 = 0.83$--$0.89$. Relative to the coherence bound of the measured signals, the Kalman form recovers $69$--$95\%$ of the predictable variance of the bending moments.
\end{itemize}

This work establishes a fixed-operator baseline, comprising a deterministic projection and a stochastic Kalman extension. The predictability horizon set by the identified damping shows that high-frequency data assimilation is needed to prevent error accumulation under unmeasured stochastic forcing. To overcome this limitation, future work should extend the present fixed-covariance Kalman realization towards fully probabilistic inference that propagates uncertainty in the identified operator, modal parameters and unmeasured forcing. Augmented Kalman filters are standard for state estimation, but they often struggle to resolve time-varying ``engineering modes'' (e.g., aero-elastic damping ratios) accurately in highly non-Gaussian environments. A promising direction is to combine Hankel-DMD with sequential Monte Carlo or variational inference on the Koopman observable function rather than on a state-space model. This would allow a rigorous quantification of the uncertainty in modal parameter estimation, which standard AKF implementations currently lack. The online updating of the operator and the use of the identified modal parameters for damage detection, neither of which is demonstrated here (Section~\ref{sec:scope}), are the two further steps required before the framework constitutes a complete digital twin.

The structural independence and flexibility of the framework also suggest a scalable pathway for fleet-level monitoring. Because the Hankel-DMD operator captures the fundamental topology of the system's dynamics independently of specific sensor placements, future work should explore population-based SHM. With transfer learning, the Koopman linear operators learned from a subset of instrumented turbines could be adapted to monitor entire wind farms. This would reduce the computational and hardware cost of managing large offshore assets.

\section*{Acknowledgments}

This work was supported by the Research Council of Finland under the project Dynamic Equation Free Project (Grant No. 130203-1). The authors gratefully acknowledge the Council's financial support, which made this research possible.


\appendix
\renewcommand{\thetable}{A.\arabic{table}}
\renewcommand{\thefigure}{A.\arabic{figure}}
\setcounter{table}{0}
\setcounter{figure}{0}

\section{Supporting results}\label{app:supporting}

\subsection{Simulation model configuration}

\begin{table}[htbp]
\centering
\caption{OpenFAST configuration used to generate the operational data and the linearized reference model.}
\label{tab:openfast_config}
\footnotesize
\begin{tabular}{p{0.27\textwidth} p{0.67\textwidth}}
\toprule
\textbf{Item} & \textbf{Setting}\\
\midrule
Software & OpenFAST v3.2.1; model based on the OC3-Hywind certification case (Test~24) \cite{jonkman2010definition}\\
Structure (ElastoDyn) & All degrees of freedom enabled: platform surge, sway, heave, roll, pitch, yaw; tower
 fore-aft and side-to-side modes 1--2 (assumed-mode polynomials of the OC3-Hywind tower file, 1\% structural damping);
 blade flap 1--2 and edge 1; drivetrain torsion; generator; nacelle yaw. 20 tower nodes; gauges at nodes
 1, 3, 5, 7, 10, 13, 15, 17, 20 (11.9--85.7~m)\\
Aerodynamics (AeroDyn v15) & Blade-element momentum theory with Prandtl tip and hub losses and tangential induction;
 Pitt/Peters skewed-wake correction; Beddoes--Leishman unsteady aerodynamics (UAMod~3); tower potential-flow
 influence and tower drag\\
Inflow (InflowWind) & One TurbSim full-field turbulent wind field, 12~m/s mean hub-height speed, used for all 12 OLCs\\
Hydrodynamics (HydroDyn) & Potential-flow radiation and diffraction (WAMIT); irregular JONSWAP waves with $H_s$, $T_p$
 and heading of Table~\ref{tab:load_cases}; wave analysis time 3600~s; no current\\
Mooring & FEAMooring (dynamic) in the time-domain simulations; MAP++ (quasi-static) in the linearization\\
Control (ServoDyn) & NREL 5-MW baseline controller (Bladed-style DLL): variable-speed torque and collective pitch control\\
Time integration & DT $=0.0125$~s; analysed outputs at $\Delta t=0.05$~s (20~Hz)\\
Duration & 3600~s per OLC; the first 300~s are discarded\\
Linearization & One operating point at $t=100$~s: steady 8~m/s inflow, still water, nearly stationary rotor
 ($\Omega=0.0189$~rad/s $\approx0.18$~rpm); 44 ElastoDyn states; outputs include the tower-gauge channels\\
\bottomrule
\end{tabular}
\end{table}

\subsection{Dimensions of the identification and virtual-sensing problem}

\begin{table}[htbp]
\centering
\caption{Dimensions of the identification and virtual-sensing problem, with the values used in this study
(per direction).}
\label{tab:dimensions}
\footnotesize
\begin{tabular}{l l l}
\toprule
\textbf{Quantity} & \textbf{Dimension} & \textbf{Value}\\
\midrule
Channels (accelerations + moments) & $p$ & 18\\
Active channels & $q$ & 1--14\\
Delay depth & $d$ & 60 samples (3~s)\\
Truncation rank & $r$ & 10--50 (identification), 24--120 (virtual sensing)\\
Samples per segment & $N$ & 66\,000 (identification), 33\,000 (virtual sensing)\\
Delay vector $\tilde{\mathbf{y}}_k$ & $pd$ & 1080\\
Hankel matrices $\mathbf{H}_1,\mathbf{H}_2$ (per segment) & $pd\times(N-d)$ & $1080\times65\,940$ or $1080\times32\,940$\\
Modes $\boldsymbol{\Phi}$, eigenvalues $\boldsymbol{\Lambda}$ & $pd\times r$, $r\times r$ & \\
Sparse mode matrix $\boldsymbol{\Phi}_{sparse}$ & $qd\times r$ & $60\times r$ to $840\times r$\\
Most-recent-delay block $\boldsymbol{\Phi}_L$ & $p\times r$ & \\
Modal amplitudes/state $\mathbf{b}_k$ & $r$ & \\
Noise covariances $\mathbf{Q}$, $\mathbf{R}$ (Kalman form) & $r\times r$, $p\times p$ & \\
Kalman gain $\mathbf{K}_a$ & $r\times q$ & \\
\bottomrule
\end{tabular}
\end{table}

\subsection{Correlation-domain identification and sensitivity to the embedding parameters}

\begin{figure}[htbp]
    \centering
    \includegraphics[width=\textwidth]{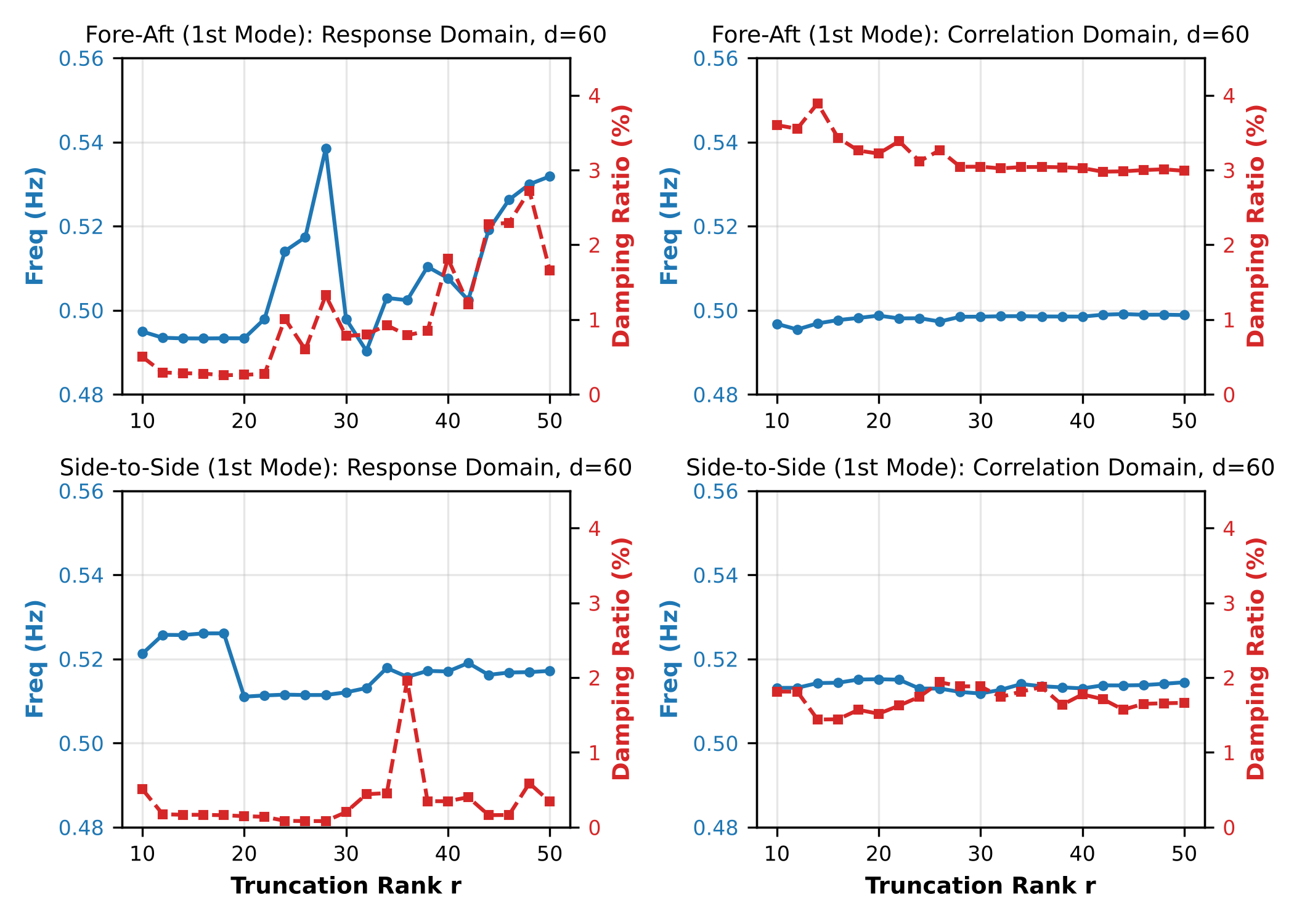}
    \caption{First-mode frequency (circles) and damping ratio (squares) versus the truncation rank, $d = 60$:
    response-domain (left) and correlation-domain (right) Hankel-DMD.}
    \label{fig:sensitivity}
\end{figure}

For a stationary forced response, the output correlation functions $\mathbf{R}_k = \mathbb{E}[\mathbf{y}_{t+k} \mathbf{y}_t^\top]$, $k \ge 1$, evolve like free decays under the conditional-mean (stochastic Koopman) dynamics of Section~\ref{sec:stochastic_koopman}. Applying the same Hankel-DMD to the columns of $\mathbf{R}_k$ ($k = 1, \dots, 400$, i.e.\ 20~s) removes the realization of the unmeasured forcing from the identification. Figure~\ref{fig:sensitivity} compares the first-mode frequency and damping identified from the response data and from the correlation functions over the truncation rank at $d = 60$. The correlation-domain estimates are stable over the whole range and agree with SSI-COV (Table~\ref{tab:sys_id_comparison}), whereas the response-domain damping is biased low and varies with the rank.

\subsection{Mode shapes and the kinematics of the simulation model}
\begin{table}[htbp]
\centering
\caption{Check of the identified acceleration mode shapes against the kinematics of the simulation model.
$R^2_{tower}$: least-squares fit with the two assumed tower modes of ElastoDyn only (fixed-base kinematics);
$R^2_{+pl}$: fit with platform translation and rotation added; MAC with the assumed tower mode of the same
order; share of platform rigid-body motion in the fitted shape. Gauge heights: ElastoDyn tower nodes 1, 3, 5, 7, 10, 13, 15, 17 and 20 of 20 (11.9--85.7~m).}
\label{tab:shape_check}
\footnotesize
\begin{tabular}{l cccc cccc}
\toprule
 & \multicolumn{4}{c}{\textbf{Hankel-DMD}} & \multicolumn{4}{c}{\textbf{SSI-COV}}\\
Mode & $R^2_{tower}$ & $R^2_{+pl}$ & MAC & platform (\%) & $R^2_{tower}$ & $R^2_{+pl}$ & MAC & platform (\%)\\
\midrule
Fore-aft 1st & 0.78 & 1.000 & 0.67 & 29 & 0.75 & 1.000 & 0.58 & 32\\
Fore-aft 2nd & 0.99 & 1.000 & 1.00 & 1 & 0.99 & 1.000 & 1.00 & 1\\
Side-to-side 1st & 0.75 & 1.000 & 0.55 & 34 & 0.75 & 1.000 & 0.55 & 34\\
Side-to-side 2nd & 0.99 & 1.000 & 1.00 & 1 & 0.99 & 1.000 & 1.00 & 1\\
\bottomrule
\end{tabular}

\vspace{1mm}
\parbox{\linewidth}{\scriptsize
The fitted acceleration shape is
$a(z)\approx A_{\rm plat}+zB_{\rm plat}+A_1\phi_1(z)+A_2\phi_2(z)$.
Here, $R^2_{tower}$ uses only the two tower-mode terms, while $R^2_{+pl}$ also includes platform translation and rotation.
}

\end{table}
\subsection{Per-channel virtual-sensing results}

\begin{table}[htbp]
\centering
\caption{Reconstruction from the nacelle accelerometer alone (layout L1): $R^2$ of selected hidden channels
(median over the 12 held-out OLCs, time-blocked protocol). Sensors 0--8 are accelerometers and
9--17 are bending moments, numbered from the tower base; sensor 8 is the active nacelle accelerometer.
The node of the first coupled mode lies between sensors 4 and 5. Best value in each column is shown
in bold; ties at the reported precision are both bold.}
\label{tab:vs_l1_channels}
\footnotesize
\resizebox{\textwidth}{!}{%
\begin{tabular}{ll ccccccc cccc}
\toprule
& &
\multicolumn{7}{c}{\textbf{Acceleration}} &
\multicolumn{4}{c}{\textbf{Bending moment}}\\
\cmidrule(lr){3-9}\cmidrule(lr){10-13}
Dir. & Estimator & 0 & 2 & 3 & 4 & 5 & 6 & 7 & 9 & 13 & 16 & 17\\
\midrule

FA & \textit{Baselines} & & & & & & & & & & & \\
& \quad Gappy POD
& 0.90 & 0.50 & 0.07 & $-0.04$ & 0.11 & 0.29 & 0.58 & 0.90 & 0.84 & 0.50 & $-0.04$\\
& \quad Static least squares
& 0.90 & 0.54 & 0.12 & 0.01 & 0.15 & 0.32 & 0.60 & 0.91 & 0.85 & 0.54 & $-0.01$\\
& \quad Wiener filter (per layout)
& 0.95 & \textbf{0.95} & \textbf{0.92} & \textbf{0.90} & \textbf{0.91} & \textbf{0.93} & \textbf{0.96}
& 0.96 & 0.96 & 0.80 & 0.22\\

& \textit{Hankel-DMD} & & & & & & & & & & & \\
& \quad Projection
& 0.87 & 0.88 & 0.84 & 0.82 & 0.83 & 0.86 & 0.92 & 0.89 & 0.87 & 0.56 & $-0.17$\\
& \quad Kalman form (stochastic extension)
& \textbf{0.96} & 0.94 & 0.90 & 0.88 & 0.89 & 0.91 & 0.95
& \textbf{0.97} & \textbf{0.98} & \textbf{0.83} & \textbf{0.27}\\

\midrule

SS & \textit{Baselines} & & & & & & & & & & & \\
& \quad Gappy POD
& 0.71 & 0.77 & 0.60 & 0.09 & 0.11 & 0.49 & 0.83 & 0.94 & 0.93 & 0.89 & 0.65\\
& \quad Static least squares
& 0.71 & 0.78 & 0.61 & 0.11 & 0.12 & 0.50 & 0.84 & 0.94 & 0.93 & 0.90 & 0.66\\
& \quad Wiener filter (per layout)
& \textbf{0.74} & \textbf{0.88} & \textbf{0.90} & \textbf{0.91} & \textbf{0.94} & \textbf{0.97} & \textbf{0.99}
& \textbf{0.99} & \textbf{0.99} & \textbf{0.98} & \textbf{0.86}\\

& \textit{Hankel-DMD} & & & & & & & & & & & \\
& \quad Projection
& 0.67 & 0.84 & 0.85 & 0.83 & 0.86 & 0.92 & 0.98 & 0.96 & 0.96 & 0.93 & 0.65\\
& \quad Kalman form (stochastic extension)
& 0.73 & 0.87 & 0.89 & 0.86 & 0.89 & 0.94 & 0.98
& \textbf{0.99} & \textbf{0.99} & \textbf{0.98} & \textbf{0.86}\\

\bottomrule
\end{tabular}}

\vspace{1mm}
\parbox{\textwidth}{\scriptsize
The Wiener filter is optimized separately for each fixed sensor layout and therefore provides a strong layout-specific benchmark. In contrast, the proposed Hankel-DMD/Kalman estimator achieves comparable accuracy while retaining the same $\mathbf{\Phi}$, $\mathbf{\Lambda}$, $\mathbf{Q}$ and $\mathbf{R}$; only the observation matrix changes when the active sensor layout changes, so the dynamical model does not require retraining.
}

\end{table}

\begin{table}[htbp]
\centering
\caption{Per-sensor reconstruction accuracy for Cases A--C ($R^2$; NRMSE in \% for the two Hankel-DMD
estimators, Eq.~\eqref{eq:metrics}).}
\label{tab:vs_per_sensor}
\footnotesize
\resizebox{\textwidth}{!}{%
\begin{tabular}{lll ccccc cc}
\toprule
 & & & \multicolumn{3}{c}{\textit{Baselines}} & \multicolumn{2}{c}{\textit{Hankel-DMD}} & & \\
\cmidrule(lr){4-6}\cmidrule(lr){7-8}
Dir. & Scenario & Hidden sensor & Gappy POD & Static LS & Wiener & Projection & Kalman form & NRMSE proj. & NRMSE Kalman\\
\midrule
FA & Case A & 2 (Lower-tower acc.) & 1.000 & 1.000 & 1.000 & 1.000 & 1.000 & 0.04 & 0.01\\
 &  & 3 (Lower-tower acc.) & 1.000 & 1.000 & 1.000 & 1.000 & 1.000 & 0.03 & 0.00\\
 &  & 5 (Mid-tower acc.) & 1.000 & 1.000 & 1.000 & 1.000 & 1.000 & 0.02 & 0.00\\
 &  & 7 (Sub-nacelle acc.) & 1.000 & 1.000 & 1.000 & 1.000 & 1.000 & 0.02 & 0.00\\
 &  & 11 (Lower-mid moment) & 1.000 & 1.000 & 1.000 & 1.000 & 1.000 & 0.02 & 0.01\\
 &  & 12 (Mid-tower moment) & 1.000 & 1.000 & 1.000 & 1.000 & 1.000 & 0.02 & 0.01\\
 &  & 14 (Upper-tower moment) & 1.000 & 1.000 & 1.000 & 1.000 & 1.000 & 0.03 & 0.01\\
 &  & 16 (Sub-nacelle moment) & 1.000 & 1.000 & 1.000 & 1.000 & 1.000 & 0.03 & 0.01\\
 & Case B & 0 (Base acc.) & 0.999 & 1.000 & 1.000 & 1.000 & 1.000 & 0.08 & 0.01\\
 &  & 8 (Nacelle acc.) & 1.000 & 1.000 & 1.000 & 1.000 & 1.000 & 0.17 & 0.00\\
 &  & 9 (Base moment) & 1.000 & 1.000 & 1.000 & 1.000 & 1.000 & 0.02 & 0.01\\
 &  & 17 (Nacelle moment) & 1.000 & 1.000 & 1.000 & 1.000 & 1.000 & 0.10 & 0.03\\
 & Case C & 0 (Base acc.) & 1.000 & 1.000 & 1.000 & 1.000 & 1.000 & 0.06 & 0.01\\
 &  & 8 (Nacelle acc.) & 1.000 & 1.000 & 1.000 & 1.000 & 1.000 & 0.18 & 0.00\\
 &  & 9 (Base moment) & 1.000 & 1.000 & 1.000 & 1.000 & 1.000 & 0.03 & 0.01\\
 &  & 17 (Nacelle moment) & 1.000 & 1.000 & 1.000 & 1.000 & 1.000 & 0.09 & 0.03\\
\midrule
SS & Case A & 2 (Lower-tower acc.) & 0.998 & 1.000 & 1.000 & 1.000 & 1.000 & 0.05 & 0.01\\
 &  & 3 (Lower-tower acc.) & 0.997 & 1.000 & 1.000 & 1.000 & 1.000 & 0.06 & 0.02\\
 &  & 5 (Mid-tower acc.) & 1.000 & 1.000 & 1.000 & 1.000 & 1.000 & 0.02 & 0.01\\
 &  & 7 (Sub-nacelle acc.) & 1.000 & 1.000 & 1.000 & 1.000 & 1.000 & 0.03 & 0.00\\
 &  & 11 (Lower-mid moment) & 1.000 & 1.000 & 1.000 & 1.000 & 1.000 & 0.02 & 0.01\\
 &  & 12 (Mid-tower moment) & 1.000 & 1.000 & 1.000 & 1.000 & 1.000 & 0.02 & 0.00\\
 &  & 14 (Upper-tower moment) & 1.000 & 1.000 & 1.000 & 1.000 & 1.000 & 0.02 & 0.01\\
 &  & 16 (Sub-nacelle moment) & 1.000 & 1.000 & 1.000 & 1.000 & 1.000 & 0.02 & 0.01\\
 & Case B & 0 (Base acc.) & 0.998 & 1.000 & 1.000 & 1.000 & 1.000 & 0.15 & 0.02\\
 &  & 8 (Nacelle acc.) & 0.999 & 1.000 & 1.000 & 1.000 & 1.000 & 0.06 & 0.01\\
 &  & 9 (Base moment) & 1.000 & 1.000 & 1.000 & 1.000 & 1.000 & 0.03 & 0.01\\
 &  & 17 (Nacelle moment) & 1.000 & 1.000 & 1.000 & 0.999 & 1.000 & 0.21 & 0.03\\
 & Case C & 0 (Base acc.) & 0.997 & 1.000 & 1.000 & 1.000 & 1.000 & 0.15 & 0.01\\
 &  & 8 (Nacelle acc.) & 0.999 & 1.000 & 1.000 & 1.000 & 1.000 & 0.06 & 0.01\\
 &  & 9 (Base moment) & 1.000 & 1.000 & 1.000 & 1.000 & 1.000 & 0.03 & 0.01\\
 &  & 17 (Nacelle moment) & 1.000 & 1.000 & 1.000 & 1.000 & 1.000 & 0.23 & 0.04\\
\bottomrule
\end{tabular}}
\end{table}

\subsection{Robustness tests}

\begin{table}[htbp]
\centering
\caption{Pre-declared robustness tests (time-blocked; hyper-parameters fixed to the values selected in the
main study). Median $R^2$ of the hidden channels over the random layouts with $q$ active sensors and the
test OLCs. Baselines: S, static least squares; W, Wiener filter. Hankel-DMD: P, projection; K, Kalman form.
The best value for each $q$ is shown in bold; when Wiener and Kalman are equal at the reported precision,
only the proposed Kalman form is highlighted.}
\label{tab:vs_robust}
\footnotesize
\resizebox{\textwidth}{!}{%
\begin{tabular}{ll cccc cccc cccc}
\toprule
 & & \multicolumn{4}{c}{$q=1$} & \multicolumn{4}{c}{$q=2$} & \multicolumn{4}{c}{$q=3$}\\
Test & Dir. & S & W & P & K & S & W & P & K & S & W & P & K\\
\midrule

Main study (Fig.~\ref{fig:vs_sensors})
& FA
& 0.59 & 0.91 & 0.78 & \textbf{0.91}
& 0.89 & 0.98 & 0.92 & \textbf{0.98}
& 0.90 & 0.99 & 0.96 & \textbf{0.99}\\

& SS
& 0.72 & 0.93 & 0.87 & \textbf{0.93}
& 0.89 & 0.97 & 0.94 & \textbf{0.97}
& 0.95 & \textbf{1.00} & 0.98 & 0.99\\

\midrule

\shortstack[l]{T1: 600 s of one OLC\\Train: 600 s, one OLC}
& FA
& 0.58 & 0.90 & 0.77 & \textbf{0.90}
& 0.88 & 0.97 & 0.92 & \textbf{0.98}
& 0.90 & 0.99 & 0.96 & \textbf{0.99}\\

& SS
& 0.71 & \textbf{0.93} & 0.83 & 0.91
& 0.89 & \textbf{0.97} & 0.94 & 0.96
& 0.94 & \textbf{1.00} & 0.98 & 0.99\\

\midrule

\shortstack[l]{T2: $H_s$ 1--2 m $\rightarrow$ 3--4 m\\Train: 1--2 m; Test: 3--4 m}
& FA
& 0.58 & 0.90 & 0.77 & \textbf{0.91}
& 0.88 & 0.97 & 0.92 & \textbf{0.98}
& 0.90 & 0.99 & 0.96 & \textbf{0.99}\\

& SS
& 0.71 & \textbf{0.93} & 0.88 & 0.92
& 0.89 & 0.97 & 0.95 & \textbf{0.97}
& 0.94 & \textbf{1.00} & 0.98 & 0.99\\

\midrule

\shortstack[l]{T2: $H_s$ 3--4 m $\rightarrow$ 1--2 m\\Train: 3--4 m; Test: 1--2 m}
& FA
& 0.58 & 0.91 & 0.78 & \textbf{0.91}
& 0.88 & 0.97 & 0.93 & \textbf{0.98}
& 0.91 & 0.99 & 0.96 & \textbf{0.99}\\

& SS
& 0.72 & 0.93 & 0.85 & \textbf{0.93}
& 0.89 & 0.97 & 0.93 & \textbf{0.97}
& 0.95 & \textbf{1.00} & 0.98 & 0.99\\

\midrule

\shortstack[l]{T3: 5\% sensor noise\\Train/test: noisy; score: noise-free}
& FA
& 0.58 & 0.85 & 0.14 & \textbf{0.86}
& 0.86 & 0.96 & 0.79 & \textbf{0.96}
& 0.90 & 0.97 & 0.95 & \textbf{0.98}\\

& SS
& 0.71 & 0.87 & $-0.25$ & \textbf{0.87}
& 0.86 & 0.94 & 0.82 & \textbf{0.94}
& 0.94 & 0.98 & 0.91 & \textbf{0.99}\\
\bottomrule
\end{tabular}
}

\par\smallskip
{\footnotesize\raggedright
\textit{Note:} For each training condition, one Hankel-DMD model
$(\mathbf{\Phi},\mathbf{\Lambda})$ is reused for all $q$ and sensor layouts;
only $\mathbf{\Phi}_{\mathrm{sparse}}^{(a)}$ or the Kalman gain
$\mathbf{K}^{(a)}$ changes with the active-sensor set $a$, active sensors.
The Wiener baseline instead need to fits layout-specific weights
$\mathbf{W}_{\mathrm{W}}^{(a)}$ from the training data.
All hyperparameters are fixed to the main-study selections.
\par}

\end{table}


\begin{thebibliography}{99}

\bibitem{krathe2025}
Krathe, V. L., Jonkman, J., \& Bachynski-Polić, E. E. (2025).
The influence of wind veer and drivetrain flexibility on fatigue loading for large floating wind turbines.
\textit{Wind Energy}, 27(1), 15--37.

\bibitem{bilbao2022}
Bilbao, J., Lourens, E. M., Schulze, A., \& Ziegler, L. (2022).
Virtual sensing in an onshore wind turbine tower using a Gaussian process latent force model.
\textit{Data-Centric Engineering}, 3, e35.

\bibitem{kim2024}
Kim, Y., Choi, Y., \& Yoo, B. (2024).
Gappy AE: A nonlinear approach for gappy data reconstruction using auto-encoder.
\textit{Computer Methods in Applied Mechanics and Engineering}, 426, 116978.

\bibitem{simpson2026}
Simpson, H. A., Chatzi, E. N., \& Chatzis, M. N. (2026).
A comparison of deterministic and Bayesian model updating frameworks for identifying offshore wind turbine foundation parameters.
\textit{Journal of Sound and Vibration}, 625, 119575.

\bibitem{ozan2025}
Ozan, D. E., Nóvoa, A., \& Magri, L. (2025).
Data Assimilated Model Based Reinforcement Learning for Partially Observed Chaotic Flows.
\textit{CoRR}.

\bibitem{xie2025}
Xie, H., Wan, L., Shi, F., Xin, J., Zhou, H., He, B., Jin, C., \& Michailides, C. (2025).
A Joint Method on Dynamic States Estimation for Digital Twin of Floating Offshore Wind Turbines.
\textit{Journal of Marine Science and Engineering}, 13(10).

\bibitem{SHM2024}
Yang, Y., Liang, F., Zhu, Q., \& Zhang, H. (2024). An overview on structural health monitoring and fault diagnosis of offshore wind turbine support structures. \textit{Journal of Marine Science and Engineering}, \textit{12}(3), 377.

\bibitem{duarte2014}
Duarte, T. M., Sarmento, A. J. N. A., \& Jonkman, J. (2014).
Effects of second order hydrodynamic forces on floating offshore wind turbines.
In \textit{32nd ASME Wind Energy Symposium} (Paper No. AIAA 2014-0361), National Harbor, MD.

\bibitem{majda2018}
Majda, A. J. \& Chen, N. (2018).
Model Error, Information Barriers, State Estimation and Prediction in Complex Multiscale Systems.
\textit{Entropy}, 20(9), 644.

\bibitem{caglio2025}
Caglio, L., Sadeqi, A., Stang, H., \& Katsanos, E. (2025).
Joint input-state estimation of structures subjected to complex loads via augmented Kalman Filter with physics informed latent force models.
\textit{Mechanical Systems and Signal Processing}, 223, 111852.

\bibitem{moynihan2023}
Moynihan, B., Mehrjoo, A., Moaveni, B., McAdam, R., Rüdinger, F., \& Hines, E. (2023).
System identification and finite element model updating of a 6 MW offshore wind turbine using vibrational response measurements.
\textit{Renewable Energy}, 219, 119430.

\bibitem{vettori2023}
Vettori, S., Di Lorenzo, E., Peeters, B., Luczak, M. M., \& Chatzi, E. (2023).
An adaptive-noise Augmented Kalman Filter approach for input-state estimation in structural dynamics.
\textit{Mechanical Systems and Signal Processing}, 184, 109654.

\bibitem{grande2026}
Grande, D., Buizza, R., \& Storto, A. (2026).
Machine learning in ocean data assimilation: Advances, gaps and the road to operations.
\textit{Ocean Modelling}, 102678.

\bibitem{shahin2025}
Shahin, M., Nasr, A. N., Malhi, A., Bauk, S., Banda, O. V., Kujala, P., ... \& Wang, S. (2025).
Hybrid Physics-AI Digital Twin Framework for Shared Mooring Systems in Deepwater Floating Offshore Wind Farms.
\textit{Authorea Preprints}.

\bibitem{taze2025}
Taze, I. E., Hoda, M. A., Miquelez, I., Maddaloni, P., \& Eftekhar Azam, S. (2025).
A Review of Digital Twinning Applications for Floating Offshore Wind Turbines: Insights, Innovations, and Implementation.
\textit{Energies}, 18(13), 3369.

\bibitem{rajic2025}
Rajić, M., Mančić, M., Glumac, A., Rossi, M., \& Rebelo, C. (2025).
Digital twins and AI integration in offshore renewable energy: A Review.
\textit{IOP Conference Series: Earth and Environmental Science}, 1552(1), 012007.

\bibitem{dang2026}
Dang, H. V., \& Nguyen, P. C. (2026).
Deep operator learning for high-fidelity fluid flow field reconstruction from sparse sensor measurements.
\textit{Journal of Computing and Information Science in Engineering}, 26(1), 011007.

\bibitem{madushele2026}
Madushele, N., Olatunji, O. O., \& Adedeji, P. A. (2026).
\textit{Digital Twin Technology in Condition Monitoring of Wind Turbines}.
CRC Press.

\bibitem{sanchis2025}
Sanchis-Agudo, M., Wang, Y., Arnau, R., Guastoni, L., Lim, J., Duraisamy, K., \& Vinuesa, R. (2025).
Easy attention: A simple attention mechanism for temporal predictions with transformers.
\textit{APL Computational Physics}, 1(1).

\bibitem{duthe2025}
Duthé, G., Evangelou, N., Liu, W., Kevrekidis, I. G., \& Chatzi, E. (2025).
A Mechanistic Analysis of Transformers for Dynamical Systems.
\textit{arXiv preprint arXiv:2512.21113}.

\bibitem{song2025}
Song, J., Zhou, K., Shu, Z. R., Cai, X. Y., \& Duan, M. G. (2025).
Automated modal identification of offshore wind turbines via an uncertainty-quantified stochastic subspace method.
\textit{Ocean Engineering}, 342, 123152.

\bibitem{oliveira2018}
Oliveira, G., Magalhães, F., Cunha, Á., \& Caetano, E. (2018).
Continuous dynamic monitoring of an onshore wind turbine.
\textit{Engineering Structures}, 164, 22--39.

\bibitem{weil2025}
Weil, M., Jurado, C. S., Weijtjens, W., \& Devriendt, C. (2025).
Machine learning and uncertainty quantification to track and monitor natural frequencies in vibration-based SHM applied to offshore wind turbines.
\textit{Data-Centric Engineering}, 6, e7.

\bibitem{liu2023}
Liu, Y., Liang, J., \& Wang, Y. (2023).
Support condition identification of wind turbines based on a statistical time-domain damping parameter.
\textit{Inverse Problems}, 39(12), 125021.

\bibitem{brijder2023}
Brijder, R., Helsen, S., \& Ompusunggu, A. P. (2023).
Switching kalman filtering-based corrosion detection and prognostics for offshore wind-turbine structures.
\textit{Wind}, 3(1), 1--13.

\bibitem{wei2025}
Wei, S., Wang, X., Gao, X., Li, Z., Huo, Z., \& Xu, C. (2025).
Motion estimation of floating wind turbines based on Kalman filter using simplified theoretical model.
\textit{Sustainable Energy Technologies and Assessments}, 82, 104522.

\bibitem{ammerman2025}
Ammerman, I., Alkarem, Y., Kimball, R. W., Huguenard, K., \& Hejrati, B. (2025).
Real-Time Structural Dynamics Estimation of Floating Wind Turbines Via Linear and Non-Linear Kalman Filtering.
\textit{SSRN 5177776}.


\bibitem{liew2022}
Liew, J., Göçmen, T., Lio, W. H., \& Larsen, G. C. (2022).
Streaming dynamic mode decomposition for short-term forecasting in wind farms.
\textit{Wind Energy}, 25(4), 719--734.

\bibitem{lydon2025}
Lydon, B., Polagye, B., \& Brunton, S. (2025).
Data-driven modeling of an oscillating surge wave energy converter using dynamic mode decomposition.
\textit{Journal of Renewable and Sustainable Energy}, 17(2).

\bibitem{serani2023}
Serani, A., Dragone, P., Stern, F., \& Diez, M. (2023).
On the use of dynamic mode decomposition for time-series forecasting of ships operating in waves.
\textit{Ocean Engineering}, 267, 113235.

\bibitem{iungo2015}
Iungo, G. V., Santoni-Ortiz, C., Abkar, M., Porté-Agel, F., Rotea, M. A., \& Leonardi, S. (2015).
Data-driven reduced order model for prediction of wind turbine wakes.
\textit{Journal of Physics: Conference Series}, 625(1), 012009.

\bibitem{annoni2016}
Annoni, J. R., Nichols, J., \& Seiler, P. J. (2016).
Wind farm modeling and control using dynamic mode decomposition.
In \textit{34th Wind Energy Symposium}, 2201.

\bibitem{liu2020}
Liu, F., Gao, S., Tian, Z., \& Liu, D. (2020).
A new time-frequency analysis method based on single mode function decomposition for offshore wind turbines.
\textit{Marine Structures}, 72, 102782.

\bibitem{Pitchforth2021}
Pitchforth, D. J., Rogers, T. J., Tygesen, U. T., \& Cross, E. J. (2021). Grey-box models for wave loading prediction. \textit{Mechanical Systems and Signal Processing}, \textit{159}, 107741.

\bibitem{Tygesen2021}
Tygesen, U. T., Cross, E. J., Gardner, P., Worden, K., Qadri, B. A., \& Rogers, T. J. (2021). Digital transformation by the implementation of the true digital twin concept and big data technology for structural integrity management. In \textit{Proceedings of Asset Integrity Management-Ageing and Life Extension Conference}.

\bibitem{McGreivy2024}
McGreivy, N., \& Hakim, A. (2024). Weak baselines and reporting biases lead to overoptimism in machine learning for fluid-related partial differential equations. \textit{Nature machine intelligence}, \textit{6}(10), 1256-1269.



\bibitem{kutz2016}
Kutz, J. N., Brunton, S. L., Brunton, B. W., \& Proctor, J. L. (2016). \textit{Dynamic mode decomposition: data-driven modeling of complex systems}. Society for Industrial and Applied Mathematics.


\bibitem{kutz2020}
Bai, Z., Kaiser, E., Proctor, J. L., Kutz, J. N., \& Brunton, S. L. (2020). Dynamic mode decomposition for compressive system identification. \textit{AIAA Journal}, \textit{58}(2), 561-574.

\bibitem{palma2025fast}
Palma, G., Bardazzi, A., Lucarelli, A., Pilloton, C., Serani, A., Lugni, C., \& Diez, M. (2025).
Analysis, forecasting, and system identification of a floating offshore wind turbine using dynamic mode decomposition.
\textit{Journal of Marine Science and Engineering}, 13(4), 656.



\bibitem{dai2022}
X. Dai, D. Xu, M. Zhang, and R. J. A. M. Stevens,
A three-dimensional dynamic mode decomposition analysis of wind farm flow aerodynamics,
\textit{Renewable Energy}, vol. 191, pp. 608--624, 2022.





\bibitem{mezic2005}
Korda, M., \& Mezić, I. (2018). Linear predictors for nonlinear dynamical systems: Koopman operator meets model predictive control. \textit{Automatica}, \textit{93}, 149-160.

\bibitem{kantamneni2024}
Kantamneni, S., Liu, Z., \& Tegmark, M. (2024). How Do Transformers Model Physics? Investigating the Simple Harmonic Oscillator. \textit{Entropy}, \textit{26}(11), 997.


\bibitem{openfast_doc}
National Renewable Energy Laboratory.
\textit{OpenFAST Documentation --- OpenFAST v4.1.2 documentation}.
Available at: \url{https://openfast.readthedocs.io/} (Accessed: December 15, 2025).

\bibitem{jonkman2005fast}
Jonkman, J.M. and Buhl, M.L. Jr. (2005).
\textit{FAST User’s Guide---Updated August 2005}.
Technical Report. National Renewable Energy Laboratory (NREL): Golden, CO, USA.

\bibitem{jonkman2010definition}
Jonkman, J. (2010).
\textit{Definition of the floating system for phase IV of OC3}.
Technical Report NREL/TP-200-47535. National Renewable Energy Laboratory (NREL).
Available at: \url{https://www.nrel.gov/docs/fy10osti/47535.pdf} (Accessed: 3 February 2025).

\bibitem{Robertson2014}
Robertson, A., Jonkman, J., Vorpahl, F., Popko, W., Qvist, J., Frøyd, L., ...  Guérinel, M. (2014, June). Offshore code comparison collaboration continuation within IEA wind task 30: Phase II results regarding a floating semisubmersible wind system. In International Conference on Offshore Mechanics and Arctic Engineering (Vol. 45547, p. V09BT09A012). American Society of Mechanical Engineers.

\bibitem{jonkman2007dynamics}
Jonkman, J.M. (2007).
\textit{Dynamics Modeling and Loads Analysis of an Offshore Floating Wind Turbine}.
PhD Thesis. University of Colorado Boulder: Boulder, CO, USA.

\bibitem{kane1985dynamics}
Kane, T.R. and Levinson, D.A. (1985).
\textit{Dynamics, Theory and Applications}.
McGraw Hill: New York, NY, USA.

\bibitem{Galvan2025}
Galván, J., Domínguez, J. M., \& Albizuri, J. (2025). Operational modal analysis applied to floating offshore wind turbine dynamics identification in wave basins in the presence of free-surface modes. \textit{Ocean Engineering}, \textit{327}, 120909.

\bibitem{mehlan2023}
Mehlan, F. C., Keller, J., \& Nejad, A. R. (2023).
Virtual sensing of wind turbine hub loads and drivetrain fatigue damage.
\textit{Forschung im Ingenieurwesen}, 87(1), 207--218.

\bibitem{SSICOV}
Peeters, B., and De Roeck, G. (1999).
Reference-based stochastic subspace identification for output-only modal analysis.
\textit{Mechanical Systems and Signal Processing}, 13(6), 855--878.

\bibitem{branlard2024}
Branlard, E., Jonkman, J., Brown, C., \& Zhang, J. (2024).
A digital twin solution for floating offshore wind turbines validated using a full-scale prototype.
\textit{Wind Energy Science}, 9(1), 1--24.

\bibitem{Gonzalez2019}
Gonzalez, E., Stephen, B., Infield, D.,  Melero, J. J. (2019). Using high-frequency SCADA data for wind turbine performance monitoring: A sensitivity study. Renewable energy, 131, 841-853.






\bibitem{koopman1931}
Koopman, B. O. (1931).
Hamiltonian systems and transformation in Hilbert space.
\textit{Proceedings of the National Academy of Sciences}, 17(5), 315--318.

\bibitem{mezic2020geometry}
Mezi\'c, I. (2020).
Spectrum of the Koopman operator, spectral expansions in functional spaces, and state-space geometry.
\textit{Journal of Nonlinear Science}, 30(5), 2091--2145.

\bibitem{rowley2009}
Rowley, C. W., Mezi\'c, I., Bagheri, S., Schlatter, P., \& Henningson, D. S. (2009).
Spectral analysis of nonlinear flows.
\textit{Journal of Fluid Mechanics}, 641, 115--127.

\bibitem{schmid2010}
Schmid, P. J. (2010).
Dynamic mode decomposition of numerical and experimental data.
\textit{Journal of Fluid Mechanics}, 656, 5--28.

\bibitem{williams2015}
Williams, M. O., Kevrekidis, I. G., \& Rowley, C. W. (2015).
A data-driven approximation of the Koopman operator: Extending dynamic mode decomposition.
\textit{Journal of Nonlinear Science}, 25, 1307--1346.

\bibitem{sidiropoulos2000}
Sidiropoulos, N. D., \& Bro, R. (2000).
On the uniqueness of multilinear decomposition of $N$-way arrays.
\textit{Journal of Chemometrics}, 14(3), 229--239.

\bibitem{mezic2005nd}
Mezi\'c, I. (2005).
Spectral properties of dynamical systems, model reduction and decompositions.
\textit{Nonlinear Dynamics}, 41, 309--325.

\bibitem{crnjaric2020}
\v{C}rnjari\'c-\v{Z}ic, N., Ma\'ce\v{s}i\'c, S., \& Mezi\'c, I. (2020).
Koopman operator spectrum for random dynamical systems.
\textit{Journal of Nonlinear Science}, 30, 2007--2056.

\bibitem{arbabi2017}
Arbabi, H., \& Mezi\'c, I. (2017).
Ergodic theory, dynamic mode decomposition, and computation of spectral properties of the Koopman operator.
\textit{SIAM Journal on Applied Dynamical Systems}, 16(4), 2096--2126.

\bibitem{stark2003}
Stark, J., Broomhead, D. S., Davies, M. E., \& Huke, J. (2003).
Delay embeddings for forced systems. II. Stochastic forcing.
\textit{Journal of Nonlinear Science}, 13, 519--577.

\bibitem{Brunton2017HAVOK}
Brunton, S. L., Brunton, B. W., Proctor, J. L., Kaiser, E., \& Kutz, J. N. (2017).
Chaos as an intermittently forced linear system.
\textit{Nature Communications}, 8, 19.

\bibitem{Brunton2015CS}
Brunton, S. L., Proctor, J. L., Tu, J. H., \& Kutz, J. N. (2015).
Compressed sensing and dynamic mode decomposition.
\textit{Journal of Computational Dynamics}, 2(2), 165--191.

\bibitem{GappyPOD}
Everson, R., \& Sirovich, L. (1995).
Karhunen--Lo\`eve procedure for gappy data.
\textit{Journal of the Optical Society of America A}, 12(8), 1657--1664.

\bibitem{Surana2016}
Surana, A., \& Banaszuk, A. (2016).
Linear observer synthesis for nonlinear systems using Koopman operator framework.
\textit{IFAC-PapersOnLine}, 49(18), 716--723.

\bibitem{Iliopoulos2016}
Iliopoulos, A., Shirzadeh, R., Weijtjens, W., Guillaume, P., Van Hemelrijck, D., \& Devriendt, C. (2016).
A modal decomposition and expansion approach for prediction of dynamic responses on a monopile offshore wind turbine
using a limited number of vibration sensors.
\textit{Mechanical Systems and Signal Processing}, 68--69, 84--104.

\bibitem{Maes2016}
Maes, K., Iliopoulos, A., Weijtjens, W., Devriendt, C., \& Lombaert, G. (2016).
Dynamic strain estimation for fatigue assessment of an offshore monopile wind turbine using filtering and modal
expansion algorithms.
\textit{Mechanical Systems and Signal Processing}, 76--77, 592--611.

\bibitem{Lourens2012}
Lourens, E.-M., Reynders, E., De Roeck, G., Degrande, G., \& Lombaert, G. (2012).
An augmented Kalman filter for force identification in structural dynamics.
\textit{Mechanical Systems and Signal Processing}, 27, 446--460.

\bibitem{Rogers2020}
Rogers, T. J., Worden, K., \& Cross, E. J. (2020).
On the application of Gaussian process latent force models for joint input-state-parameter estimation: With a view to
Bayesian operational identification.
\textit{Mechanical Systems and Signal Processing}, 140, 106580.

\bibitem{Brincker2001FDD}
Brincker, R., Zhang, L., \& Andersen, P. (2001).
Modal identification of output-only systems using frequency domain decomposition.
\textit{Smart Materials and Structures}, 10(3), 441--445.

\bibitem{Burton2011}
Burton, T., Jenkins, N., Sharpe, D., \& Bossanyi, E. (2011).
\textit{Wind Energy Handbook} (2nd ed.). Wiley.

\bibitem{AndersonMoore1979}
Anderson, B. D. O., \& Moore, J. B. (1979).
\textit{Optimal Filtering}. Prentice-Hall.

\bibitem{Wright2020DT}
Wright, L., \& Davidson, S. (2020).
How to tell the difference between a model and a digital twin.
\textit{Advanced Modeling and Simulation in Engineering Sciences}, 7, 13.

\bibitem{Kapteyn2021DT}
Kapteyn, M. G., Pretorius, J. V. R., \& Willcox, K. E. (2021).
A probabilistic graphical model foundation for enabling predictive digital twins at scale.
\textit{Nature Computational Science}, 1, 337--347.

\bibitem{NASEM2024DT}
National Academies of Sciences, Engineering, and Medicine (2024).
\textit{Foundational Research Gaps and Future Directions for Digital Twins}.
The National Academies Press, Washington, DC. doi:10.17226/26894.

\bibitem{Fowler2023FOCAL}
Fowler, M., Lenfest, E., Viselli, A., Goupee, A., Kimball, R., Bergua, R., Wang, L., Zalkind, D., Wright, A., \&
Robertson, A. (2023).
Wind/wave testing of a 1:70-scale performance-matched model of the IEA Wind 15 MW reference wind turbine with
real-time ROSCO control and floating feedback.
\textit{Machines}, 11(9), 865.

\bibitem{Fowler2023ISOPE}
Fowler, M. L., Lenfest, E., Viselli, A., Goupee, A. J., Kimball, R. W., Zalkind, D., Wright, A., Bergua, R., Wang, L.,
\& Robertson, A. (2023).
1:70-scale model testing of the Reference OpenSource Controller (ROSCO) on the IEA-Wind 15MW reference wind turbine
including floating feedback.
In \textit{Proceedings of the 33rd International Ocean and Polar Engineering Conference (ISOPE)}. NREL/CP-5000-85444.

\bibitem{FOCALdata}
National Renewable Energy Laboratory \& Pacific Northwest National Laboratory.
\textit{FOCAL Campaign IV: Integrated System Control: Turbine + Hull}.
Dataset, Atmosphere to Electrons (A2e) Data Archive and Portal / Wind Data Hub, U.S. Department of Energy, OSTI ID
1972267. Available at: \url{https://www.osti.gov/biblio/1972267} (Campaign IV runs m02d01t01.ac01, wind w02, waves
e11, e13, e14, e15; accessed October 2026).

\end{thebibliography}
\end{document}